\documentclass[showpacs,aps,prx,amsmath,amssymb,floatfix,twocolumn]{revtex4-1}

\usepackage{graphicx,epstopdf}
\usepackage{bm}
\usepackage{mathrsfs}
\usepackage{color}
\usepackage[bookmarks,colorlinks,linkcolor=blue,anchorcolor=black,urlcolor=blue,citecolor=blue]{hyperref}
\usepackage{natbib}
\usepackage{CJK}
\usepackage{lipsum}

\allowdisplaybreaks[3]

\begin{document}
\begin{CJK*}{UTF8}{gbsn}

\title{Chiral Quantum-Optical Elements for Waveguide-QED with Sub-wavelength Rydberg-Atom Arrays}
\author{Lida Zhang~(张理达)$^{1}$}
\author{Fan Yang~(杨帆)$^{2}$}
\author{Klaus M{\o}lmer$^{2}$}
\author{Thomas Pohl$^{3}$}
\affiliation{$^{1}$School of Physics, East China University of Science and Technology, Shanghai, 200237, China}
\affiliation{$^{2}$Center for Hybrid Quantum Networks, Niels Bohr Institute,
University of Copenhagen, Blegdamsvej 17, DK-2100 Copenhagen, Denmark}
\affiliation{$^{3}$Institute for Theoretical Physics, TU Wien, Wiedner Hauptstra\ss e 8-10/136, A-1040 Vienna, Austria}

\begin{abstract}
We describe an approach to achieve near-perfect unidirectional light-matter coupling to an effective quantum emitter that is formed by a subwavelength array of atoms in the Rydberg-blockade regime. The nonlinear reflection and transmission of such two-dimensional superatoms are exploited in different interferometric setups for the deterministic generation of tunable single photons and entangling two-photon operations with high fidelities, $\mathcal{F}\gtrsim0.999$. The described setup can  function as a versatile nonlinear optical element in a free-space photonic quantum network  with  simple linear elements and without the need of additional mode confinement, optical resonators, or optical isolators.  
\end{abstract}

\maketitle
\end{CJK*}
\section{Introduction}\label{sec:intro}
While the use of photons for long-distance communication is well established, their application in quantum technologies for quantum computing \cite{Slussarenko2019APR,Bartlett2021Optica}, simulation \cite{Walther2012,Hartmann_2016}, and metrology \cite{Giovannetti2011NPhoton,Polino2020AVS} has been rapidly advancing in recent years. All such applications typically require a light-matter interface with a nonlinear optical response in order to generate and manipulate nonclassical states of light. Realizing the strongest possible nonlinearity that acts at the level of single light-quanta would enable the deterministic processing of photonic quantum information.
These prospects have motivated substantial research efforts towards reaching this extreme limit of quantum nonlinear optics \cite{chang2014}, using single atoms trapped in high-finesse cavities~\cite{Birnbaum2005Nature,Reiserer2015PRL,Welte2018PRX}, quantum dots coupled to photonic nanostructures~\cite{Tiecke2014Nature,Petersen2014Science,Lodahl2015RMP,Chang2018RMP,Yu2019PNAS,Prasad2020NPhoton}, or ensembles of strongly interacting Rydberg atoms~\cite{Peyronel2012Nature,Paris2010PRX,Murray2016Adv,Firstenberg2016JPB,Thompson2017Nature,yang2020atom}. While strong optical nonlinearities have been demonstrated in these systems, the  suppression of  decoherence and photon loss to enable the high-fidelity control and manipulation of single-photon quantum states remains an outstanding challenge. 

Coherent light-matter interfacing, on the other hand, can be achieved with two-dimensional atomic arrays \cite{Facchinetti2016PRL,Bettles2016PRL,Shahmoon2017PRL}, in which photon scattering is suppressed collectively for subwavelength spacing between the atoms. The cooperative response of two-dimensional emitter arrays enables applications as linear optical elements, e.g, for  efficient optical memories and for high-fidelity wavefront shaping \cite{Facchinetti2016PRL,Perczel2017PRL,Guimond2019PRL,Ballantine2020PRL,Grankin2018PRA,BaSSler:23}. The strong collective photon coupling also permits to generate strong nonlinearities by using multiple arrays \cite{pedersen2023} or by exploiting the strong interaction between atomic Rydberg states in the array \cite{Bekenstein2020NPhys,Moreno2021PRL,Zhang2022Quantum}.

\begin{figure}[t!]
 \centering 
\includegraphics[width=0.49\textwidth]{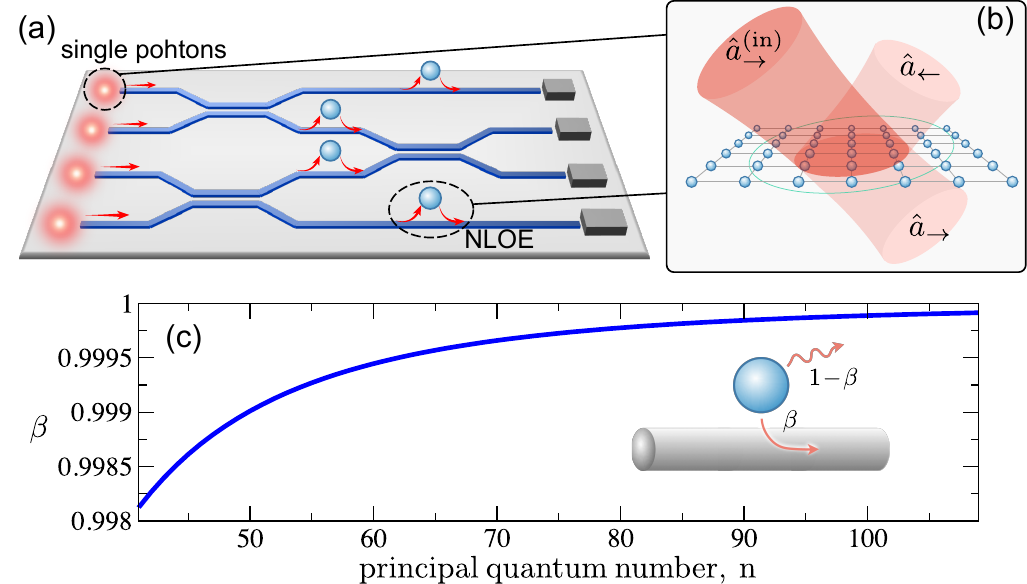} 
 \caption{(a) Illustration of a photonic network in which arrays of Rydberg atoms form nonlinear optical elements (NLOE) that are capable of generating tuneable single photons and performing entangling two-photon operations. Illuminating the array at a finite angle offers a simple means for unidirectional coupling whereby an incident quantum state in an input mode ($\hat{a}_\rightarrow^{({\rm in})}$) is transferred into to two output modes, $\hat{a}_\rightarrow$ and $\hat{a}_\leftarrow$, that all propagate along the same direction. Such a mode-selective coupling is possible with a high efficiency $\beta\sim1$, which is shown in (c) as function of the  principal quantum number of $nS$-states of Rubidium atoms in an array with subwavelength spacing of $a=0.65\lambda$.}
 \label{fig1}
\end{figure}

Here, we combine these features and propose a complete toolbox for the high-fidelity deterministic generation and processing of photonic quantum states, based on sub-wavelength arrays of three-level Rydberg atoms. We identify optimal driving schemes and parameters that yield strong single-mode photon-coupling with  virtually no losses, while maximizing the effect of Rydberg-Rydberg atom interactions. The extended geometry of the array can be used to achieve unidirectional photon coupling, and thereby approach the ideal limit of chiral waveguide-QED \cite{Sheremet23}, in which a single saturable quantum emitter is perfectly coupled to a single propagating photonic mode without any scattering losses. Detailed calculations for finite arrays that include relevant loss and decoherence sources show that unidirectional coupling efficiencies of $\beta\sim0.9993$ should be achievable with typical Rydberg states of ongoing experiments [see Fig.~\ref{fig1}(c)]. We show that this system can work as a versatile nonlinear optical element that offers all functionalities needed  in a photonic quantum network, from the generation of tunable single-photon pulses to high-fidelity two-photon operations [see Figs.~\ref{fig1}(a) and \ref{fig1}(b)], such as photon sorting and conditional phase-gates. Importantly, the Rydberg atom array is intrinsically mode-matched to any incident transverse photonic mode and it can realize nearly perfect entangling two-photon operations with  infidelities below the per mille level.

The article is organized as follows. In Sec.~\ref{sec:3level}, we describe the basic setup of three-level Rydberg-atom arrays and discuss  their linear optical response to an incident continuous wave (cw)-field. In Sec.~\ref{sec:2level}, we outline a simplified description of the array in terms of two-level emitters and discuss conditions to obtain narrow reflection resonances and strong interactions. The optical nonlinearity that results from the Rydberg-blockade between excited atoms is characterized through emerging two-photon correlations in Sec.~\ref{sec:blockade}. In section \ref{sec:chiral}, we  propose a setup to implement unidirectional photon coupling and we consider the interaction with propagating photon pulses in the ensuing section \ref{sec:pulsed}. The application of this setup to generate single photons and perform two-photon operations is detailed  in Sec.~\ref{sec:applications} along with a characterization of its performance for these tasks.

\section{Three-level Rydberg-atom arrays}\label{sec:3level}
\begin{figure}[t!]
 \centering 
\includegraphics[width=0.49\textwidth]{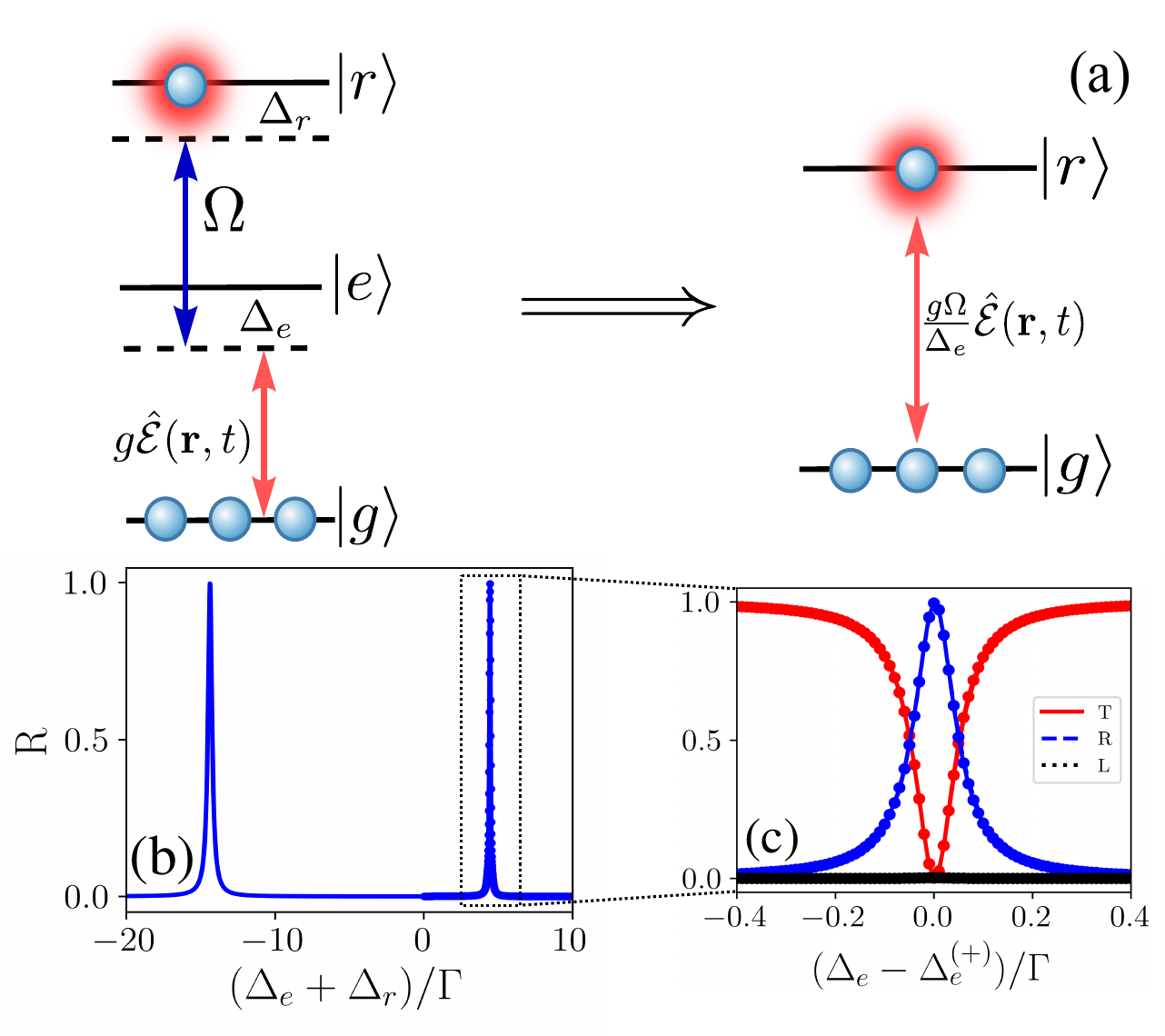} 
 \caption{(a) Illustration of the atomic three-level system and its approximation by a two-level configuration. (b) Comparison of the steady-state linear reflection spectrum as a function of detuning $\Delta_{e}$ determined by the full three-level model and the effective two-level approximation. (c) Detailed comparison of the linear transmission (T), reflection (R), and loss (L) spectrum around the right absorption peak. The dots and the solid lines are obtained with the three-level configuration and the effective two-level approximation, respectively. Here a rounded $21\times21$ array is chosen, where atoms  in the corners are removed to form a circular disc with an atom number of 317. The remaining parameters are $\Omega_p(\mathbf{0})/\Gamma = 0.001, w_{0}=3.0\lambda$, and $\Omega/\Gamma = 8.0, a = 0.75\lambda, \Delta_{r}/\Gamma = -10.0$.
 }
 \label{fig2}
\end{figure}

The nonlinear optical elements discussed above are formed by two-dimensional arrays of $N$ three-level atoms at positions ${\bf r}_i$, $i=1,...,N$. The level scheme of the atoms is illustrated in Fig.\ref{fig2}(a).
The incident probe light is described by the quantum field $\hat{\mathcal{E}}(\mathbf{r})$ that satisfies bosonic commutation relations, $[\hat{\mathcal{E}}(\mathbf{r}), \hat{\mathcal{E}}^{\dagger}(\mathbf{r}^\prime)]=\delta(\mathbf{r}-\mathbf{r}^\prime)$, and defines the spatial photon density as $\langle\hat{\mathcal{E}}^{\dagger}(\mathbf{r})\hat{\mathcal{E}}(\mathbf{r})\rangle$.  The quantum field drives the transition between the ground state $|g\rangle$ and  excited state $|e\rangle$ of each atom in the array. The probe light excites the atoms with a frequency detuning $\Delta_e$ and coupling strength $g=\lambda\sqrt{3 \Gamma c/(8\pi)}$ that is determined by the decay rate $\Gamma$ and resonant wavelength $\lambda$ of the  $|e\rangle$-$|g\rangle$ transition. In addition, a classical control laser field  couples the state $|e\rangle$ to a high-lying Rydberg state $|r\rangle$ with a corresponding control Rabi frequency $\Omega$ and frequency detuning $\Delta_r$. The Hamiltonian 
 \begin{align}\label{eq:H0}
 \hat{H}_{0} =& -\sum^{N}_{j=1}\big(\Delta_{e}\hat{\sigma}^{(j)}_{ee} + (\Delta_{e}+\Delta_{r})\hat{\sigma}^{(j)}_{rr} \nonumber\\
 & \quad\quad\quad + [g\hat{\mathcal{E}}(\mathbf{r}_{j})\hat{\sigma}^{(j)}_{eg} + \Omega \hat{\sigma}^{(j)}_{re} + h.c.]\big) 
\end{align}
describes the coupling of the atoms to the two applied light fields within the rotating wave approximation. 

Moreover, the atoms interact with the free-space radiation field, whereby multiple scattering of photons on the $|g\rangle-|e\rangle$ transition of atoms across the array gives rise to collective light-matter interactions. Such processes can be described effectively by integrating out the photonic dynamics within the Born-Markov approximation \cite{Gross1982PhysRep,Dung2002PRA,Ficek2002PhysRep}. This yields a master equation 
\begin{align}\label{eq:ME}
\partial_t \hat{\rho} =-i\left[\hat{H}_0+\hat{H}_{\rm dd},\hat{\rho}\right]+\mathcal{L}_{\rm dd}(\hat{\rho})+\mathcal{L}_{\rm Ryd}(\hat{\rho})
\end{align}
for the many-body density matrix of the atoms, where 
\begin{align}\label{eq:Hdd}
\hat{H}_{\text{dd}} = &- \sum^{N}_{j,k\neq j}J_{jk} \hat{\sigma}^{(j)}_{eg}\hat{\sigma}^{(k)}_{ge} 
\end{align}
describes the coherent exchange of atomic excitations due to emission and re-absorption of photons, while the Lindblad operator
\begin{align}\label{eq:Ldd}
\mathcal{L}_{\rm dd}(\rho)=& \sum^{N}_{j,k}\frac{\Gamma_{jk}}{2}(2\hat{\sigma}^{(k)}_{ge}\rho\hat{\sigma}^{(j)}_{eg} - \{\hat{\sigma}^{(j)}_{eg}\hat{\sigma}^{(k)}_{ge},\rho\})
\end{align}
captures the dynamics of corresponding collective decay processes that emerges from the light-induced dipole-dipole interactions between the atoms. The interaction coefficients
\begin{align}
J_{jk} + i\frac{\Gamma_{jk}}{2} = \frac{\mu_0 \omega^{2}_{eg}}{\hbar}{\bm{d}}^{*}_{eg}{\bm{G}}(\mathbf{r}_{j},\mathbf{r}_{k},\omega_{eg}) {\bm{d}}_{eg}
\end{align}
are related to the Green's function tensor, $\bm{G}$, of the free-space electromagnetic field at the resonance frequency, $\omega_{eg}=2\pi c/\lambda$, and to the transition dipole moment $\bm{d}_{eg}$ of the $|e\rangle-|g\rangle$ transition. While the transition dipole moment of the $|e\rangle-|r\rangle$ Rydberg-state transition is too small to cause significant atomic excitation-exchange interactions, we include the spontaneous decay of the Rydberg-state with a rate $\gamma$ via the Lindbladian 
\begin{align}
\mathcal{L}_{\rm Ryd}(\rho)=& \sum_{j}\frac{\gamma}{2}(2\hat{\sigma}^{(j)}_{er}\hat{\rho}\hat{\sigma}^{(j)}_{re} - \{\hat{\sigma}^{(j)}_{re}\hat{\sigma}^{(j)}_{er},\hat{\rho}\}).
\end{align}
 Having obtained the solution of Eq.(\ref{eq:ME}) for the density matrix of the atoms, we can reconstruct the quantum state of the light scattered by the atomic array. Let's consider a single-mode input field, whereby all incident photons occupy a single spatial mode $u_\rightarrow({\bf r})$ or $u_\leftarrow({\bf r})$ that obeys the paraxial wave equation $[4\pi i\partial_z+\lambda(\partial_x^2+\partial_y^2)+8\pi^2\lambda^{-1}]u_\rightarrow=0$ and $[4\pi i\partial_z-\lambda(\partial_x^2+\partial_y^2)-8\pi^2\lambda^{-1}]u_\leftarrow=0$ for forward ($u_\rightarrow$) and backward ($u_\leftarrow$) propagating light. This equation describes the propagation of light along the $z$-axis and the operators $\hat{a}^{({\rm in})}_{\rightarrow}=\int{\rm d}{\bf r}\:u^*_\rightarrow({\bf r})\hat{\mathcal{E}}({\bf r})$ and $\hat{a}^{({\rm in})}_{\leftarrow}=\int{\rm d}{\bf r}\:u^*_\leftarrow({\bf r})\hat{\mathcal{E}}({\bf r})$ shall define the occupation of photons in the incident field, propagating in the forward and backward direction, respectively. One can then obtain simple input-output relations \cite{Manzoni2018NJP}
\begin{subequations}\label{eq:photons}
 \begin{align}
\label{eq:photons_a}  \hat{a}_{\rightarrow}(t)=&\hat{a}^{(\text{in})}_{\rightarrow}(t)+i\frac{g}{c}\sqrt{\ell}\sum_{j}u^{*}(\mathbf{r}_j)\hat{\sigma}_{ge}^{(j)}(t),\\
\label{eq:photons_b}\hat{a}_{\leftarrow}(t)=&\hat{a}^{(\text{in})}_{\leftarrow}(t)+i\frac{g}{c}\sqrt{\ell}\sum_{j}u^{*}(\mathbf{r}_j)\hat{\sigma}_{ge}^{(j)}(t)
 \end{align}
\end{subequations}
for the total light field in the modes $u_\rightarrow$ and $u_\leftarrow$, with $u_\rightarrow({\bf r}_j)=u_\leftarrow({\bf r}_j)=u({\bf r}_j)/\sqrt{\ell}$ in the plane of the array. Here, $\ell^{-1}=\int{\rm d}{\bf r}_\perp |u_\rightarrow({\bf r})|^2$ defines the inverse length of the quantization volume for the free-space field modes along the propagation direction, such that the transverse mode function is normalized as $\int{\rm d}{\bf r}_\perp |u({\bf r})|^2 =1$, where we have used ${\bf r}_\perp=(x,y)$ for the transverse coordinates.
We consider coherent driving such that one can replace $\hat{\mathcal{E}}$ by $\mathcal{E}=\langle\hat{a}_\rightarrow^{({\rm in)}}\rangle u_\rightarrow({\bf r})+\langle\hat{a}_\leftarrow^{({\rm in)}}\rangle u_\leftarrow({\bf r})$, which defines the probe Rabi frequency $\Omega_p({\bf r})=g\mathcal{E}({\bf r})$.
The operators $ \hat{a}_{\rightarrow}$ and  $\hat{a}_{\leftarrow}$, respectively, describe the transmitted ($z>0$) and reflected ($z<0$) photons in the incident spatial modes. 
For an exclusively forward propagating input field ($\langle\hat{a}^{\dagger({\rm in})}_\leftarrow\hat{a}^{({\rm in})}_\leftarrow\rangle=0$), we can define the reflectivity $R = \langle\hat{a}^{\dagger}_{\leftarrow}\hat{a}_{\leftarrow}\rangle/\langle\hat{a}^{\dagger({\rm in})}_\rightarrow\hat{a}^{({\rm in})}_\rightarrow\rangle$ and transmission $T = \langle\hat{a}^{\dagger}_{\rightarrow}\hat{a}_{\rightarrow}\rangle/\langle\hat{a}^{\dagger({\rm in})}_\rightarrow\hat{a}^{({\rm in})}_\rightarrow\rangle$ of photons in the mode $u_\rightarrow({\bf r})$ and $u_\leftarrow({\bf r})$. 
 Correspondingly, the loss coefficient is given by $L=1-T-R$ and accounts for the scattering of photons into other transversal modes. 

For a single incident photon, Eq.(\ref{eq:ME}) only involves a single atomic excitation and can be readily solved. The resulting expression \cite{Bekenstein2020NPhys,Zhang2022Quantum}
\begin{align}\label{eq:R3l}
 R = \left|\frac{g^{2}}{ca^2}\frac{1}{\Delta_{e}+\Delta_{c}-\frac{\Omega^2}{\Delta_{e}+\Delta_{r} + i\gamma/2}+i\frac{\Gamma_{c}}{2}}\right|^2,
\end{align}
for the reflection coefficient involves the collective level shift, $\Delta_c$, and collective decay rate, $\Gamma_c$, of the atomic array respectively, which arise from the dipole-dipole interaction in Eqs.(\ref{eq:Hdd}) and (\ref{eq:Ldd}).
For large atomic lattices and broad input beams, the collective Lamb shift is given by $\Delta_c=\sum_{j\neq i} J_{i,j}$, while the collective decay rate approaches $\Gamma_{\rm c}=\frac{3}{4\pi}\frac{\lambda^2}{a^2}\Gamma=\frac{2g^2}{c a^2}$ for high atom numbers, $N\rightarrow\infty$, and large beam waists, $w_0\gg\lambda$. 

Equation (\ref{eq:R3l}) is valid for an infinitely extended array and a vanishing transverse photon momentum across the array. In the numerical calculations we consider incident beams with a Gaussian mode profile
\begin{equation}
u({\bf r}_\perp,z=0)=\sqrt{\frac{2}{\pi w_0^2}}\exp\left(-\frac{r_\perp^2}{w_0^2}\right)
\end{equation}
where ${\bf r}_\perp=(x,y)$, $z=0$ defines the position within the plane of the array, and $w_0$ denotes the Gaussian waist of the beam. We consider square lattices, where all atoms outside a given radius are removed to form a circularly shaped array. This reduces the atom number at a given system size and thereby facilitates the numerical simulations of larger lattices.
  
A typical reflection spectrum obtained from such a calculation is shown in Fig.\ref{fig2}(b). The spectrum features two Autler-Townes reflection peaks and a minimum around two-photon resonance, $\Delta_e+\Delta_r=0$, where electromagnetically induced transparency~\cite{Fleischhauer2005RMP} enhances transmission. At this point transmission is only limited by  Rydberg-state decay which yields a small reflection coefficient $R\approx (\frac{g^{2}}{ca^2} \frac{\gamma}{2\Omega^2})^2$ that vanishes with increasing principal quantum number $n$ as the Rydberg-state decay rate, $\gamma$, decreases. In addition one finds two reflection resonances at the probe detunings 
\begin{equation}
\Delta_{e}^{(\pm)}=[-(\Delta_{r}+\Delta_{c})\pm\sqrt{(\Delta_{r}-\Delta_{c})^2+4\Omega^2})]/2,
\end{equation}
which are determined by the Rydberg-state detuning $\Delta_r$ and the collective level shift $\Delta_c$. The two probe detunings $\Delta_e=\Delta_e^{(\pm)}$ correspond to the Autler-Townes resonances of the three-level atoms, which become well separated for large control Rabi frequencies $\Omega\gg \Gamma_c$.

\section{Reflection resonances and two-level limit}\label{sec:2level}
The two Autler-Townes resonances correspond to the energies of the dressed two-level system, formed by the two atomic states $|e\rangle$ and $|r\rangle$ that are coupled by the control-field Rabi frequency $\Omega$ [see Fig.\ref{fig2}(a)]. For large Rydberg-state detunings $\Delta_r$, this coupling is far off resonant such that state-mixing is small. Consequently, one of the two dressed states is primarily composed of the intermediate state $|e\rangle$, which yields a broad reflection resonance, while the other predominantly consists of the Rydberg-state state $|r\rangle$ and generates a narrow reflection resonance. Hence, the narrow resonance is most sensitive to Rydberg-Rydberg atom interactions and therefore yields ideal conditions to exploit the Rydberg blockade for generating large optical nonlinearities.

For large Rabi frequencies $\Omega$ and correspondingly large values of $\Delta_e$, one can adiabatically eliminate the $|e\rangle$-state dynamics around the narrow Autler-Townes resonance. This yields an approximate master equation 
\begin{align}\label{eq:ME2}
\partial_t \hat{\rho} =-i\left[\hat{H},\hat{\rho}\right]+\mathcal{L}(\hat{\rho})
\end{align}
for the driven dynamics of the ground and Rydberg state with an effective two-level Hamiltonian (see Appendix \ref{app:A})
\begin{align}\label{eq:H2l}
\hat{H}\approx&-\sum_i \bar\Delta\hat{\sigma}^{(i)}_{rr} + [\bar{g}\hat{\mathcal{E}}(\mathbf{r}_{i})\hat{\sigma}^{(i)}_{rg} + h.c. ]- \sum_{j\neq i}\bar{J}_{ij}\hat{\sigma}^{(i)}_{gr}\hat{\sigma}^{(j)}_{rg}, \nonumber\\
\end{align}
and Lindblad operator
\begin{align}\label{eq:L2l}
 \mathcal{L}(\rho)=&\sum_{i,j}\frac{\bar{\Gamma}_{ij}+\gamma\delta_{ij}}{2}(2\hat{\sigma}^{(j)}_{gr}\rho\hat{\sigma}^{(i)}_{rg} - \{\hat{\sigma}^{(i)}_{rg}\hat{\sigma}^{(j)}_{gr},\rho\}).
 \end{align}
As indicated in Fig.\ref{fig2}(b), the effective frequency detuning, single-photon coupling strength and induced dipole-dipole interactions are, respectively, given by
$\bar{\Delta}=\Delta_{e}+\Delta_{r} - \frac{\Omega^{2}}{\Delta_{e}}$, $\bar{g} = -g\Omega/\Delta_{e}$, $\bar{J}_{ij}=J_{ij}\Omega^{2}/\Delta^2_{e}$, and $\bar{\Gamma}_{ij}=\Gamma_{ij}\Omega^2/\Delta^2_{e}$. Similarly, the photon field can now be obtained as (see Appendix \ref{app:A})
\begin{subequations}\label{eq:IO2l}
 \begin{align}
\hat{a}_{\rightarrow}(t)=&\hat{a}^{(\text{in})}_{\rightarrow}(t)+i\frac{\bar{g}}{c}\sqrt{\ell}\sum_{j}u^{*}(\mathbf{r}_j)\hat{\sigma}_{gr}^{(j)}(t),\\
\hat{a}_{\leftarrow}(t)=&\hat{a}^{(\text{in})}_{\leftarrow}(t)+i\frac{\bar{g}}{c}\sqrt{\ell}\sum_{j}u^{*}(\mathbf{r}_j)\hat{\sigma}_{gr}^{(j)}(t),
 \end{align}
\end{subequations}
from the collective polarization between the ground state and Rydberg state. Figure \ref{fig2}(c) compares the linear transmission and reflection obtained from  the exact three-level calculation with Eqs.(\ref{eq:H2l})-(\ref{eq:IO2l}) and demonstrates excellent agreement with the two-level approximation around the narrow Autler-Townes resonance. Here the reflection spectrum can be written 
\begin{align}\label{eq:Rnarrow2l}
 R_0 = \left|\frac{\bar{\Gamma}_{c}/2}{\bar{\Delta}+\bar{\Delta}_{c}+i\frac{\bar{\Gamma}_{c}+\gamma}{2}}\right|^2,
\end{align}
as a function of the detuning $\bar{\Delta}$, the small linewidth $\bar{\Gamma}_c=(\Omega/\Delta_e)^2\Gamma_c$ and the collective level shift $\bar{\Delta}_c=\sum_{j\neq i} \bar{J}_{ij}$. When the photon detuning $\Delta_e$ is varied over the narrow range, $\bar{\Gamma}_c$, of the reflection resonance, both $\bar{\Gamma}_c$ and $\bar{\Delta}_c$ are almost unchanged and Eq.(\ref{eq:Rnarrow2l}) is well approximated by a Lorentzian.  For a fixed value of the ratio $\Omega/\Delta_e$ that controls the values of the effective parameters in Eqs.(\ref{eq:H2l}) and (\ref{eq:L2l}), the accuracy of the effective two-level description improves upon increasing $\Omega$ and $\Delta_r$, and is, therefore, only limited by the experimentally available Rydberg-laser power. For the parameters considered here, the reflection spectra agree to within less than $1\%$. These conditions, yield effective two-level Rydberg-atom mirrors with near-perfect reflection resonances for typical parameters of current experiments with cold Rubidium Rydberg atoms \cite{Rui2020Nature,Browaeys2020NPhys,Ebadi2021Nature,Scholl2021Nature,Semeghini2021Science,Satzinger2021Science}.

\section{Rydberg blockade and single-photon saturation}\label{sec:blockade}
Using this effective description in terms of  Rydberg-state two-level systems, we can now explore the effects of interactions between the excited Rydberg states on the nonlinear optical response of the atomic array. The static polarizability of the atoms increases rapidly with their principal quantum number, $n$, such that Rydberg states feature strong van der Waals interactions 
$U(\mathbf{r}_i,\mathbf{r}_j)=C_6/|\mathbf{r}_j-\mathbf{r}_i|^6$, with an interaction coefficient, $C_6\sim n^{11}$, which can exceed that of ground state atoms by many orders of magnitude \cite{Saffman2010RMP}. The resulting level shifts of interacting pairs of Rydberg atoms can be sufficiently large to inhibit the excitation of more than a single atom within a diameter of several $\mu$m \cite{Lukin2001PRL,Jaksch2000PRL}. This Rydberg blockade has been demonstrated and exploited in several experiments and a range of different applications from quantum information processing \cite{Saffman2010RMP} to nonlinear optics \cite{Firstenberg2016JPB,Murray2016Adv} and quantum simulations \cite{Browaeys2020NPhys,Wu2021CPB}.

In the current setting, the Rydberg blockade causes a nonlinear reflection by the atomic array that is highly sensitive to the number of incident photons \cite{Dudin2012Science,Peyronel2012Nature,Mandoki2017PRX,Ripka2018Science}. Under the conditions outline above, a single resonant photon ($\bar{\Delta}=-\bar{\Delta}_c$) will undergo reflection with near-unit probability, $R\sim1$ [cf. Eq.(\ref{eq:Rnarrow2l})]. During this process the photon generates a Rydberg excitation which, for a time $\sim \bar{\Gamma}_c^{-1}$, blocks the coupling of other atoms to the Rydberg state. The blockade, therefore, exposes all subsequently incident photons to the optical response of the bare $|g\rangle-|e\rangle$ transition. The corresponding reflection coefficient for the blockaded arrays is thus given by
\begin{align}\label{eq:R2l}
 R_{bl} = \left|\frac{g^{2}}{ca^2}\frac{1}{\Delta_{e}+\Delta_{c}+i\frac{\Gamma_{c}}{2}}\right|^2.
\end{align}

The condition $\bar{\Delta}=-\bar{\Delta}_c$, for which $R_0\sim1$, implies a large single-photon detuning $|\Delta_e|>\Omega$ for large control Rabi frequencies $\Omega$. Consequently, the blockade generates a low reflection coefficient $R_{bl}< (\Gamma_c/\Omega)^2/4\ll1$. One thus obtains an efficient and nearly lossless mechanism for switching between near-unit reflection and near-unit transmission by a single photon. 

\begin{figure}[t!]
 \centering
 \includegraphics[width=0.49\textwidth]{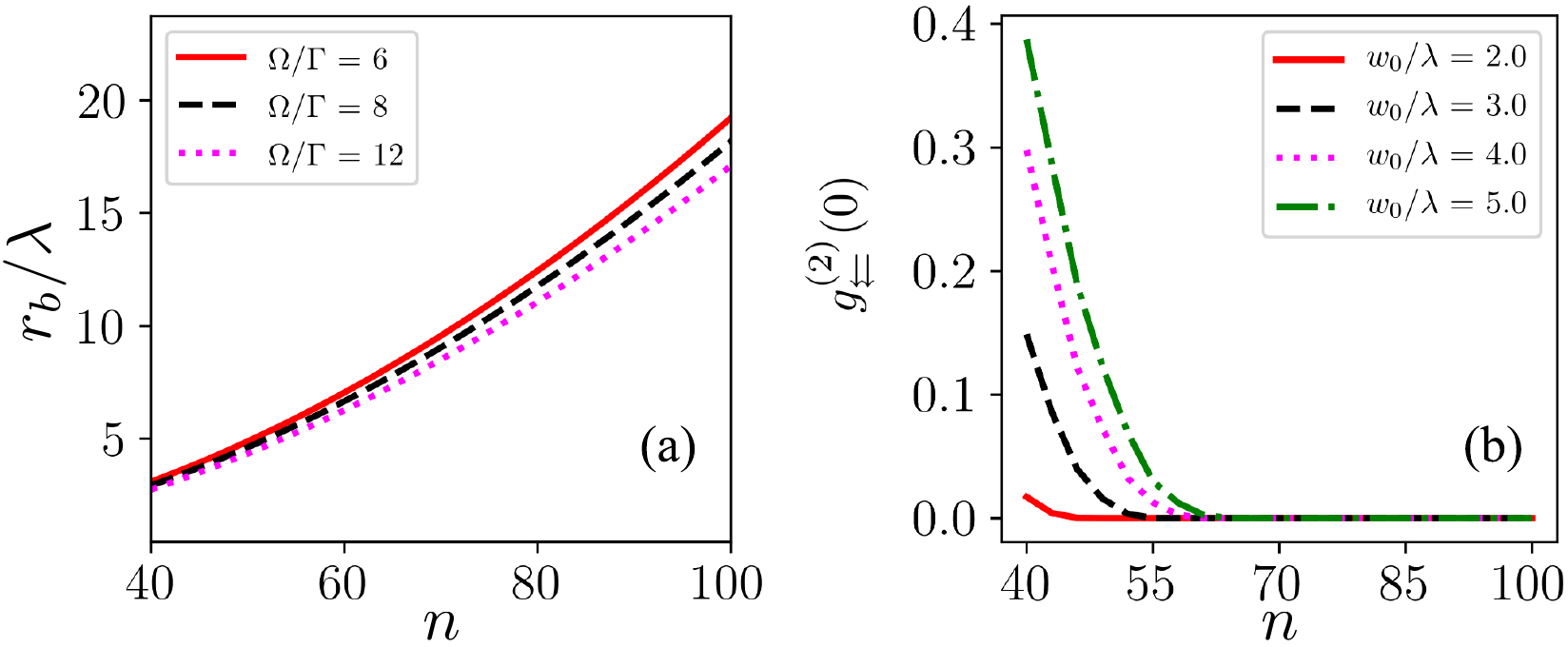} 
 \caption{(a) The Rydberg blockade radius for different principal quantum numbers $n$ of the Rydberg-state. The steady-state correlation function $g_{\leftleftarrows}^{(2)}(0)$ for simultaneously reflected photons is shown in panel (b) as a function of $n$ and for different values $w_0$ of the waist of the incident probe beam. The rapid drop of $g_{\leftleftarrows}^{(2)}$ indicates the onset of an efficient blockade when the blockade radius $r_b$ exceeds $w_0$. Parameters are $\Omega_p({\bf 0})/\Gamma = 0.001$, $\Omega/\Gamma = 8.0$, $a = 0.75\lambda$, and $\Delta_{r}/\Gamma = -10.0$.}
 \label{fig3}
\end{figure}

The blockade is effective if the van der Waals interaction exceeds the linewidth $\bar{\Gamma}_c+\gamma$ of the effective two-level mirror described by the reflection spectrum $R_0$, Eq.(\ref{eq:Rnarrow2l}). 
This defines the blockade radius, $r_{b}$ as 
\begin{align}\label{eq:Rb}
\frac{C_6}{r_b^6}=\bar{\Gamma}_c,
\end{align}
which we show in Fig.\ \ref{fig3}(a) as a function of the principal quantum number for $nS$-states of Rubidium atoms \cite{Li2003PRA,Beterov2009PRA}. For typical values of $n$ and $\Omega$, the narrow linewidth of the reflection resonances yields sizeable blockade radii that significantly exceed the wavelength $\lambda=780$nm of the atomic $D_2$-line in Rubidium and, therefore, covers a large number of atoms in the subwavelength array. Such large blockade radii  permit to maintain the Rydberg blockade for probe fields with a sufficiently large beam waist $w_0$ to suppress losses and imperfections due to finite transverse momenta of the incident field.

We can employ the equal-time correlation function
\begin{align}\label{eq:g20}
g_{\alpha \atop \beta}^{(2)}=&\lim_{t\rightarrow\infty} \frac{\langle\hat{a}_{\alpha}^\dagger(t)\hat{a}_{\beta}^\dagger(t)\hat{a}_{\beta}(t)\hat{a}_{\alpha}(t)\rangle}{\langle\hat{a}_{\alpha}^\dagger(t)\hat{a}_{\alpha}(t)\rangle \langle\hat{a}_{\beta}^\dagger(t)\hat{a}_{\beta}(t)\rangle}
\end{align}
of the outgoing light as an optical probe to analyze the efficiency of the Rydberg-excitation blockade. Here, the field operators describe the outgoing photons that propagate in the forward ($\alpha,\beta=\rightarrow$) and backward ($\alpha,\beta=\leftarrow$) direction. Figure \ref{fig3}(b) shows the equal-time correlation function of simultaneously reflected photons as a function of $n$ for different values of $w_0$. The value of $g_{\leftleftarrows}^{(2)}$ drops rapidly with increasing $n$ and vanishes once the blockade radius, $r_b$, exceeds the waist of the Gaussian beam profile. Indeed, this indicates a full excitation blockade of the array, leading to perfect antibunching of the reflected light. Under these conditions, the atomic array turns the Rydberg blockade into a photon blockade, where the inhibition of simultaneous Rydberg excitation suppresses the simultaneous reflection of two or more photons. 

\begin{figure}[t!]
 \centering
 \includegraphics[width=0.48\textwidth]{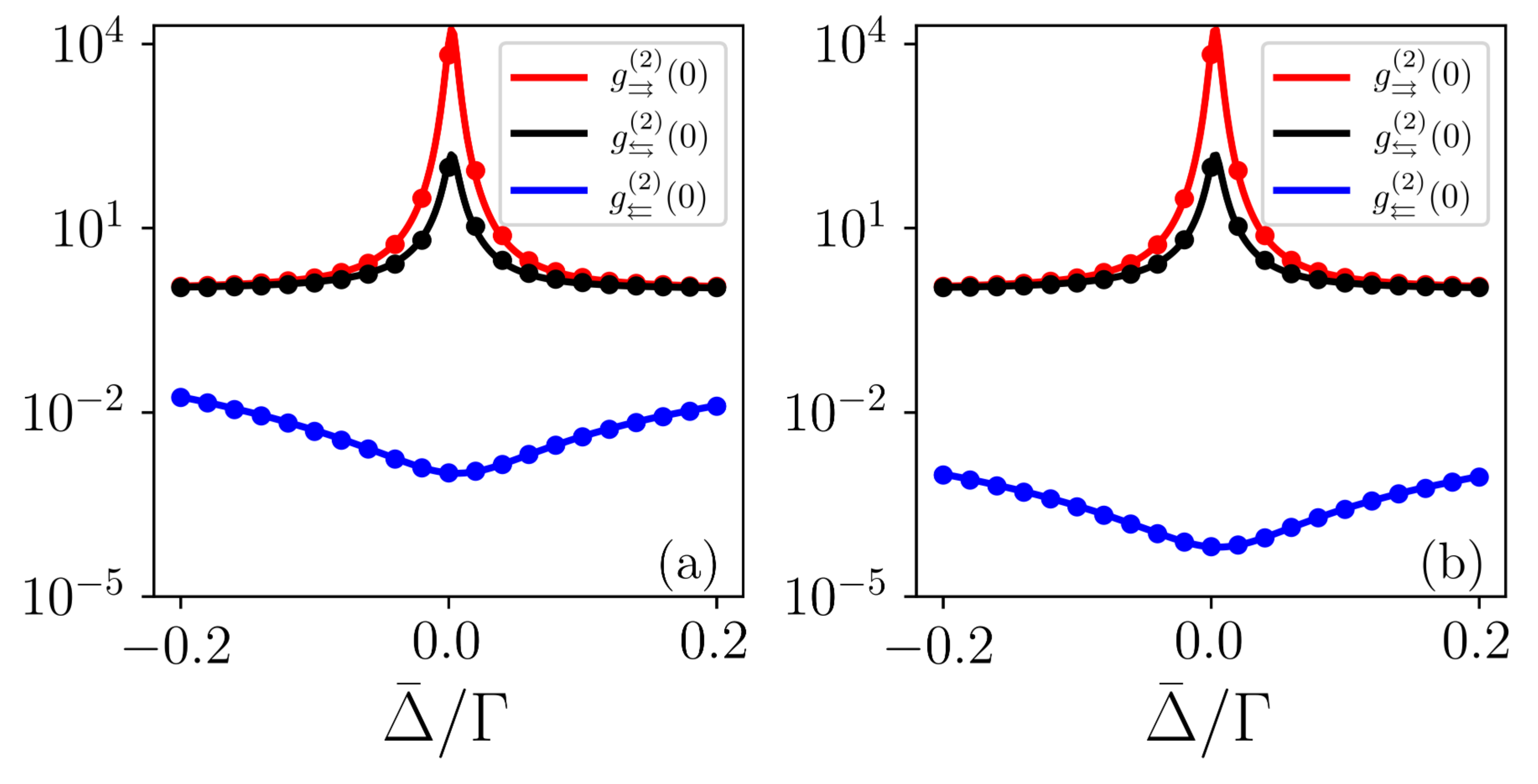} 
 \caption{The pair correlation function for transmitted and reflected photons is shown as a function of the effective probe detuning $\bar{\Delta}$. The  simulation results for three-level atoms (points) agree excellently well with approximate two-level limit (solid lines). The used parameters  are $\Delta_{r}/\Gamma=-10,\Omega/\Gamma = 8$ for (a) and $\Delta_{r}/\Gamma=-40,\Omega/\Gamma = 32$ for (b), as well as $w_{0} = 1.125\lambda$, $n=100 $, $\Omega_p({\bf 0})/\Gamma = 0.001$, and $a = 0.75\lambda$.}
 \label{fig4}
\end{figure}

On the other hand, the transmitted light shows strong photon bunching, i.e. $g_{\rightrightarrows}^{(2)}\gg1$ and $g_{\leftrightarrows}^{(2)}\gg1$ (see Fig.~\ref{fig4}). Since the linear transmission is highly suppressed on resonance, significant photon transmission is only possible due to the Rydberg-blockade induced by multiple coincident photons. 
Figure \ref{fig4} compares the two-photon correlations obtained from the exact three-level simulations of Eqs.(\ref{eq:H0})-(\ref{eq:photons}) with the effective two-level description according to Eqs.(\ref{eq:H2l})-(\ref{eq:IO2l}) for an array of $7\times7$ atoms. The results are virtually identical, and we shall use the more efficient two-level calculations for analyses of larger systems and wider probe beams. Under such conditions, the atomic array can act like a giant saturable quantum emitter with near-perfect single-mode coupling to light and very low photon losses.

\begin{figure}[b!]
\centering
\includegraphics[width=0.4\textwidth]{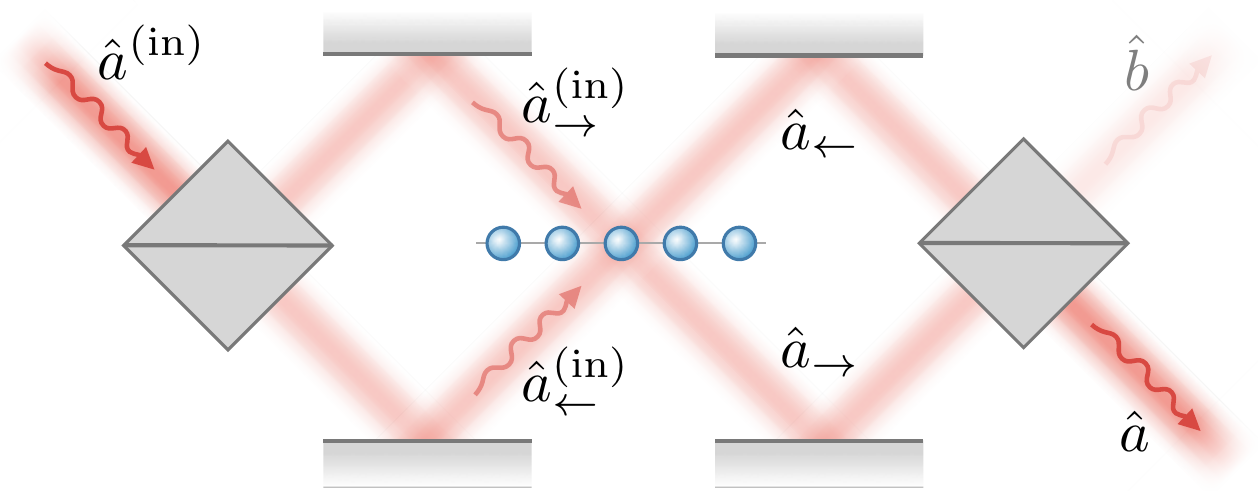} 
\caption{Beam splitter configuration to achieve unidirectional coupling between an incident photon mode $\hat{a}_{\rm in}$ and the atomic array. For perfect alignment one transfers the photonic input state to a single output mode $\hat{a}$, while any signal in the second output port $\hat{b}$ provides an  interferometric measure of imperfections that may permit fine-tuning of the setup.}
\label{fig5}
\end{figure}

\section{Unidirectional photon interactions}\label{sec:chiral}
The collective optical response of small atomic arrays together with the Rydberg blockade makes it possible to reach the ideal limit of waveguide-QED \cite{Roy2017RMP,Chang2018RMP}, akin to ongoing experiments with cold atoms near nano-fibers or photonic crystals \cite{Thompson2013Science,Tiecke2014Nature,Petersen2014Science}, quantum dots \cite{Arcari2014PRL,Tiranov2023Science}, or super conducting circuits \cite{Kannan2020Nature}. As we have seen above, the Rydberg-atom array implements a fully saturable quantum emitter which can be strongly coupled to a single propagating mode with an effective coupling efficiency, $\beta$, of that can approach record values of $\beta\sim1$ [see Fig.\ref{fig1}(c)], only limited by the weak Rydberg-state decay. As opposed to point-like scatterers and waveguide-QED, the atomic array provides a flexible optical interface that can yield such high $\beta$-factors for any incident spatial mode.

These features can be exploited by shining the probe light at a non-orthogonal angle, which prevents the photons from being reflected back into the initial spatial mode [see Fig.\ref{fig1}(a)]. The resulting splitting of the incident photons into two distinct spatial modes can then be used to generate a directional coupling between the photons and the array, whereby the incident light is scattered only in the forward direction of the original input mode [see Fig.\ref{fig5}]. 

The proposed setup is illustrated in Fig.~\ref{fig5} and only involves a few additional linear optical elements. First, the incident photon field, $\hat{a}^{(\rm in)}(t)$, enters a symmetric beam splitter at one of its two input ports and is converted into two modes, $\hat{a}^{(\rm in)}_{\rightarrow}$ and $ \hat{a}^{(\rm in)}_{\leftarrow}$, that subsequently hit the array from the two opposite sides as indicated in Fig.~\ref{fig5}.
The interaction with the array generates the following output fields
\begin{subequations}
 \begin{align}
  \hat{a}_{\rightarrow}(t) =& \hat{a}^{(\rm in)}_{\rightarrow}(t) + \frac{ig}{c}\sqrt{\ell}\sum_{j} u^{*}(\mathbf{r}_{j})\hat{\sigma}^{(j)}_{ge},\\
  \hat{a}_{\leftarrow}(t) =& \hat{a}^{(\rm in)}_{\leftarrow}(t) + \frac{ig}{c}\sqrt{\ell}\sum_{j} u^{*}(\mathbf{r}_{j})\hat{\sigma}^{(j)}_{ge},
 \end{align}
\end{subequations}
where the mode function $ u(\mathbf{r}_{j})$ depends on the angle of incidence $\theta$, but is identical for $\hat{a}_{\rightarrow}^{(\rm in)}$ and $\hat{a}_{\leftarrow}^{(\rm in)}$ under perfectly symmetric conditions. Finally, a second beam splitter is used to recombine the two fields, and yields 
 \begin{align}
  \hat{a}(t) =& \frac{\hat{a}^{(\rm in)}_{\rightarrow}(t)+\hat{a}^{(\rm in)}_{\leftarrow}(t)}{\sqrt{2}} + \frac{i\sqrt{2}g}{c}\sqrt{\ell}\sum_{j} u^{*}(\mathbf{r}_{j})\hat{\sigma}^{(j)}_{ge}
 \end{align}
on one of its output ports, and
 \begin{align}
  \hat{b}(t) =& \frac{\hat{a}^{(\rm in)}_{\rightarrow}(t)-\hat{a}^{(\rm in)}_{\leftarrow}(t)}{\sqrt{2}} 
 \end{align}
on the other. Overall, this converts a single forward-propagating input field $\hat{a}^{(\rm in)}(t)$ into a single output field
 \begin{align}\label{eq:aout}
  \hat{a}(t)  =&\hat{a}^{(\rm in)}(t) + \frac{i\sqrt{2}g}{c}\sqrt{\ell}\sum_{j} u^{*}(\mathbf{r}_{j})\hat{\sigma}^{(j)}_{ge}
 \end{align}
propagating in the same direction. The underlying mechanism is based on the parity symmetry of the atomic array, which interacts strongly with the symmetric superposition, $\hat{a}^{(\rm in)}_{\rightarrow}+\hat{a}^{(\rm in)}_{\leftarrow}$, and is completely decoupled from the anti-symmetric mode, $\hat{a}^{(\rm in)}_{\rightarrow}-\hat{a}^{(\rm in)}_{\leftarrow}$. Yet,  misalignment of the optical elements may break the overall symmetry of the setup and affect the ideal photon output Eq.~(\ref{eq:aout}). Hereby, however, the setup can act as a sensitive interferometer to control and compensate such imperfections, since any harmful misalignment will lead to a finite signal in the second output field $\hat{b}$. For example a small displacement, $\Delta z$, of the atomic array, generates a signal 
 \begin{align}
\langle\hat{b}^\dagger \hat{b}\rangle= 4\pi^2\langle\hat{a}^\dagger \hat{a}\rangle \frac{\Delta z^2}{\lambda^2},
 \end{align}
that can be employed for accurate phase control or adjustment of the beam splitter arrangement.

To describe the interaction with photons at non-orthogonal incidence, we consider linearly polarized light with the polarization vector oriented in the plane of the atomic array irrespective of the angle $\theta$. Let us further assume an additional magnetic field, oriented along the same direction, which lifts the degeneracy between magnetic sub-levels and is tuned in such a way that the incident field is near-resonant with a chosen $\pi$-transition. If the Zeeman shifts of the atomic states exceed the atomic dipole-dipole interaction, the magnetic field prevents the population  of other excited Zeeman states. This efficiently suppresses the coupling between different polarization states of emitted photons and enables a simplified description in terms of a single $|g\rangle-|e\rangle$ transition. Choosing a linear polarization in the plane of the array, however, breaks rotational symmetry and leads to an anisotropic interaction between the atoms as opposed to the case of circular in-plane polarization. In addition, a finite incident angle $\theta$ sets slightly more stringent constraints on the lattice constant
\begin{align}
\label{Bragg-scattering}
 a\leq\frac{\lambda}{1+\sin\theta}
\end{align}
in order to suppress higher order Bragg-scattering into other modes and ensure the collective suppression of photon losses.
\begin{figure}[b!]
 \centering
 \includegraphics[width=0.48\textwidth]{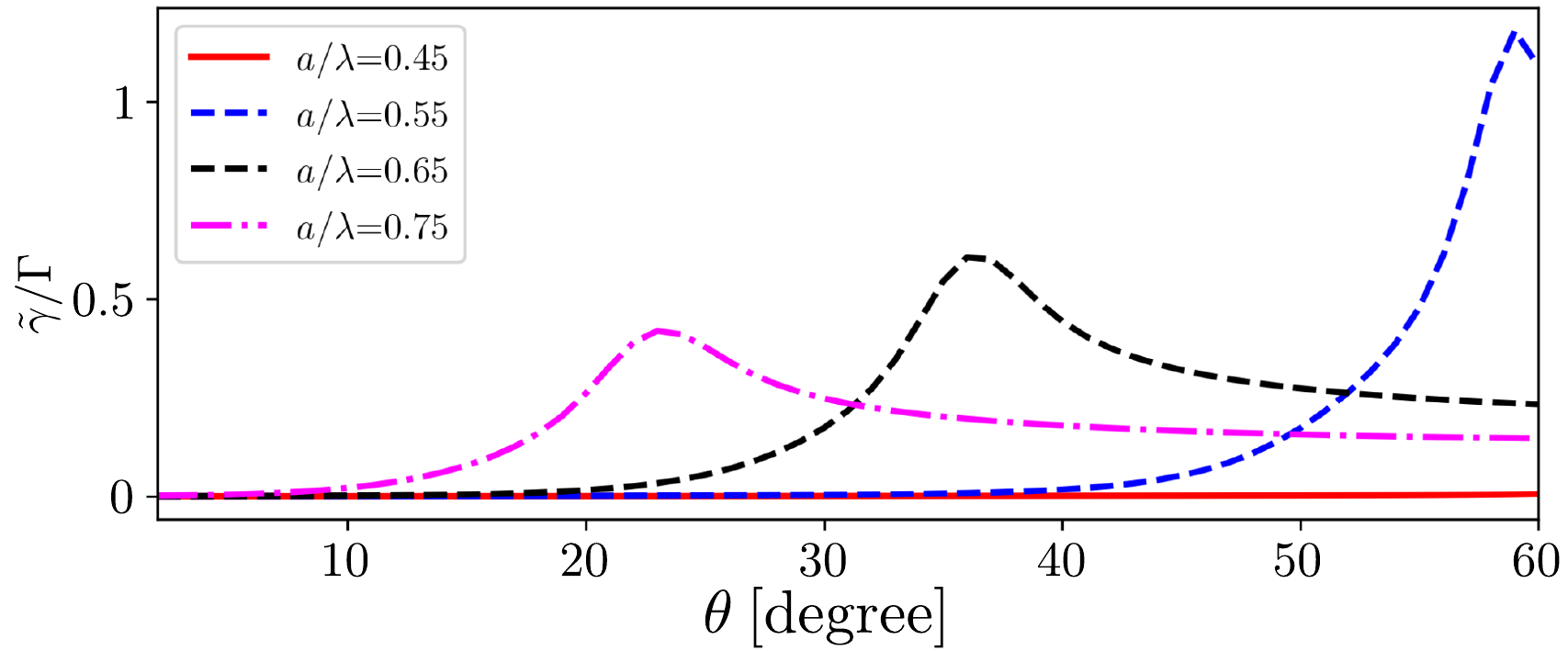} 
 \caption{Photon loss rate $\tilde{\gamma}$ in units of $\Gamma$ as a function of the incident angle $\theta$ for different lattice constants $a$. Results are shown for a rounded $31\times 31$ atomic array and a probe-beam waist of $w_{0}=2\lambda$ to minimize  finite-size effects for all considered  values of $a$. The applied laser fields resonantly ($\bar{\Delta}+\bar{\Delta}_{c}=0$) excited $100S$ Rydberg states of Rb atoms with $\Omega/\Gamma = 8.0$ and $\Delta_{r}/\Gamma = -10.0$.}
 \label{fig6}
\end{figure}
Photon losses out of the incident mode can be detected as a loss rate $\tilde{\gamma}$, which generally yields a useful characterization of the photon-coupling efficiency \cite{Solomons24}.
Here, we quantify photon losses by comparing the transmission spectrum of the beam splitter setup with the transmission 
\begin{align}
 T(\delta) = \bigg|\frac{i(\tilde{\gamma} - \tilde{\Gamma})}{2\delta + i(\tilde{\gamma} + \tilde{\Gamma})}\bigg|^{2}
 \label{eq:tDelta}
\end{align}
of a single two-level emitter that is coupled to a single propagating photon mode \cite{Sheremet23}. By analogy with such an effective waveguide-QED configuration and by fitting the corresponding expression Eq.\ (\ref{eq:tDelta}) to the calculated transmission spectrum, we can extract the photon loss rate $\tilde{\gamma}$ and the effective emission rate $\tilde{\Gamma}$ into the effective waveguide mode $\hat{a}$ (c.f. Fig.\ref{fig5}). Figure \ref{fig6} shows $\tilde{\gamma}$ as a function of the incident angle $\theta$. Increasing $\theta$ can significantly diminish the coupling efficiency due to the onset of  Bragg scattering into other modes. Such losses are suppressed for smaller lattice constants, $a$, and vanish for $a<\lambda/2$, as discussed above [see Eq.(\ref{Bragg-scattering})]. In particular, one finds that experimentally achievable lattice constants permit very low photon losses at sufficiently large angles to implement the beam splitter setup of Fig.\ref{fig5}. We can quantify its coupling efficiency by the effective $\beta$-factor
\begin{align}
\beta = \frac{\tilde{\Gamma}}{\tilde{\Gamma} + \tilde{\gamma}},
\end{align}
which is shown in Fig.\ref{fig7}. Increasing the transverse beam waist $w_0$ reduces imperfections due to finite transverse-momentum contributions ${\bf k}_\perp$ \cite{Shahmoon2017PRL} of the incident mode $u_\rightarrow({\bf r})$ and $u_\leftarrow({\bf r})$. Consequently, the coupling efficiency $\beta$ increases with the beam waist, but eventually saturates at a value $\beta_{\text{opt}}\lesssim1$. Importantly, one can reach this optimal value already for rather small beam waists, $w_0$, which lie well below the typical Rydberg blockade radius for (cf. Fig.\ref{fig3}).

Under these conditions, the asymptotic value 
\begin{equation}\label{eq:beta}
\beta_{\text{opt}}=\frac{\bar{\Gamma}_c}{\bar{\Gamma}_c+\gamma}
\end{equation}
is only limited by the decay rate, $\gamma$, of the Rydberg state. It therefore increases rapidly with the principal quantum number $n$, as illustrated in Fig.\ref{fig7}(b). Since the value of $\beta\sim1-\gamma/\Gamma_c$ is ultimately limited by the ratio of the Rydberg-state decay rate $\gamma$ and the collective decay rate $\Gamma_c$ of $|e\rangle$, it can in principle be further increased by working on the superradiative reflection resonance of the atomic mirror that appears for smaller lattice constants $a\sim0.2\lambda$. 

\begin{figure}[t!]
 \centering
 \includegraphics[width=0.48\textwidth]{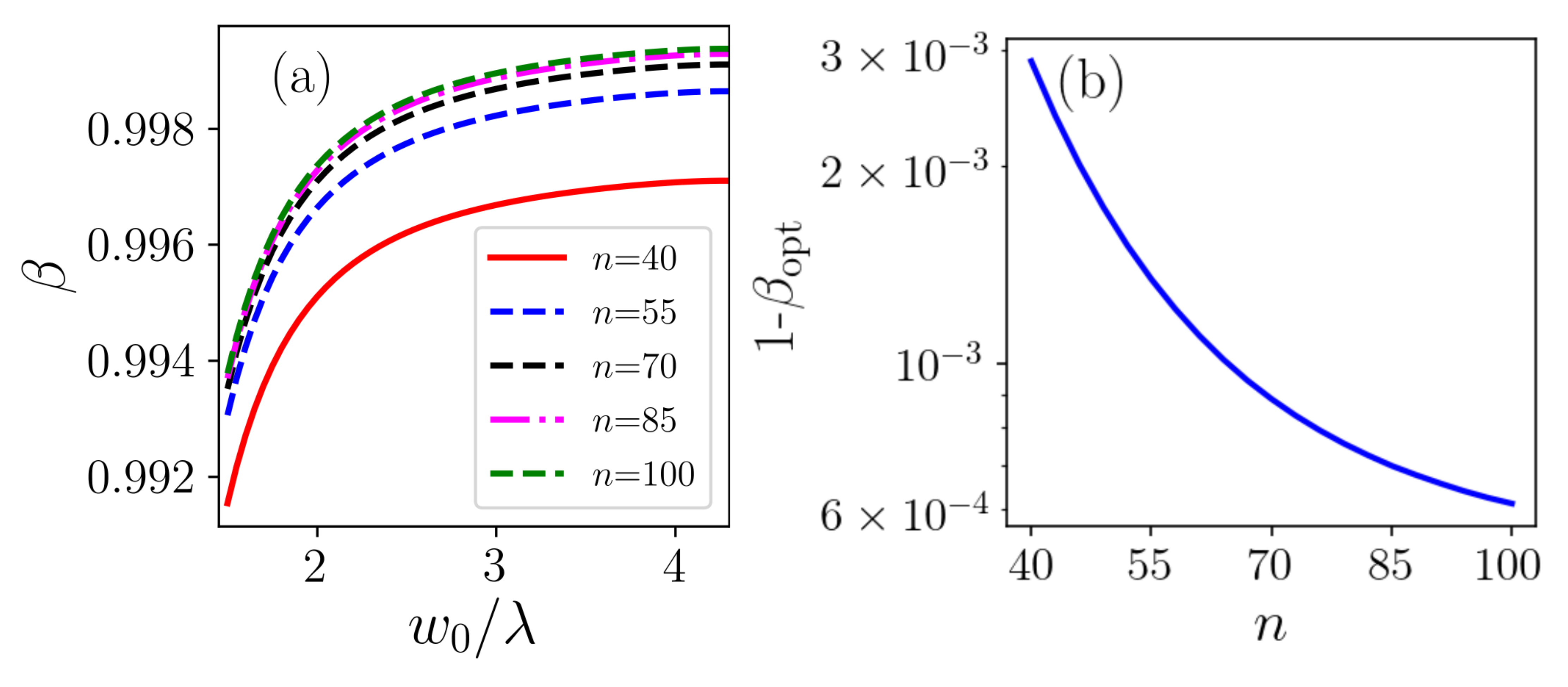} 
 \caption{(a) The coupling efficiency $\beta$ is shown as a function of the probe-beam waist $w_{0}$ for different principal quantum numbers $n$. Panel (b) shows the optimal $\beta_{\text{opt}}$ as a function of the Rydberg state $n$. The calculations have been performed for a rounded $75\times75$ array with a lattice constant $a=0.65\lambda$ that is illuminated by a Gaussian beam at an angle $\theta=5^{\circ}$. In panel (b) we have chosen a large beam waist $w_{0}=8\lambda$. The remaining parameters are $\Omega_p/\Gamma = 0.001$,  $\Omega/\Gamma = 8.0$, and $\Delta_{r}/\Gamma = -10.0$.}
 \label{fig7}
\end{figure}

\begin{figure}[b!]
 \centering
 \includegraphics[width=0.4\textwidth]{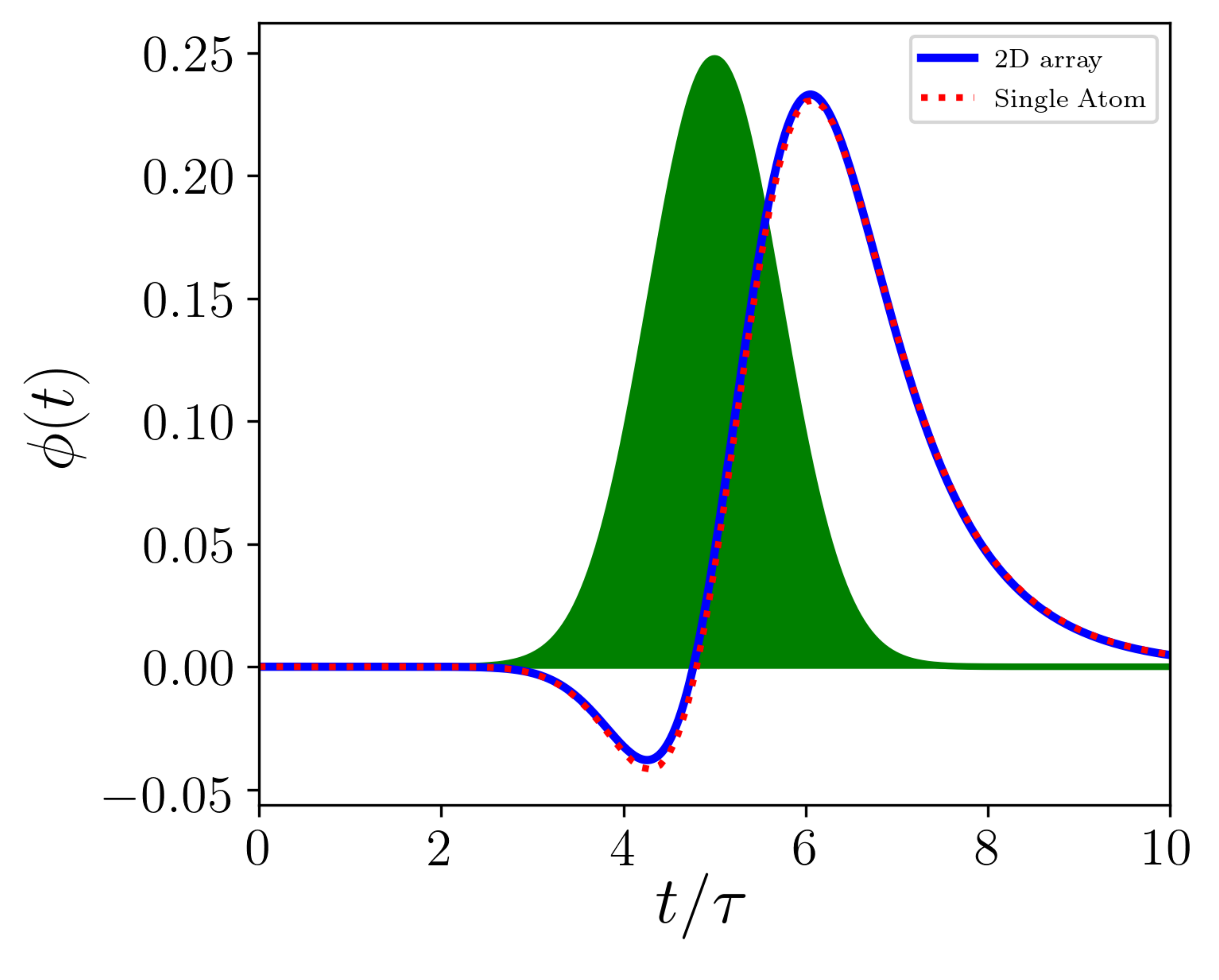} 
 \caption{Comparison of the output single photon amplitude, $\phi(t)$ between the case of a single 2-level atom coupled to a single waveguide mode and the case of the extended Rydberg array coupled to a free-space pulse with Gaussian transverse spatial beam profile. The temporal Gaussian shape of the incident pulse is shown in green. It has a  beam waist of $w_{0}=2.1\lambda$, a pulse length of $\tau = 2.24/\tilde{\Gamma}$, and resonantly ($\bar{\Delta}+\bar{\Delta}_{c}=0$) excites Rydberg states  with $n=100$. The remaining parameters are $\Omega/\Gamma = 8.0$, $a = 0.75\lambda$, and $\Delta_{r}/\Gamma = -10.0$.}
 \label{fig8}
\end{figure}

\section{Photon pulse dynamics} \label{sec:pulsed}
Thus far, we have focused on cw-fields and studied the optical response of the atomic array for a fixed frequency of the incident probe photons. This is readily generalized  by considering the operators 
\begin{align}\label{eq:propfield}
    \hat{a}_{\phi}(t)&=\int{\rm d}{\bf r}\:u^*({\bf r})\phi^*({\bf e}  {\bf r}-ct)\hat{\mathcal{E}}({\bf r},t)
\end{align}
for a spatiotemporal mode of finite-length photon pulses that propagate along the unit vector ${\bf e}$ with a pulse envelope $\phi(t)$ and a transverse beam profile $u({\bf r})$. The incident fields are then described by annihilation operators
\begin{align}
    \hat{a}^{({\rm in})}_{\rightarrow}(t)&=\sqrt{\ell}\int{\rm d}{\bf r}\:u^*_\rightarrow({\bf r})\phi^*({\bf e}_{\rightarrow}  {\bf r}-ct)\hat{\mathcal{E}}({\bf r},t)\nonumber\\
    \hat{a}^{({\rm in})}_{\leftarrow}(t)&=\sqrt{\ell}\int{\rm d}{\bf r}\:u^*_\leftarrow({\bf r})\phi^*({\bf e}_\leftarrow {\bf r}-ct)\hat{\mathcal{E}}({\bf r},t)
\end{align}
for forward and backward propagating pulses along the antiparallel directions,  ${\bf e}_\rightarrow$ and ${\bf e}_\leftarrow$, respectively, using the mode functions introduced in Sec.~\ref{sec:3level}. 
The dynamics of an incident single-photon pulse can then be studied, by simulating the atomic lattice under weak coherent driving and projecting the state of the outgoing light onto the single-photon sector. 
In this case, one can solve the master equation (\ref{eq:ME2})-(\ref{eq:L2l}) of the atomic lattice by propagating a single quantum trajectory \cite{Molmer1993JOSAB}, $|\Psi(t)\rangle$ that describes the state of the $N$ atoms. The amplitude of the transmitted photon after propagating through the beam splitter setup can then be obtained from $\psi(t)=\langle 0|\hat{a}(t)|\Psi(t)\rangle$, where $|0\rangle$ denotes the initial state of the array with all atoms in their ground state. Here, the  field operator $\hat{a}(t)$ for the output light that defines the photon density along the propagation direction as
\begin{equation}
\hat{a}(t)=a^{({\rm in})}\phi(ct)+i\frac{\sqrt{2}g}{c}\sum_j u^*({\bf r}_j)\hat{\sigma}_{ge}^{(j)}(t)
\end{equation}
only acts on the atoms and corresponds to Eq.(\ref{eq:aout}), but for a time dependent coherent-state input field with an amplitude $a^{({\rm in})}=\langle\hat{a}^{({\rm in})}\rangle\ll1$.
Figure \ref{fig8} shows the simulated photon amplitude for a Gaussian input pulse $\phi(z)=e^{-\frac{z^2}{2c^2\tau^2}}/(\pi c^2 \tau^2)^{1/4}$. For the chosen duration the spectral pulse width is comparable to the collective linewidth of the array ($\tilde{\Gamma}\tau\sim1$), such that the photon undergoes significant pulse distortion. The result is in excellent agreement with the pulse transmission by a single two-level emitter in a one-dimensional  waveguide-QED setting under perfect-coupling conditions, $\beta=1$.

\begin{figure}[t!]
 \centering
 \includegraphics[width=0.49\textwidth]{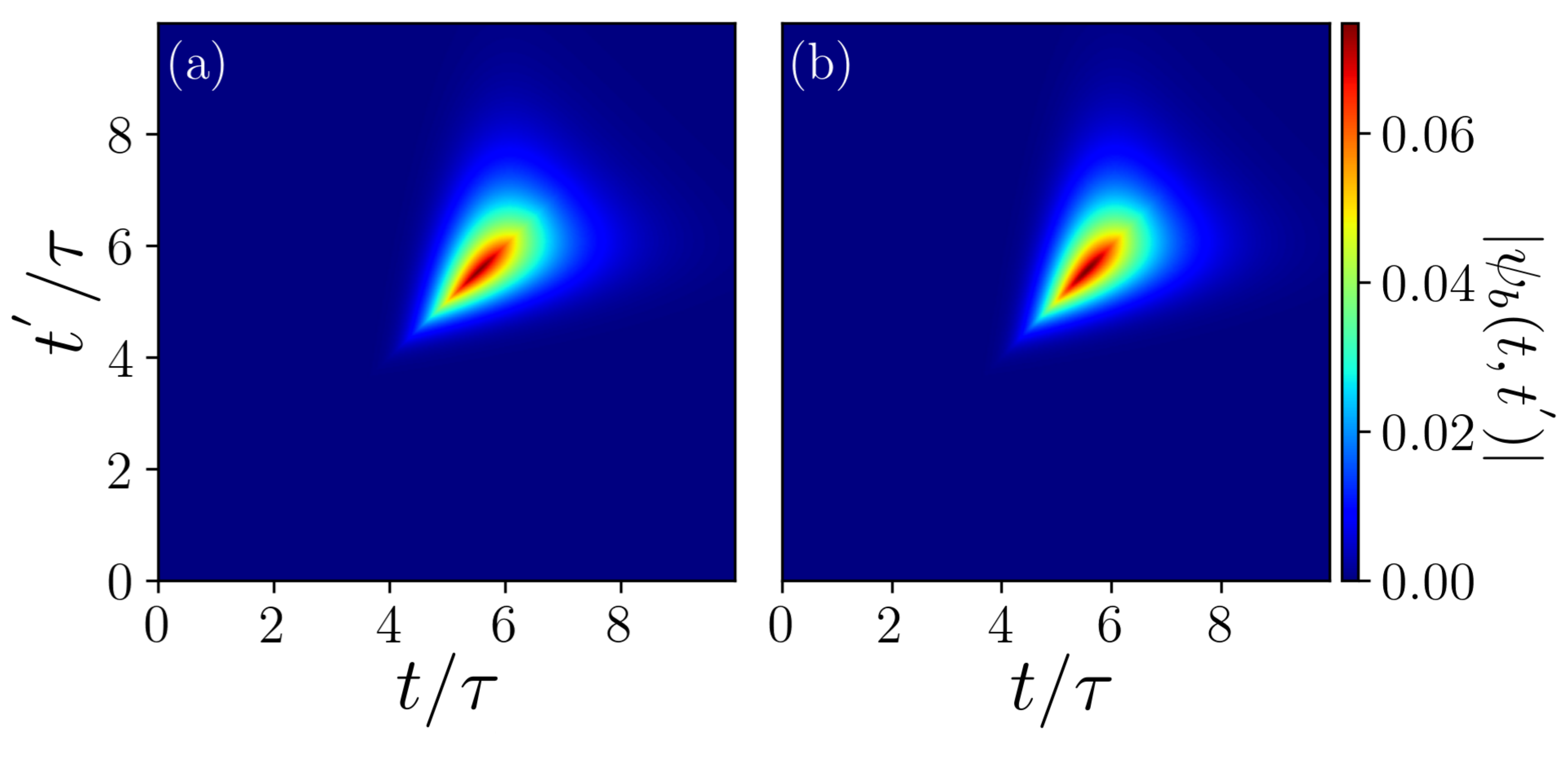} 
 \caption{Comparison of the wave function of the two-photon bound state generated by (a) the Rydberg array and (b) a waveguide-coupled 2-level quantum emitter for the same parameters as in Fig.~\ref{fig8}.}
 \label{fig9}
\end{figure}

Similarly, we can study the interaction with an incident two-photon pulse by simulating the atomic array under weak coherent driving ($a^{({\rm in})}\ll1)$ and calculating the two-photon output from $\psi_2(t,t^\prime) = \langle 0|\hat{a}(t)\hat{a}(t^\prime)|\Psi\rangle$. Also here, the obtained two-photon output shown in Fig.\ref{fig9} under perfect coupling conditions ($\beta\sim1$) follows closely the behavior of the effective waveguide system with a single saturable emitter and $\beta=1$. In this case \cite{Shen2007PRA,Shen2007PRL}, the correlated two-photon amplitude 
\begin{align}\label{eq:bound_state}
\psi_2(t,t^\prime)=\psi(t)\psi(t^\prime)+\psi_b(t,t^\prime)
\end{align}
can be interpreted in terms of a photonic bound state  $\psi_b(t,t^\prime)$ that is exponentially localized with respect to the time delay $|t-t^\prime|$ between the two transmitted photons. Using Eq.(\ref{eq:bound_state}), we can formally extract $\psi_b$ from the numerical two-photon amplitude and indeed find good agreement with the known analytical form   \cite{Shen2007PRA,Shen2007PRL}. For a more quantitative comparison of the dynamical nonlinear response, we have calculated the overlap between the two-photon output from the ideal waveguide-QED setting and the Rydberg-atom array, and find and infidelity of less than $10^{-3}$ for all studied parameters and principal quantum numbers $40\leq n\leq100$.

\section{Applications}\label{sec:applications}
The obtained nonlinearity and directional photon coupling suggest a range of applications, tapping into the capabilities of waveguide-QED with strongly coupled  quantum emitters \cite{Sheremet23}. Below, we discuss two examples: the generation of single photons and two-photon gate operations.
\subsection{Single photon generation}
The availability of tunable single-photon pulses is an essential prerequisite for a wide range of applications. Our three-level Rydberg array suggests a direct approach for  the deterministic generation of single-photon pulses via photon storage and retrieval protocols with coherent input light. 

\begin{figure}[t!]
 \centering
 \includegraphics[width=0.49\textwidth]{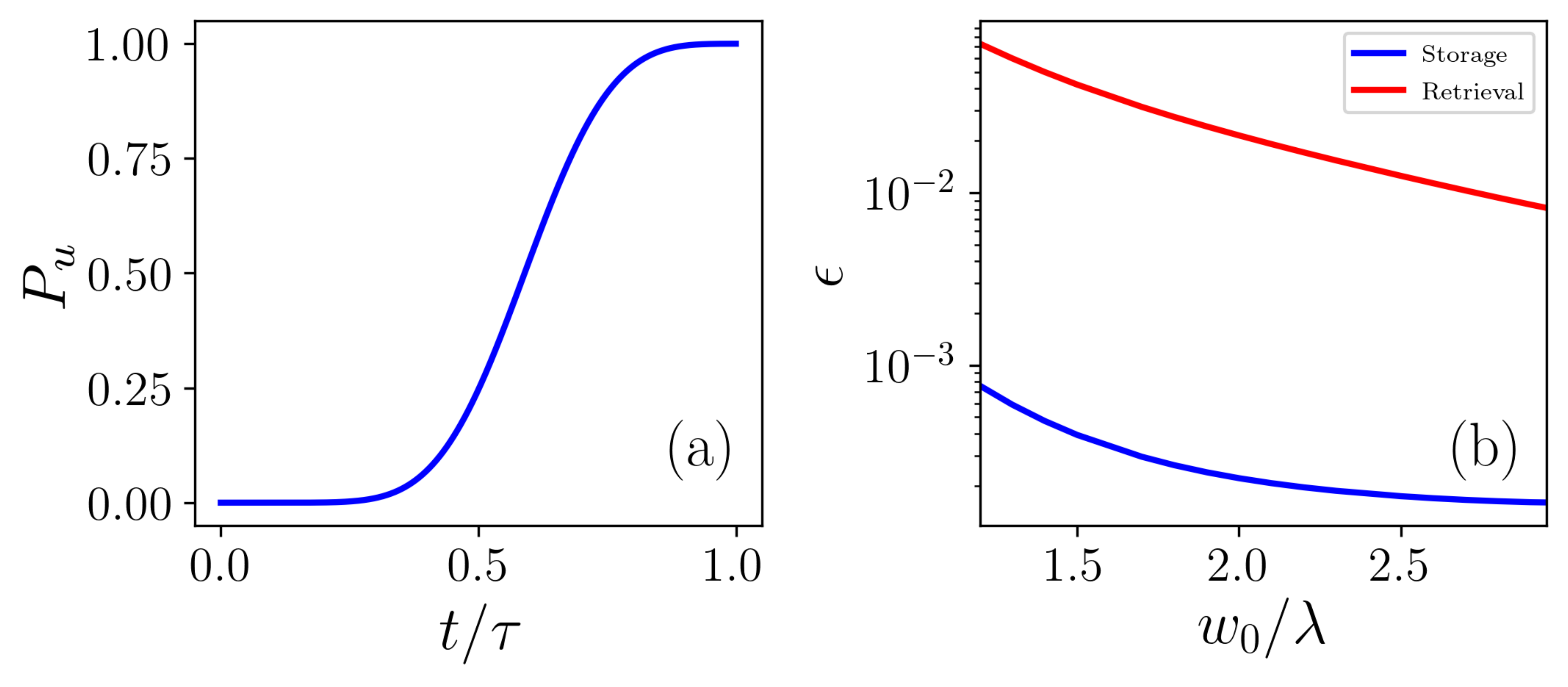} 
 \caption{(a) Temporal dynamics of the population in the driven mode $\mathcal{E}(\mathbf{r})$ for a $\pi$ pulse. (b) The corresponding storage and retrieval infidelity as a function of the beam size $w_{0}$. Here we have chosen a $21\times 21$ array with $a = 0.75\lambda$. In panel (a), the two laser pulses generate a single Rydberg excitation in the blockaded array with a control Rabi frequency $\Omega/\Gamma = 4$, and detuning $\Delta_{r}/\Gamma=-20$ on the Raman resonance, where $\bar{\Delta}+\bar{\Delta}_c$=0. The probe pulse has a transverse width of $w_{0}=3.0\lambda$, and temporal square-pulse shape to generate a two-photon $\pi$-pulse with a duration $\Gamma\tau=10$. Panel (b) shows the storage and retrieval infidelity of a single photon for these parameters but a varying beam waist $w_0$ of the probe pulse.}
 \label{fig10}
\end{figure}

\begin{figure*}[t!]
	\centering
	\includegraphics[width=\linewidth]{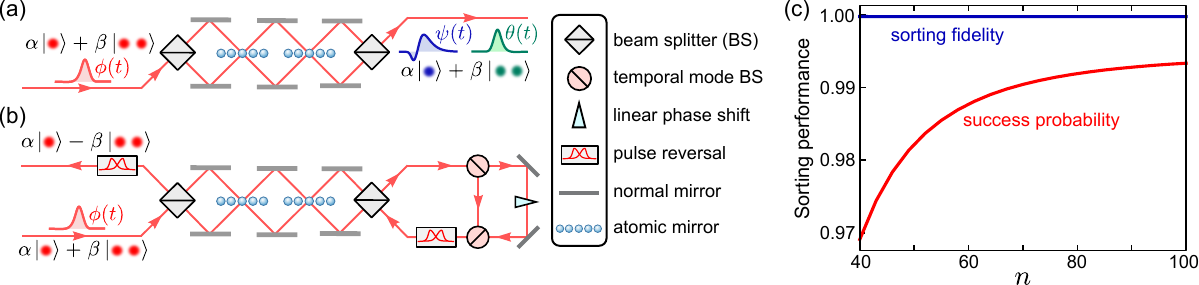}
	\caption{(a) Schematic of the photon sorting scheme based on an interferometer setup that contains two Rydberg-atom arrays. Panel (b) shows a circuit that utilizes this nonlinear element in combination with further linear operations to implement an NS gate. The achievable sorting fidelity and success probability is shown in (c) as a function of principal quantum number of the Rydberg state.}
	\label{fig11}
\end{figure*}

First, one generates a single de-localized Rydberg excitation by illuminating the array with classical light and exploiting the interaction blockade which prevents excitation of more than a single Rydberg atom. In practice, this can be accomplished in different ways, e.g., by adiabatically chirped excitation of the effective two-level system, by light-storage on single-photon resonance under EIT conditions, or by applying a $\pi$-pulse to invert the Rydberg-blockaded superatom.

Here, we focus on the latter strategy. Since we are working far off resonance (cf. Fig.\ref{fig2}), we can consider the effective 2-level limit. While this suppresses the attainable single-atom Rabi frequency, the Rydberg-blockade collectively enhances the Rabi frequency of the $N$-atom array by a factor of $\sim\sqrt{N}$. One can, thus, consider a $\pi$-pulse that generates a single Rydberg excitation on a time scale that is much shorter than the typical lifetime time of the excitation. We can, therefore, write the $N$-atom state as $|\Psi\rangle = c_{0}(t)|0\rangle + \sum^{N}_{j=1}c_{j}(t)\hat{\sigma}^{(j)}_{rg}|0\rangle$, whose amplitudes evolve as
\begin{subequations}
	\label{coeff-eqs-main}
	\begin{align}
   \dot{c}_{0} =& i\sum_{j} \bar{\Omega}_{j}c_{j}\\[2ex]
   \dot{c}_j =& i(\bar{\Delta} + i\frac{\gamma}{2})c_{j} + i\bar{\Omega}c_{0} +i\sum_{k\neq j}(\bar{J}_{kj} + \frac{i}{2}\bar{\Gamma}_{kj})c_{k}.
	\end{align}
\end{subequations}
with the time dependent two-photon Rabi frequency $\bar{\Omega}_{j}=\bar{g}u(\mathbf{r}_{j})\phi(\bm{e}\mathbf{r}_{j}-ct)$. As shown in Fig.\ref{fig10}(a), one can obtain a near unit population
	\begin{align}\label{eq:Pe}
P_{u}=|\sum_{j}u^{*}(\mathbf{r}_{j})c_{j}|^{2}/\sum_{j}|u(\mathbf{r}_{j})|^{2},
	\end{align}
of Rydberg excitations in the spatial atomic mode defined by the  excitation pulse. In the depicted example, the chosen pulse length is $\tau=10\Gamma^{-1}$, which is sufficiently short to suppress decoherence from spontaneous decay and corresponds to an experimentally achievable value of $\tau=260$ns for Rb atoms. Note that the dipole-dipole interaction tends to distort the spatial profile of the spin wave excitation with respect to the incident light. This effect plays only a minor role for moderate beam waists of a few optical wavelengths and can be further suppressed by use of a wider excitation beam. Figure\ \ref{fig10}(b) demonstrates that high storage efficiencies with $1-P_u\sim 10^{-4}$ are possible with typical parameters.

The generated Rydberg-spinwave excitation can then be retrieved efficiently, via an adiabatic readout under EIT conditions, i.e., by turning up the control field $\Omega(t)$ on the $|r\rangle-|e\rangle$ transition. As shown in \cite{Manzoni2018NJP}, sub-wavelength arrays can yield extraordinarily high retrieval efficiencies, as the  photon losses can be exponentially suppressed compared to optical memories based on disordered atomic ensembles. The retrieval dynamics is described by the simple set of evolution equations
\begin{align}
\frac{de_{j}}{dt} =& i\Delta_{e}e_{j} + i\Omega^{*}(t)c_{j} + i\sum_{k}(J_{kj} + \frac{i}{2}\Gamma_{kj})e_{k}\\
\frac{dc_{j}}{dt} =& i(\Delta_{e} + \Delta_{r})c_{j} + i\Omega(t)c_{j} 
\label{eqs_retrival}
\end{align} 
for the amplitudes $e_j$ and $c_j$ of an $|e\rangle$-state and Rydberg-state excitation of the $j$th atoms, respectively. 
The polarization field of the array yields the emission into the incident mode $\mathcal{E}({\bf r})$ and defines the retrieval efficiency  \cite{Manzoni2018NJP}
\begin{align}
\eta= \frac{g^{2}}{c}\sum_{jl}u^{*}(\mathbf{r}_{j})u(\mathbf{r}_{l})\int^{\infty}_{0}e_{j}(t)e^{*}_{l}(t){\rm d} t.
\end{align}
Also here, the dipole-dipole interaction slightly distorts the mode of the emitted field. This effect is reduced by employing a larger beam waist. For typical experimental parameters, we find large fidelities of $\sim 0.99$ for generating single photons in a single and fully controllable spatio-temporal mode. For a homogeneous control field $\Omega$, the transverse mode function is perfectly matched to that of the incident  field. In addition, the temporal mode profile can be shaped upon read-out by the time dependence of $\Omega(t)$. Since this read-out field operates on single-photon resonance, the relevant line width is given by $\Gamma_c$. Hence $\Omega(t)$ can be varied rapidly compared to timescales $\bar{\Gamma}^{-1}_c\gg{\Gamma}^{-1}_c$ while fully maintaining adiabaticity for the retrieval process. This in turn enables precise temporal control of the generated single-photon pulse. The  $\pi$-pulse excitation followed by resonant retrieval thus yields an efficient source of single photons with full spatio-temporal tunability, making them useful for controlled photon interactions with the nonlinear beam splitter setup described in section \ref{sec:chiral}.

\subsection{Photon sorting and quantum gates}
	The high degree of tunability can be used for applications of the Rydberg-mediated nonlinearity, as we shall discuss in this section. When two atom arrays are integrated into the described beam splitter configuration of Fig.~\ref{fig11}(a), the system can be thought of a chiral waveguide QED system containing two cascaded two-level emitters. It has been shown in Ref.~\cite{yang2022determin} that such an emitter pair is capable of performing a nearly deterministic  separation of single- and two-photon components into orthogonal modes of a given input temporal pulse mode $\phi(t)$ [see Fig.~\ref{fig11}(a)], i.e., 
	\begin{equation}
	c_1 \hat{a}_{\phi}^\dagger|0\rangle + 
	c_2 \frac{ (\hat{a}_{\phi}^\dagger)^2}{\sqrt{2}}|0\rangle
	~ \rightarrow ~ c_1 \hat{a}_{\psi}^\dagger|0\rangle + 
	c_2 \frac{(\hat{a}_{\theta}^\dagger)^2 }{\sqrt{2}}|0\rangle, \label{eq:eq1}
	\end{equation}
where $\hat{a}_{f}^\dagger=\int dt f(t)\hat{a}^\dagger(t)$ creates a single photon in the temporal mode $f(t)$ [c.f. Eq.(\ref{eq:propfield})], with $\psi(t)$ and $\theta(t)$ being two orthogonal ones [$\int dt \psi^*(t) \theta(t)=0$]. The intuitive understanding of the above process is that for a single-photon input state, its dispersion will accumulate when passing through two emitters, resulting in a distorted output wavefunction $\psi(t)$; while for a two-photon input, an approximate time reversal process is established: the second emitter disentangles the correlated output wavefunction from the first one and generates two uncorrelated photons occupying an undistorted but time-reversed pulse $\theta(t)\approx \phi(T-t)$, with $T$ the delay time induced by the emitters.

Perfect sorting of one- and two-photon components  permits implementation of a deterministic nonlinear-sign (NS) gate \cite{witthaut2012photon,ralph2015photon}, which can be further used to make the all-optical KLM quantum computing  protocol \cite{Knill2001Nature} deterministic. An NS gate flips the sign of the two-photon component, while leaving the vacuum and single-photon parts unaltered, i.e.,
\begin{equation}
c_0 |0\rangle + c_1 \hat{a}_{\phi}^\dagger|0\rangle + 
c_2 \frac{ (\hat{a}_{\phi}^\dagger)^2}{\sqrt{2}}|0\rangle
~ \rightarrow ~  c_0 |0\rangle + c_1 \hat{a}_{\phi}^\dagger|0\rangle - 
c_2 \frac{ (\hat{a}_{\phi}^\dagger)^2}{\sqrt{2}}|0\rangle, \label{eq:eq2}
\end{equation}
and it can be realized by using the interferometer setup shown in Fig.~\ref{fig11}(b). First, the same nonlinear element as in Fig.~\ref{fig11}(a) is used to separate the single-mode input field into photon-number dependent orthogonal temporal modes. In a second step, these distinct temporal modes are transferred to spatially separated modes. Such a temporal-mode beam splitter can be implemented via sum frequency generation \cite{eckstein2011quantum,ansari2018tomography}. A linear phase shifter can then be placed in the two-photon path to induce a sign-flip of the two-photon amplitude, as illustrated in Figs.~\ref{fig11}(b). 
Following another temporal-mode beam splitter and a linear time reversal operation, the photon pulses are sent through the other, thus far unused, ports of the Rydberg-array interferometer to restore the original single-mode nature of the input pulses, however, with an added sign-flip on the two-photon component. The unidirectionally of the nonlinear element makes it possible to use the Rydberg arrays twice in the quantum circuit, provided one ensures a sufficient pulse delay  to avoid interference of photons from the different input ports of the Rydberg beam splitter setup.

In such a configuration, the Rydberg-array photon sorter is the only (passive) nonlinear device, and all remaining single-qubit operations, such as  temporal-mode beam splitting \cite{eckstein2011quantum,ansari2018tomography} and forward/backward time reversal \cite{chumak2010all,sivan2011time,minkov2018localization}, can be performed with high fidelity. Figure ~\ref{fig11}(c) shows the attainable sorting fidelity and the corresponding success probability for different principal quantum numbers $n$ of the Rydberg state. For a $15\times 15$ atom array with $n=100$, the optimal sorting fidelity can reach $ \mathcal{F} \gtrsim 0.9998$ with a high success probability of $\mathcal{P}> 0.99$ for the quantum gate operation discussed above.
This high value is obtained for an array with lattice spacing $a/\lambda=0.65$. However, further improvements are possible for denser arrays since this would increase the effective coupling strength $\tilde{\Gamma}$. With an even higher $\mathcal{P}\sim1$, it may then become feasible to implement deterministic Bell-state measurements and nonlinear phase gate operations on large photonic quantum networks.

\section{Discussions and Conclusions}\label{sec:conclusion}
In this article, we have analyzed an approach to deploying subwavelength atomic arrays as nonlinear optical elements in photonic quantum circuits. Here, one exploits the near lossless collective coupling of extended arrays to a single free-space optical mode, while large nonlinearities are obtained from the strong interactions between highly excited Rydberg states of the atoms. Since the direct coupling to high-lying states features intrinsically small transition dipole moments, the combination of both effects requires three- or multi-level coupling schemes, which leaves different possibilities to introduce atomic Rydberg states. While Rydberg-dressing schemes \cite{Moreno2021PRL} limit the range and strength of generated interactions \cite{gil14,zeiher16} and resonant 3-level excitation \cite{Zhang2022Quantum,walther22,Solomons23} reduces the achievable nonlinear response \cite{Zhang2022Quantum}, we find here that driving a Rydberg Raman resonance under large single-photon detunings offers optimal conditions for realising strong effective photon interactions. This configuration yields a narrow reflection resonance at which the nonlinear effect of Rydberg-state interactions is maximized and can turn the array into a saturable 2-level quantum emitter with near-perfect coupling to a free-space photonic mode.

This mode-insensitive response and the extended geometry of the array enable the design of interferometric setups that behave like nonlinear unidirectional elements. We illustrate potential applications, by showing that such elements permit the deterministic generation of highly tunable single-photon pulses and photon number sorting operations with high fidelities. 

For two counterpropagating photons, the proposed element mediates a V-type three-level non-linearity, which may be used to implement a controlled phase gate \cite{PhysRevLett.129.130502}. The expected fidelities enabled by the strong collective light-matter coupling, thus, give promising perspectives for implementations of deterministic gates for photonic quantum information processing ~\cite{Graham2022Nature,Scholl2023Nature,Singh2023Science,Evered2023Nature,Bluvstein2024Nature,Shah2024PRA} and quantum simulation~\cite{Nishad2023PRA,Chen2023Nature,Shaw2024Nature}. Effective multi-emitter waveguide QED interactions could further enable the preparation of useful non-Gaussian states of light for quantum communication,  metrology and sensing~\cite{Eckner2023Nature,Bornet2023Nature,Bassler2024PRL,Schaffner2024PRXQuantum,PhysRevA.107.013717,lund2024subtraction}. 

Cascading multiple arrays, the system may be used to explore many-body dynamics in chiral waveguide QED, such as the propagation of multiphoton bound states \cite{Mohmoodian2020PRX}, repulsive photons \cite{Iversen2021PRL,Iversen2022PRR}, and superradiance bursts \cite{PhysRevX.14.011020}. 
Hereby, adjusting the transmittance of the beam splitters permits continuous tuning of the chirality, which may facilitate studies of spontaneous symmetry breaking in superradiance processes \cite{PhysRevLett.131.033605}.

Utilizing the Rydberg blockade in atomic arrays has enabled remarkable recent breakthroughs in quantum computing~\cite{Graham2022Nature,Scholl2023Nature,Singh2023Science,Evered2023Nature,Bluvstein2024Nature,Shah2024PRA}, quantum simulation~\cite{Nishad2023PRA,Chen2023Nature,Shaw2024Nature}, quantum metrology and sensing~\cite{Eckner2023Nature,Bornet2023Nature,Bassler2024PRL,Schaffner2024PRXQuantum} using atomic qubits, and the combination of the Rydberg-states and optical dipole interactions offers broad perspectives for exploring such applications with photonic quantum states. 

\begin{acknowledgements}
This work was supported by the Carlsberg
Foundation through the ``Semper Ardens'' Research Project QCooL, by the Danish National Research Foundation through the Center of Excellence for
Hybrid Quantum Networks (Grant agreement No. DNRF139) and the Center of Excellence ``CCQ" (Grant agreement No.: DNRF152), by the Austrian Science Fund (Grant No. 10.55776/COE1) and the European Union (NextGenerationEU), and by the Horizon Europe ERC Synergy Grant SuperWave (Grant No. 101071882).
\end{acknowledgements}

\appendix

\section{Two-level approximation}\label{app:A}
To derive the effective two-level description in the limit of $|\Delta_{e}|\gg |J_{ij}|, |\Gamma_{ij}|$, we start from the Heisenberg equation for the transition operator 
\begin{align}
\label{eq:sigma_ge_j}
\frac{d\hat{\sigma}^{(j)}_{ge}}{dt} = & ig\hat{\mathcal{E}}(\mathbf{r}_{j})+i\big(\Delta_{e} + i\frac{\Gamma}{2}\big)\hat{\sigma}^{(j)}_{ge} + i\Omega^{*}\hat{\sigma}^{(j)}_{gr}\nonumber\\
& + i\sum_{m,j\neq m}[J_{mj}+\frac{i}{2}\Gamma_{mj}]\hat{\sigma}^{(m)}_{ge}
\end{align} 
where $\hat{\sigma}^{(j)}_{gg}\approx\hat{I}$ in the the weak-driving limit and we have omitted Langevin noise terms since we are exclusively considering observables of normal ordered operators. For large $\Delta_e$, the equation is readily solved perturbatively and yields to leading order in $1/\Delta_e$ 
\begin{align}
\label{eq:addiabatic_elimination}
\hat{\sigma}^{(j)}_{ge}\simeq -\frac{g\hat{\mathcal{E}}(\mathbf{r}_{j}) + \Omega^{*}\hat{\sigma}^{(j)}_{gr}}{\Delta_{e}}.
\end{align}
Equation\ (\ref{eq:H0}) then becomes
\begin{align}
\hat{H}_{0} \simeq -\sum_{j}\bar{\Delta}\hat{\sigma}^{(j)}_{rr} + [\bar{g}\hat{\mathcal{E}}(\mathbf{r}_{j})\hat{\sigma}^{(j)}_{rg} + h.c.]
\end{align}
with the effective atom-light coupling strength $\bar{g}=-\Omega g/\Delta_{e}$. 
Similarly, the dipole-dipole interaction can be rewritten as 
\begin{align}
\hat{H}_{\text{dd}} &\simeq \sum_{j,k\neq j}\frac{J_{jk}}{\Delta^{2}_{e}}[g\mathcal{E}^{*}(\mathbf{r}_{j}) + \Omega\hat{\sigma}^{(j)}_{rg}][g\mathcal{E}(\mathbf{r}_{k}) + \Omega^{*}\hat{\sigma}^{(k)}_{gr}]\\
&\simeq \sum_{j,k\neq j}\bar{J}_{jk}\hat{\sigma}^{(j)}_{rg}\hat{\sigma}^{(k)}_{gr}\nonumber\\
\mathcal{L}(\rho) &\simeq \sum_{i,j}\frac{\bar{\Gamma}_{ij} +\gamma \delta_{ij}}{2}(2\hat{\sigma}^{(i)}_{gr})\rho\hat{\sigma}^{(j)}_{rg}-\{\hat{\sigma}^{(j)}_{rg}\hat{\sigma}^{(j)}_{gr}),\rho\})\nonumber
\end{align}
where we have substituted Eq.(\ref{eq:addiabatic_elimination}) and  $\bar{J}_{jk}+i\bar{\Gamma}_{jk}/2=|\Omega|^{2}( J_{jk}+i\Gamma_{jk}/2)/\Delta^{2}_{e}$. Throughout this work, assume coherent input fields for which we can replace $\hat{a}^{(\text{in})}_{\rightarrow}$ and $\hat{a}^{(\text{in})}_{\leftarrow}$ by $a^{(\text{in})}_{\rightarrow}$ and $a^{(\text{in})}_{\leftarrow}$. Substituting Eq.(\ref{eq:addiabatic_elimination}) into the input-output relations (\ref{eq:photons}) then gives
%Eqs.(\ref{eq:photons})
\begin{subequations}\label{eq:photons}
 \begin{align}
\label{eq:photons_a}  \hat{a}_{\rightarrow}(t)=&\hat{a}^{(\text{in})}_{\rightarrow}(t)+i\frac{\bar{g}}{c}\sqrt{\ell}\sum_{j}u^{*}(\mathbf{r}_j) \hat{\sigma}^{(j)}_{gr}+i\varepsilon  ,\\
\label{eq:photons_b}  \hat{a}_{\leftarrow}(t)=&\hat{a}^{(\text{in})}_{\leftarrow}(t)+i\frac{\bar{g}}{c}\sqrt{\ell}\sum_{j}u^{*}(\mathbf{r}_j) \hat{\sigma}^{(j)}_{gr}+i\varepsilon  ,
 \end{align}
\end{subequations}
where $\varepsilon=-\frac{g^2(a^{(\text{in})}_{\rightarrow}+a^{(\text{in})}_{\leftarrow})}{c\Delta_e}\ell\sum_{j}|u(\mathbf{r}_j)|^2$ is a negligibly small correction to the total field that stems from the weak photon emission from the far off-resonantly excited intermediate states. Omitting this term, yields the input output relations (\ref{eq:IO2l}) quoted in the main text.

\bibliographystyle{apsrev4-1}
\bibliography{references}

\begin{thebibliography}{98}%
\makeatletter
\providecommand \@ifxundefined [1]{%
 \@ifx{#1\undefined}
}%
\providecommand \@ifnum [1]{%
 \ifnum #1\expandafter \@firstoftwo
 \else \expandafter \@secondoftwo
 \fi
}%
\providecommand \@ifx [1]{%
 \ifx #1\expandafter \@firstoftwo
 \else \expandafter \@secondoftwo
 \fi
}%
\providecommand \natexlab [1]{#1}%
\providecommand \enquote  [1]{``#1''}%
\providecommand \bibnamefont  [1]{#1}%
\providecommand \bibfnamefont [1]{#1}%
\providecommand \citenamefont [1]{#1}%
\providecommand \href@noop [0]{\@secondoftwo}%
\providecommand \href [0]{\begingroup \@sanitize@url \@href}%
\providecommand \@href[1]{\@@startlink{#1}\@@href}%
\providecommand \@@href[1]{\endgroup#1\@@endlink}%
\providecommand \@sanitize@url [0]{\catcode `\\12\catcode `\$12\catcode
  `\&12\catcode `\#12\catcode `\^12\catcode `\_12\catcode `\%12\relax}%
\providecommand \@@startlink[1]{}%
\providecommand \@@endlink[0]{}%
\providecommand \url  [0]{\begingroup\@sanitize@url \@url }%
\providecommand \@url [1]{\endgroup\@href {#1}{\urlprefix }}%
\providecommand \urlprefix  [0]{URL }%
\providecommand \Eprint [0]{\href }%
\providecommand \doibase [0]{http://dx.doi.org/}%
\providecommand \selectlanguage [0]{\@gobble}%
\providecommand \bibinfo  [0]{\@secondoftwo}%
\providecommand \bibfield  [0]{\@secondoftwo}%
\providecommand \translation [1]{[#1]}%
\providecommand \BibitemOpen [0]{}%
\providecommand \bibitemStop [0]{}%
\providecommand \bibitemNoStop [0]{.\EOS\space}%
\providecommand \EOS [0]{\spacefactor3000\relax}%
\providecommand \BibitemShut  [1]{\csname bibitem#1\endcsname}%
\let\auto@bib@innerbib\@empty
%</preamble>
\bibitem [{\citenamefont {Slussarenko}\ and\ \citenamefont
  {Pryde}(2019)}]{Slussarenko2019APR}%
  \BibitemOpen
  \bibfield  {author} {\bibinfo {author} {\bibfnamefont {S.}~\bibnamefont
  {Slussarenko}}\ and\ \bibinfo {author} {\bibfnamefont {G.~J.}\ \bibnamefont
  {Pryde}},\ }\href {\doibase 10.1063/1.5115814} {\bibfield  {journal}
  {\bibinfo  {journal} {Applied Physics Reviews}\ }\textbf {\bibinfo {volume}
  {6}},\ \bibinfo {pages} {041303} (\bibinfo {year} {2019})}\BibitemShut
  {NoStop}%
\bibitem [{\citenamefont {Bartlett}\ \emph {et~al.}(2021)\citenamefont
  {Bartlett}, \citenamefont {Dutt},\ and\ \citenamefont
  {Fan}}]{Bartlett2021Optica}%
  \BibitemOpen
  \bibfield  {author} {\bibinfo {author} {\bibfnamefont {B.}~\bibnamefont
  {Bartlett}}, \bibinfo {author} {\bibfnamefont {A.}~\bibnamefont {Dutt}}, \
  and\ \bibinfo {author} {\bibfnamefont {S.}~\bibnamefont {Fan}},\ }\href
  {\doibase 10.1364/OPTICA.424258} {\bibfield  {journal} {\bibinfo  {journal}
  {Optica}\ }\textbf {\bibinfo {volume} {8}},\ \bibinfo {pages} {1515}
  (\bibinfo {year} {2021})}\BibitemShut {NoStop}%
\bibitem [{\citenamefont {Aspuru-Guzik}\ and\ \citenamefont
  {Walther}(2012)}]{Walther2012}%
  \BibitemOpen
  \bibfield  {author} {\bibinfo {author} {\bibfnamefont {A.}~\bibnamefont
  {Aspuru-Guzik}}\ and\ \bibinfo {author} {\bibfnamefont {P.}~\bibnamefont
  {Walther}},\ }\href {\doibase 10.1038/nphys2253} {\bibfield  {journal}
  {\bibinfo  {journal} {Nature Physics}\ }\textbf {\bibinfo {volume} {8}},\
  \bibinfo {pages} {285} (\bibinfo {year} {2012})}\BibitemShut {NoStop}%
\bibitem [{\citenamefont {Hartmann}(2016)}]{Hartmann_2016}%
  \BibitemOpen
  \bibfield  {author} {\bibinfo {author} {\bibfnamefont {M.~J.}\ \bibnamefont
  {Hartmann}},\ }\href {\doibase 10.1088/2040-8978/18/10/104005} {\bibfield
  {journal} {\bibinfo  {journal} {Journal of Optics}\ }\textbf {\bibinfo
  {volume} {18}},\ \bibinfo {pages} {104005} (\bibinfo {year}
  {2016})}\BibitemShut {NoStop}%
\bibitem [{\citenamefont {Giovannetti}\ \emph {et~al.}(2011)\citenamefont
  {Giovannetti}, \citenamefont {Lloyd},\ and\ \citenamefont
  {Maccone}}]{Giovannetti2011NPhoton}%
  \BibitemOpen
  \bibfield  {author} {\bibinfo {author} {\bibfnamefont {V.}~\bibnamefont
  {Giovannetti}}, \bibinfo {author} {\bibfnamefont {S.}~\bibnamefont {Lloyd}},
  \ and\ \bibinfo {author} {\bibfnamefont {L.}~\bibnamefont {Maccone}},\ }\href
  {\doibase 10.1038/nphoton.2011.35} {\bibfield  {journal} {\bibinfo  {journal}
  {Nature Photonics}\ }\textbf {\bibinfo {volume} {5}},\ \bibinfo {pages} {222}
  (\bibinfo {year} {2011})}\BibitemShut {NoStop}%
\bibitem [{\citenamefont {Polino}\ \emph {et~al.}(2020)\citenamefont {Polino},
  \citenamefont {Valeri}, \citenamefont {Spagnolo},\ and\ \citenamefont
  {Sciarrino}}]{Polino2020AVS}%
  \BibitemOpen
  \bibfield  {author} {\bibinfo {author} {\bibfnamefont {E.}~\bibnamefont
  {Polino}}, \bibinfo {author} {\bibfnamefont {M.}~\bibnamefont {Valeri}},
  \bibinfo {author} {\bibfnamefont {N.}~\bibnamefont {Spagnolo}}, \ and\
  \bibinfo {author} {\bibfnamefont {F.}~\bibnamefont {Sciarrino}},\ }\href
  {\doibase 10.1116/5.0007577} {\bibfield  {journal} {\bibinfo  {journal} {AVS
  Quantum Science}\ }\textbf {\bibinfo {volume} {2}},\ \bibinfo {pages}
  {024703} (\bibinfo {year} {2020})}\BibitemShut {NoStop}%
\bibitem [{\citenamefont {Chang}\ \emph {et~al.}(2014)\citenamefont {Chang},
  \citenamefont {Vuleti{\'c}},\ and\ \citenamefont {Lukin}}]{chang2014}%
  \BibitemOpen
  \bibfield  {author} {\bibinfo {author} {\bibfnamefont {D.~E.}\ \bibnamefont
  {Chang}}, \bibinfo {author} {\bibfnamefont {V.}~\bibnamefont {Vuleti{\'c}}},
  \ and\ \bibinfo {author} {\bibfnamefont {M.~D.}\ \bibnamefont {Lukin}},\
  }\href {\doibase 10.1038/nphoton.2014.192} {\bibfield  {journal} {\bibinfo
  {journal} {Nature Photonics}\ }\textbf {\bibinfo {volume} {8}},\ \bibinfo
  {pages} {685} (\bibinfo {year} {2014})}\BibitemShut {NoStop}%
\bibitem [{\citenamefont {Birnbaum}\ \emph {et~al.}(2005)\citenamefont
  {Birnbaum}, \citenamefont {Boca}, \citenamefont {Miller}, \citenamefont
  {Boozer}, \citenamefont {Northup},\ and\ \citenamefont
  {Kimble}}]{Birnbaum2005Nature}%
  \BibitemOpen
  \bibfield  {author} {\bibinfo {author} {\bibfnamefont {K.~M.}\ \bibnamefont
  {Birnbaum}}, \bibinfo {author} {\bibfnamefont {A.}~\bibnamefont {Boca}},
  \bibinfo {author} {\bibfnamefont {R.}~\bibnamefont {Miller}}, \bibinfo
  {author} {\bibfnamefont {A.~D.}\ \bibnamefont {Boozer}}, \bibinfo {author}
  {\bibfnamefont {T.~E.}\ \bibnamefont {Northup}}, \ and\ \bibinfo {author}
  {\bibfnamefont {H.~J.}\ \bibnamefont {Kimble}},\ }\href {\doibase
  10.1038/nature03804} {\bibfield  {journal} {\bibinfo  {journal} {Nature}\
  }\textbf {\bibinfo {volume} {436}},\ \bibinfo {pages} {87} (\bibinfo {year}
  {2005})}\BibitemShut {NoStop}%
\bibitem [{\citenamefont {Reiserer}\ and\ \citenamefont
  {Rempe}(2015)}]{Reiserer2015PRL}%
  \BibitemOpen
  \bibfield  {author} {\bibinfo {author} {\bibfnamefont {A.}~\bibnamefont
  {Reiserer}}\ and\ \bibinfo {author} {\bibfnamefont {G.}~\bibnamefont
  {Rempe}},\ }\href {\doibase 10.1103/RevModPhys.87.1379} {\bibfield  {journal}
  {\bibinfo  {journal} {Rev. Mod. Phys.}\ }\textbf {\bibinfo {volume} {87}},\
  \bibinfo {pages} {1379} (\bibinfo {year} {2015})}\BibitemShut {NoStop}%
\bibitem [{\citenamefont {Welte}\ \emph {et~al.}(2018)\citenamefont {Welte},
  \citenamefont {Hacker}, \citenamefont {Daiss}, \citenamefont {Ritter},\ and\
  \citenamefont {Rempe}}]{Welte2018PRX}%
  \BibitemOpen
  \bibfield  {author} {\bibinfo {author} {\bibfnamefont {S.}~\bibnamefont
  {Welte}}, \bibinfo {author} {\bibfnamefont {B.}~\bibnamefont {Hacker}},
  \bibinfo {author} {\bibfnamefont {S.}~\bibnamefont {Daiss}}, \bibinfo
  {author} {\bibfnamefont {S.}~\bibnamefont {Ritter}}, \ and\ \bibinfo {author}
  {\bibfnamefont {G.}~\bibnamefont {Rempe}},\ }\href {\doibase
  10.1103/PhysRevX.8.011018} {\bibfield  {journal} {\bibinfo  {journal} {Phys.
  Rev. X}\ }\textbf {\bibinfo {volume} {8}},\ \bibinfo {pages} {011018}
  (\bibinfo {year} {2018})}\BibitemShut {NoStop}%
\bibitem [{\citenamefont {Tiecke}\ \emph {et~al.}(2014)\citenamefont {Tiecke},
  \citenamefont {Thompson}, \citenamefont {de~Leon}, \citenamefont {Liu},
  \citenamefont {Vuleti{\'{c}}},\ and\ \citenamefont
  {Lukin}}]{Tiecke2014Nature}%
  \BibitemOpen
  \bibfield  {author} {\bibinfo {author} {\bibfnamefont {T.~G.}\ \bibnamefont
  {Tiecke}}, \bibinfo {author} {\bibfnamefont {J.~D.}\ \bibnamefont
  {Thompson}}, \bibinfo {author} {\bibfnamefont {N.~P.}\ \bibnamefont
  {de~Leon}}, \bibinfo {author} {\bibfnamefont {L.~R.}\ \bibnamefont {Liu}},
  \bibinfo {author} {\bibfnamefont {V.}~\bibnamefont {Vuleti{\'{c}}}}, \ and\
  \bibinfo {author} {\bibfnamefont {M.~D.}\ \bibnamefont {Lukin}},\ }\href
  {\doibase 10.1038/nature13188} {\bibfield  {journal} {\bibinfo  {journal}
  {Nature}\ }\textbf {\bibinfo {volume} {508}},\ \bibinfo {pages} {241}
  (\bibinfo {year} {2014})}\BibitemShut {NoStop}%
\bibitem [{\citenamefont {Petersen}\ \emph {et~al.}(2014)\citenamefont
  {Petersen}, \citenamefont {Volz},\ and\ \citenamefont
  {Rauschenbeutel}}]{Petersen2014Science}%
  \BibitemOpen
  \bibfield  {author} {\bibinfo {author} {\bibfnamefont {J.}~\bibnamefont
  {Petersen}}, \bibinfo {author} {\bibfnamefont {J.}~\bibnamefont {Volz}}, \
  and\ \bibinfo {author} {\bibfnamefont {A.}~\bibnamefont {Rauschenbeutel}},\
  }\href {\doibase 10.1126/science.1257671} {\bibfield  {journal} {\bibinfo
  {journal} {Science}\ }\textbf {\bibinfo {volume} {346}},\ \bibinfo {pages}
  {67} (\bibinfo {year} {2014})}\BibitemShut {NoStop}%
\bibitem [{\citenamefont {Lodahl}\ \emph {et~al.}(2015)\citenamefont {Lodahl},
  \citenamefont {Mahmoodian},\ and\ \citenamefont {Stobbe}}]{Lodahl2015RMP}%
  \BibitemOpen
  \bibfield  {author} {\bibinfo {author} {\bibfnamefont {P.}~\bibnamefont
  {Lodahl}}, \bibinfo {author} {\bibfnamefont {S.}~\bibnamefont {Mahmoodian}},
  \ and\ \bibinfo {author} {\bibfnamefont {S.}~\bibnamefont {Stobbe}},\ }\href
  {\doibase 10.1103/RevModPhys.87.347} {\bibfield  {journal} {\bibinfo
  {journal} {Rev. Mod. Phys.}\ }\textbf {\bibinfo {volume} {87}},\ \bibinfo
  {pages} {347} (\bibinfo {year} {2015})}\BibitemShut {NoStop}%
\bibitem [{\citenamefont {Chang}\ \emph {et~al.}(2018)\citenamefont {Chang},
  \citenamefont {Douglas}, \citenamefont {Gonz\'alez-Tudela}, \citenamefont
  {Hung},\ and\ \citenamefont {Kimble}}]{Chang2018RMP}%
  \BibitemOpen
  \bibfield  {author} {\bibinfo {author} {\bibfnamefont {D.~E.}\ \bibnamefont
  {Chang}}, \bibinfo {author} {\bibfnamefont {J.~S.}\ \bibnamefont {Douglas}},
  \bibinfo {author} {\bibfnamefont {A.}~\bibnamefont {Gonz\'alez-Tudela}},
  \bibinfo {author} {\bibfnamefont {C.-L.}\ \bibnamefont {Hung}}, \ and\
  \bibinfo {author} {\bibfnamefont {H.~J.}\ \bibnamefont {Kimble}},\ }\href
  {\doibase 10.1103/RevModPhys.90.031002} {\bibfield  {journal} {\bibinfo
  {journal} {Rev. Mod. Phys.}\ }\textbf {\bibinfo {volume} {90}},\ \bibinfo
  {pages} {031002} (\bibinfo {year} {2018})}\BibitemShut {NoStop}%
\bibitem [{\citenamefont {Yu}\ \emph {et~al.}(2019)\citenamefont {Yu},
  \citenamefont {Muniz}, \citenamefont {Hung},\ and\ \citenamefont
  {Kimble}}]{Yu2019PNAS}%
  \BibitemOpen
  \bibfield  {author} {\bibinfo {author} {\bibfnamefont {S.-P.}\ \bibnamefont
  {Yu}}, \bibinfo {author} {\bibfnamefont {J.~A.}\ \bibnamefont {Muniz}},
  \bibinfo {author} {\bibfnamefont {C.-L.}\ \bibnamefont {Hung}}, \ and\
  \bibinfo {author} {\bibfnamefont {H.~J.}\ \bibnamefont {Kimble}},\ }\href
  {\doibase 10.1073/pnas.1822110116} {\bibfield  {journal} {\bibinfo  {journal}
  {Proceedings of the National Academy of Sciences}\ }\textbf {\bibinfo
  {volume} {116}},\ \bibinfo {pages} {12743} (\bibinfo {year}
  {2019})}\BibitemShut {NoStop}%
\bibitem [{\citenamefont {Prasad}\ \emph {et~al.}(2020)\citenamefont {Prasad},
  \citenamefont {Hinney}, \citenamefont {Mahmoodian}, \citenamefont {Hammerer},
  \citenamefont {Rind}, \citenamefont {Schneeweiss}, \citenamefont
  {S{\o}rensen}, \citenamefont {Volz},\ and\ \citenamefont
  {Rauschenbeutel}}]{Prasad2020NPhoton}%
  \BibitemOpen
  \bibfield  {author} {\bibinfo {author} {\bibfnamefont {A.~S.}\ \bibnamefont
  {Prasad}}, \bibinfo {author} {\bibfnamefont {J.}~\bibnamefont {Hinney}},
  \bibinfo {author} {\bibfnamefont {S.}~\bibnamefont {Mahmoodian}}, \bibinfo
  {author} {\bibfnamefont {K.}~\bibnamefont {Hammerer}}, \bibinfo {author}
  {\bibfnamefont {S.}~\bibnamefont {Rind}}, \bibinfo {author} {\bibfnamefont
  {P.}~\bibnamefont {Schneeweiss}}, \bibinfo {author} {\bibfnamefont {A.~S.}\
  \bibnamefont {S{\o}rensen}}, \bibinfo {author} {\bibfnamefont
  {J.}~\bibnamefont {Volz}}, \ and\ \bibinfo {author} {\bibfnamefont
  {A.}~\bibnamefont {Rauschenbeutel}},\ }\href {\doibase
  10.1038/s41566-020-0692-z} {\bibfield  {journal} {\bibinfo  {journal} {Nature
  Photonics}\ }\textbf {\bibinfo {volume} {14}},\ \bibinfo {pages} {719}
  (\bibinfo {year} {2020})}\BibitemShut {NoStop}%
\bibitem [{\citenamefont {Peyronel}\ \emph {et~al.}(2012)\citenamefont
  {Peyronel}, \citenamefont {Firstenberg}, \citenamefont {Liang}, \citenamefont
  {Hofferberth}, \citenamefont {Gorshkov}, \citenamefont {Pohl}, \citenamefont
  {Lukin},\ and\ \citenamefont {Vuleti{\'{c}}}}]{Peyronel2012Nature}%
  \BibitemOpen
  \bibfield  {author} {\bibinfo {author} {\bibfnamefont {T.}~\bibnamefont
  {Peyronel}}, \bibinfo {author} {\bibfnamefont {O.}~\bibnamefont
  {Firstenberg}}, \bibinfo {author} {\bibfnamefont {Q.-Y.}\ \bibnamefont
  {Liang}}, \bibinfo {author} {\bibfnamefont {S.}~\bibnamefont {Hofferberth}},
  \bibinfo {author} {\bibfnamefont {A.~V.}\ \bibnamefont {Gorshkov}}, \bibinfo
  {author} {\bibfnamefont {T.}~\bibnamefont {Pohl}}, \bibinfo {author}
  {\bibfnamefont {M.~D.}\ \bibnamefont {Lukin}}, \ and\ \bibinfo {author}
  {\bibfnamefont {V.}~\bibnamefont {Vuleti{\'{c}}}},\ }\href {\doibase
  10.1038/nature11361} {\bibfield  {journal} {\bibinfo  {journal} {Nature}\
  }\textbf {\bibinfo {volume} {488}},\ \bibinfo {pages} {57} (\bibinfo {year}
  {2012})}\BibitemShut {NoStop}%
\bibitem [{\citenamefont {Paris-Mandoki}\ \emph
  {et~al.}(2017{\natexlab{a}})\citenamefont {Paris-Mandoki}, \citenamefont
  {Braun}, \citenamefont {Kumlin}, \citenamefont {Tresp}, \citenamefont
  {Mirgorodskiy}, \citenamefont {Christaller}, \citenamefont {B\"uchler},\ and\
  \citenamefont {Hofferberth}}]{Paris2010PRX}%
  \BibitemOpen
  \bibfield  {author} {\bibinfo {author} {\bibfnamefont {A.}~\bibnamefont
  {Paris-Mandoki}}, \bibinfo {author} {\bibfnamefont {C.}~\bibnamefont
  {Braun}}, \bibinfo {author} {\bibfnamefont {J.}~\bibnamefont {Kumlin}},
  \bibinfo {author} {\bibfnamefont {C.}~\bibnamefont {Tresp}}, \bibinfo
  {author} {\bibfnamefont {I.}~\bibnamefont {Mirgorodskiy}}, \bibinfo {author}
  {\bibfnamefont {F.}~\bibnamefont {Christaller}}, \bibinfo {author}
  {\bibfnamefont {H.~P.}\ \bibnamefont {B\"uchler}}, \ and\ \bibinfo {author}
  {\bibfnamefont {S.}~\bibnamefont {Hofferberth}},\ }\href {\doibase
  10.1103/PhysRevX.7.041010} {\bibfield  {journal} {\bibinfo  {journal} {Phys.
  Rev. X}\ }\textbf {\bibinfo {volume} {7}},\ \bibinfo {pages} {041010}
  (\bibinfo {year} {2017}{\natexlab{a}})}\BibitemShut {NoStop}%
\bibitem [{\citenamefont {Murray}\ and\ \citenamefont
  {Pohl}(2016)}]{Murray2016Adv}%
  \BibitemOpen
  \bibfield  {author} {\bibinfo {author} {\bibfnamefont {C.}~\bibnamefont
  {Murray}}\ and\ \bibinfo {author} {\bibfnamefont {T.}~\bibnamefont {Pohl}},\
  }\href {\doibase https://doi.org/10.1016/bs.aamop.2016.04.005} {\bibfield
  {journal} {\bibinfo  {journal} {Adv. At. Mol. Opt. Phys.}\ }\textbf {\bibinfo
  {volume} {65}},\ \bibinfo {pages} {321} (\bibinfo {year} {2016})}\BibitemShut
  {NoStop}%
\bibitem [{\citenamefont {Firstenberg}\ \emph {et~al.}(2016)\citenamefont
  {Firstenberg}, \citenamefont {Adams},\ and\ \citenamefont
  {Hofferberth}}]{Firstenberg2016JPB}%
  \BibitemOpen
  \bibfield  {author} {\bibinfo {author} {\bibfnamefont {O.}~\bibnamefont
  {Firstenberg}}, \bibinfo {author} {\bibfnamefont {C.~S.}\ \bibnamefont
  {Adams}}, \ and\ \bibinfo {author} {\bibfnamefont {S.}~\bibnamefont
  {Hofferberth}},\ }\href {\doibase 10.1088/0953-4075/49/15/152003} {\bibfield
  {journal} {\bibinfo  {journal} {Journal of Physics B: Atomic, Molecular and
  Optical Physics}\ }\textbf {\bibinfo {volume} {49}},\ \bibinfo {pages}
  {152003} (\bibinfo {year} {2016})}\BibitemShut {NoStop}%
\bibitem [{\citenamefont {Thompson}\ \emph {et~al.}(2017)\citenamefont
  {Thompson}, \citenamefont {Nicholson}, \citenamefont {Liang}, \citenamefont
  {Cantu}, \citenamefont {Venkatramani}, \citenamefont {Choi}, \citenamefont
  {Fedorov}, \citenamefont {Viscor}, \citenamefont {Pohl}, \citenamefont
  {Lukin},\ and\ \citenamefont {Vuleti{\'{c}}}}]{Thompson2017Nature}%
  \BibitemOpen
  \bibfield  {author} {\bibinfo {author} {\bibfnamefont {J.~D.}\ \bibnamefont
  {Thompson}}, \bibinfo {author} {\bibfnamefont {T.~L.}\ \bibnamefont
  {Nicholson}}, \bibinfo {author} {\bibfnamefont {Q.-Y.}\ \bibnamefont
  {Liang}}, \bibinfo {author} {\bibfnamefont {S.~H.}\ \bibnamefont {Cantu}},
  \bibinfo {author} {\bibfnamefont {A.~V.}\ \bibnamefont {Venkatramani}},
  \bibinfo {author} {\bibfnamefont {S.}~\bibnamefont {Choi}}, \bibinfo {author}
  {\bibfnamefont {I.~A.}\ \bibnamefont {Fedorov}}, \bibinfo {author}
  {\bibfnamefont {D.}~\bibnamefont {Viscor}}, \bibinfo {author} {\bibfnamefont
  {T.}~\bibnamefont {Pohl}}, \bibinfo {author} {\bibfnamefont {M.~D.}\
  \bibnamefont {Lukin}}, \ and\ \bibinfo {author} {\bibfnamefont
  {V.}~\bibnamefont {Vuleti{\'{c}}}},\ }\href {\doibase 10.1038/nature20823}
  {\bibfield  {journal} {\bibinfo  {journal} {Nature}\ }\textbf {\bibinfo
  {volume} {542}},\ \bibinfo {pages} {206} (\bibinfo {year}
  {2017})}\BibitemShut {NoStop}%
\bibitem [{\citenamefont {Yang}\ \emph {et~al.}(2020)\citenamefont {Yang},
  \citenamefont {Liu},\ and\ \citenamefont {You}}]{yang2020atom}%
  \BibitemOpen
  \bibfield  {author} {\bibinfo {author} {\bibfnamefont {F.}~\bibnamefont
  {Yang}}, \bibinfo {author} {\bibfnamefont {Y.-C.}\ \bibnamefont {Liu}}, \
  and\ \bibinfo {author} {\bibfnamefont {L.}~\bibnamefont {You}},\ }\href
  {\doibase 10.1103/PhysRevLett.125.143601} {\bibfield  {journal} {\bibinfo
  {journal} {Phys. Rev. Lett.}\ }\textbf {\bibinfo {volume} {125}},\ \bibinfo
  {pages} {143601} (\bibinfo {year} {2020})}\BibitemShut {NoStop}%
\bibitem [{\citenamefont {Facchinetti}\ \emph {et~al.}(2016)\citenamefont
  {Facchinetti}, \citenamefont {Jenkins},\ and\ \citenamefont
  {Ruostekoski}}]{Facchinetti2016PRL}%
  \BibitemOpen
  \bibfield  {author} {\bibinfo {author} {\bibfnamefont {G.}~\bibnamefont
  {Facchinetti}}, \bibinfo {author} {\bibfnamefont {S.~D.}\ \bibnamefont
  {Jenkins}}, \ and\ \bibinfo {author} {\bibfnamefont {J.}~\bibnamefont
  {Ruostekoski}},\ }\href {\doibase 10.1103/PhysRevLett.117.243601} {\bibfield
  {journal} {\bibinfo  {journal} {Phys. Rev. Lett.}\ }\textbf {\bibinfo
  {volume} {117}},\ \bibinfo {pages} {243601} (\bibinfo {year}
  {2016})}\BibitemShut {NoStop}%
\bibitem [{\citenamefont {Bettles}\ \emph {et~al.}(2016)\citenamefont
  {Bettles}, \citenamefont {Gardiner},\ and\ \citenamefont
  {Adams}}]{Bettles2016PRL}%
  \BibitemOpen
  \bibfield  {author} {\bibinfo {author} {\bibfnamefont {R.~J.}\ \bibnamefont
  {Bettles}}, \bibinfo {author} {\bibfnamefont {S.~A.}\ \bibnamefont
  {Gardiner}}, \ and\ \bibinfo {author} {\bibfnamefont {C.~S.}\ \bibnamefont
  {Adams}},\ }\href {\doibase 10.1103/PhysRevLett.116.103602} {\bibfield
  {journal} {\bibinfo  {journal} {Phys. Rev. Lett.}\ }\textbf {\bibinfo
  {volume} {116}},\ \bibinfo {pages} {103602} (\bibinfo {year}
  {2016})}\BibitemShut {NoStop}%
\bibitem [{\citenamefont {Shahmoon}\ \emph {et~al.}(2017)\citenamefont
  {Shahmoon}, \citenamefont {Wild}, \citenamefont {Lukin},\ and\ \citenamefont
  {Yelin}}]{Shahmoon2017PRL}%
  \BibitemOpen
  \bibfield  {author} {\bibinfo {author} {\bibfnamefont {E.}~\bibnamefont
  {Shahmoon}}, \bibinfo {author} {\bibfnamefont {D.~S.}\ \bibnamefont {Wild}},
  \bibinfo {author} {\bibfnamefont {M.~D.}\ \bibnamefont {Lukin}}, \ and\
  \bibinfo {author} {\bibfnamefont {S.~F.}\ \bibnamefont {Yelin}},\ }\href
  {\doibase 10.1103/PhysRevLett.118.113601} {\bibfield  {journal} {\bibinfo
  {journal} {Phys. Rev. Lett.}\ }\textbf {\bibinfo {volume} {118}},\ \bibinfo
  {pages} {113601} (\bibinfo {year} {2017})}\BibitemShut {NoStop}%
\bibitem [{\citenamefont {Perczel}\ \emph {et~al.}(2017)\citenamefont
  {Perczel}, \citenamefont {Borregaard}, \citenamefont {Chang}, \citenamefont
  {Pichler}, \citenamefont {Yelin}, \citenamefont {Zoller},\ and\ \citenamefont
  {Lukin}}]{Perczel2017PRL}%
  \BibitemOpen
  \bibfield  {author} {\bibinfo {author} {\bibfnamefont {J.}~\bibnamefont
  {Perczel}}, \bibinfo {author} {\bibfnamefont {J.}~\bibnamefont {Borregaard}},
  \bibinfo {author} {\bibfnamefont {D.~E.}\ \bibnamefont {Chang}}, \bibinfo
  {author} {\bibfnamefont {H.}~\bibnamefont {Pichler}}, \bibinfo {author}
  {\bibfnamefont {S.~F.}\ \bibnamefont {Yelin}}, \bibinfo {author}
  {\bibfnamefont {P.}~\bibnamefont {Zoller}}, \ and\ \bibinfo {author}
  {\bibfnamefont {M.~D.}\ \bibnamefont {Lukin}},\ }\href {\doibase
  10.1103/PhysRevLett.119.023603} {\bibfield  {journal} {\bibinfo  {journal}
  {Phys. Rev. Lett.}\ }\textbf {\bibinfo {volume} {119}},\ \bibinfo {pages}
  {023603} (\bibinfo {year} {2017})}\BibitemShut {NoStop}%
\bibitem [{\citenamefont {Guimond}\ \emph {et~al.}(2019)\citenamefont
  {Guimond}, \citenamefont {Grankin}, \citenamefont {Vasilyev}, \citenamefont
  {Vermersch},\ and\ \citenamefont {Zoller}}]{Guimond2019PRL}%
  \BibitemOpen
  \bibfield  {author} {\bibinfo {author} {\bibfnamefont {P.-O.}\ \bibnamefont
  {Guimond}}, \bibinfo {author} {\bibfnamefont {A.}~\bibnamefont {Grankin}},
  \bibinfo {author} {\bibfnamefont {D.~V.}\ \bibnamefont {Vasilyev}}, \bibinfo
  {author} {\bibfnamefont {B.}~\bibnamefont {Vermersch}}, \ and\ \bibinfo
  {author} {\bibfnamefont {P.}~\bibnamefont {Zoller}},\ }\href {\doibase
  10.1103/PhysRevLett.122.093601} {\bibfield  {journal} {\bibinfo  {journal}
  {Phys. Rev. Lett.}\ }\textbf {\bibinfo {volume} {122}},\ \bibinfo {pages}
  {093601} (\bibinfo {year} {2019})}\BibitemShut {NoStop}%
\bibitem [{\citenamefont {Ballantine}\ and\ \citenamefont
  {Ruostekoski}(2020)}]{Ballantine2020PRL}%
  \BibitemOpen
  \bibfield  {author} {\bibinfo {author} {\bibfnamefont {K.~E.}\ \bibnamefont
  {Ballantine}}\ and\ \bibinfo {author} {\bibfnamefont {J.}~\bibnamefont
  {Ruostekoski}},\ }\href {\doibase 10.1103/PhysRevLett.125.143604} {\bibfield
  {journal} {\bibinfo  {journal} {Phys. Rev. Lett.}\ }\textbf {\bibinfo
  {volume} {125}},\ \bibinfo {pages} {143604} (\bibinfo {year}
  {2020})}\BibitemShut {NoStop}%
\bibitem [{\citenamefont {Grankin}\ \emph {et~al.}(2018)\citenamefont
  {Grankin}, \citenamefont {Guimond}, \citenamefont {Vasilyev}, \citenamefont
  {Vermersch},\ and\ \citenamefont {Zoller}}]{Grankin2018PRA}%
  \BibitemOpen
  \bibfield  {author} {\bibinfo {author} {\bibfnamefont {A.}~\bibnamefont
  {Grankin}}, \bibinfo {author} {\bibfnamefont {P.~O.}\ \bibnamefont
  {Guimond}}, \bibinfo {author} {\bibfnamefont {D.~V.}\ \bibnamefont
  {Vasilyev}}, \bibinfo {author} {\bibfnamefont {B.}~\bibnamefont {Vermersch}},
  \ and\ \bibinfo {author} {\bibfnamefont {P.}~\bibnamefont {Zoller}},\ }\href
  {\doibase 10.1103/PhysRevA.98.043825} {\bibfield  {journal} {\bibinfo
  {journal} {Phys. Rev. A}\ }\textbf {\bibinfo {volume} {98}},\ \bibinfo
  {pages} {043825} (\bibinfo {year} {2018})}\BibitemShut {NoStop}%
\bibitem [{\citenamefont {Ba{\ss}ler}\ \emph {et~al.}(2023)\citenamefont
  {Ba{\ss}ler}, \citenamefont {Reitz}, \citenamefont {Schmidt},\ and\
  \citenamefont {Genes}}]{BaSSler:23}%
  \BibitemOpen
  \bibfield  {author} {\bibinfo {author} {\bibfnamefont {N.~S.}\ \bibnamefont
  {Ba{\ss}ler}}, \bibinfo {author} {\bibfnamefont {M.}~\bibnamefont {Reitz}},
  \bibinfo {author} {\bibfnamefont {K.~P.}\ \bibnamefont {Schmidt}}, \ and\
  \bibinfo {author} {\bibfnamefont {C.}~\bibnamefont {Genes}},\ }\href
  {\doibase 10.1364/OE.476830} {\bibfield  {journal} {\bibinfo  {journal} {Opt.
  Express}\ }\textbf {\bibinfo {volume} {31}},\ \bibinfo {pages} {6003}
  (\bibinfo {year} {2023})}\BibitemShut {NoStop}%
\bibitem [{\citenamefont {Pedersen}\ \emph {et~al.}(2023)\citenamefont
  {Pedersen}, \citenamefont {Zhang},\ and\ \citenamefont
  {Pohl}}]{pedersen2023}%
  \BibitemOpen
  \bibfield  {author} {\bibinfo {author} {\bibfnamefont {S.~P.}\ \bibnamefont
  {Pedersen}}, \bibinfo {author} {\bibfnamefont {L.}~\bibnamefont {Zhang}}, \
  and\ \bibinfo {author} {\bibfnamefont {T.}~\bibnamefont {Pohl}},\ }\href
  {\doibase 10.1103/PhysRevResearch.5.L012047} {\bibfield  {journal} {\bibinfo
  {journal} {Phys. Rev. Res.}\ }\textbf {\bibinfo {volume} {5}},\ \bibinfo
  {pages} {L012047} (\bibinfo {year} {2023})}\BibitemShut {NoStop}%
\bibitem [{\citenamefont {Bekenstein}\ \emph {et~al.}(2020)\citenamefont
  {Bekenstein}, \citenamefont {Pikovski}, \citenamefont {Pichler},
  \citenamefont {Shahmoon}, \citenamefont {Yelin},\ and\ \citenamefont
  {Lukin}}]{Bekenstein2020NPhys}%
  \BibitemOpen
  \bibfield  {author} {\bibinfo {author} {\bibfnamefont {R.}~\bibnamefont
  {Bekenstein}}, \bibinfo {author} {\bibfnamefont {I.}~\bibnamefont
  {Pikovski}}, \bibinfo {author} {\bibfnamefont {H.}~\bibnamefont {Pichler}},
  \bibinfo {author} {\bibfnamefont {E.}~\bibnamefont {Shahmoon}}, \bibinfo
  {author} {\bibfnamefont {S.~F.}\ \bibnamefont {Yelin}}, \ and\ \bibinfo
  {author} {\bibfnamefont {M.~D.}\ \bibnamefont {Lukin}},\ }\href {\doibase
  10.1038/s41567-020-0845-5} {\bibfield  {journal} {\bibinfo  {journal} {Nature
  Physics}\ }\textbf {\bibinfo {volume} {16}},\ \bibinfo {pages} {676}
  (\bibinfo {year} {2020})}\BibitemShut {NoStop}%
\bibitem [{\citenamefont {Moreno-Cardoner}\ \emph {et~al.}(2021)\citenamefont
  {Moreno-Cardoner}, \citenamefont {Goncalves},\ and\ \citenamefont
  {Chang}}]{Moreno2021PRL}%
  \BibitemOpen
  \bibfield  {author} {\bibinfo {author} {\bibfnamefont {M.}~\bibnamefont
  {Moreno-Cardoner}}, \bibinfo {author} {\bibfnamefont {D.}~\bibnamefont
  {Goncalves}}, \ and\ \bibinfo {author} {\bibfnamefont {D.~E.}\ \bibnamefont
  {Chang}},\ }\href {\doibase 10.1103/PhysRevLett.127.263602} {\bibfield
  {journal} {\bibinfo  {journal} {Phys. Rev. Lett.}\ }\textbf {\bibinfo
  {volume} {127}},\ \bibinfo {pages} {263602} (\bibinfo {year}
  {2021})}\BibitemShut {NoStop}%
\bibitem [{\citenamefont {Zhang}\ \emph {et~al.}(2022)\citenamefont {Zhang},
  \citenamefont {Walther}, \citenamefont {M{\o{}}lmer},\ and\ \citenamefont
  {Pohl}}]{Zhang2022Quantum}%
  \BibitemOpen
  \bibfield  {author} {\bibinfo {author} {\bibfnamefont {L.}~\bibnamefont
  {Zhang}}, \bibinfo {author} {\bibfnamefont {V.}~\bibnamefont {Walther}},
  \bibinfo {author} {\bibfnamefont {K.}~\bibnamefont {M{\o{}}lmer}}, \ and\
  \bibinfo {author} {\bibfnamefont {T.}~\bibnamefont {Pohl}},\ }\href {\doibase
  10.22331/q-2022-03-30-674} {\bibfield  {journal} {\bibinfo  {journal}
  {{Quantum}}\ }\textbf {\bibinfo {volume} {6}},\ \bibinfo {pages} {674}
  (\bibinfo {year} {2022})}\BibitemShut {NoStop}%
\bibitem [{\citenamefont {Sheremet}\ \emph {et~al.}(2023)\citenamefont
  {Sheremet}, \citenamefont {Petrov}, \citenamefont {Iorsh}, \citenamefont
  {Poshakinskiy},\ and\ \citenamefont {Poddubny}}]{Sheremet23}%
  \BibitemOpen
  \bibfield  {author} {\bibinfo {author} {\bibfnamefont {A.~S.}\ \bibnamefont
  {Sheremet}}, \bibinfo {author} {\bibfnamefont {M.~I.}\ \bibnamefont
  {Petrov}}, \bibinfo {author} {\bibfnamefont {I.~V.}\ \bibnamefont {Iorsh}},
  \bibinfo {author} {\bibfnamefont {A.~V.}\ \bibnamefont {Poshakinskiy}}, \
  and\ \bibinfo {author} {\bibfnamefont {A.~N.}\ \bibnamefont {Poddubny}},\
  }\href {\doibase 10.1103/RevModPhys.95.015002} {\bibfield  {journal}
  {\bibinfo  {journal} {Rev. Mod. Phys.}\ }\textbf {\bibinfo {volume} {95}},\
  \bibinfo {pages} {015002} (\bibinfo {year} {2023})}\BibitemShut {NoStop}%
\bibitem [{\citenamefont {Gross}\ and\ \citenamefont
  {Haroche}(1982)}]{Gross1982PhysRep}%
  \BibitemOpen
  \bibfield  {author} {\bibinfo {author} {\bibfnamefont {M.}~\bibnamefont
  {Gross}}\ and\ \bibinfo {author} {\bibfnamefont {S.}~\bibnamefont
  {Haroche}},\ }\href {\doibase https://doi.org/10.1016/0370-1573(82)90102-8}
  {\bibfield  {journal} {\bibinfo  {journal} {Physics Reports}\ }\textbf
  {\bibinfo {volume} {93}},\ \bibinfo {pages} {301} (\bibinfo {year}
  {1982})}\BibitemShut {NoStop}%
\bibitem [{\citenamefont {Dung}\ \emph {et~al.}(2002)\citenamefont {Dung},
  \citenamefont {Kn\"oll},\ and\ \citenamefont {Welsch}}]{Dung2002PRA}%
  \BibitemOpen
  \bibfield  {author} {\bibinfo {author} {\bibfnamefont {H.~T.}\ \bibnamefont
  {Dung}}, \bibinfo {author} {\bibfnamefont {L.}~\bibnamefont {Kn\"oll}}, \
  and\ \bibinfo {author} {\bibfnamefont {D.-G.}\ \bibnamefont {Welsch}},\
  }\href {\doibase 10.1103/PhysRevA.66.063810} {\bibfield  {journal} {\bibinfo
  {journal} {Phys. Rev. A}\ }\textbf {\bibinfo {volume} {66}},\ \bibinfo
  {pages} {063810} (\bibinfo {year} {2002})}\BibitemShut {NoStop}%
\bibitem [{\citenamefont {Ficek}\ and\ \citenamefont
  {Tana\'{s}}(2002)}]{Ficek2002PhysRep}%
  \BibitemOpen
  \bibfield  {author} {\bibinfo {author} {\bibfnamefont {Z.}~\bibnamefont
  {Ficek}}\ and\ \bibinfo {author} {\bibfnamefont {R.}~\bibnamefont
  {Tana\'{s}}},\ }\href {\doibase
  https://doi.org/10.1016/S0370-1573(02)00368-X} {\bibfield  {journal}
  {\bibinfo  {journal} {Physics Reports}\ }\textbf {\bibinfo {volume} {372}},\
  \bibinfo {pages} {369} (\bibinfo {year} {2002})}\BibitemShut {NoStop}%
\bibitem [{\citenamefont {Manzoni}\ \emph {et~al.}(2018)\citenamefont
  {Manzoni}, \citenamefont {Moreno-Cardoner}, \citenamefont {Asenjo-Garcia},
  \citenamefont {Porto}, \citenamefont {Gorshkov},\ and\ \citenamefont
  {Chang}}]{Manzoni2018NJP}%
  \BibitemOpen
  \bibfield  {author} {\bibinfo {author} {\bibfnamefont {M.~T.}\ \bibnamefont
  {Manzoni}}, \bibinfo {author} {\bibfnamefont {M.}~\bibnamefont
  {Moreno-Cardoner}}, \bibinfo {author} {\bibfnamefont {A.}~\bibnamefont
  {Asenjo-Garcia}}, \bibinfo {author} {\bibfnamefont {J.~V.}\ \bibnamefont
  {Porto}}, \bibinfo {author} {\bibfnamefont {A.~V.}\ \bibnamefont {Gorshkov}},
  \ and\ \bibinfo {author} {\bibfnamefont {D.~E.}\ \bibnamefont {Chang}},\
  }\href {\doibase 10.1088/1367-2630/aadb74} {\bibfield  {journal} {\bibinfo
  {journal} {New Journal of Physics}\ }\textbf {\bibinfo {volume} {20}},\
  \bibinfo {pages} {083048} (\bibinfo {year} {2018})}\BibitemShut {NoStop}%
\bibitem [{\citenamefont {Fleischhauer}\ \emph {et~al.}(2005)\citenamefont
  {Fleischhauer}, \citenamefont {Imamoglu},\ and\ \citenamefont
  {Marangos}}]{Fleischhauer2005RMP}%
  \BibitemOpen
  \bibfield  {author} {\bibinfo {author} {\bibfnamefont {M.}~\bibnamefont
  {Fleischhauer}}, \bibinfo {author} {\bibfnamefont {A.}~\bibnamefont
  {Imamoglu}}, \ and\ \bibinfo {author} {\bibfnamefont {J.~P.}\ \bibnamefont
  {Marangos}},\ }\href {\doibase 10.1103/RevModPhys.77.633} {\bibfield
  {journal} {\bibinfo  {journal} {Rev. Mod. Phys.}\ }\textbf {\bibinfo {volume}
  {77}},\ \bibinfo {pages} {633} (\bibinfo {year} {2005})}\BibitemShut
  {NoStop}%
\bibitem [{\citenamefont {Rui}\ \emph {et~al.}(2020)\citenamefont {Rui},
  \citenamefont {Wei}, \citenamefont {Rubio-Abadal}, \citenamefont {Hollerith},
  \citenamefont {Zeiher}, \citenamefont {Stamper-Kurn}, \citenamefont {Gross},\
  and\ \citenamefont {Bloch}}]{Rui2020Nature}%
  \BibitemOpen
  \bibfield  {author} {\bibinfo {author} {\bibfnamefont {J.}~\bibnamefont
  {Rui}}, \bibinfo {author} {\bibfnamefont {D.}~\bibnamefont {Wei}}, \bibinfo
  {author} {\bibfnamefont {A.}~\bibnamefont {Rubio-Abadal}}, \bibinfo {author}
  {\bibfnamefont {S.}~\bibnamefont {Hollerith}}, \bibinfo {author}
  {\bibfnamefont {J.}~\bibnamefont {Zeiher}}, \bibinfo {author} {\bibfnamefont
  {D.~M.}\ \bibnamefont {Stamper-Kurn}}, \bibinfo {author} {\bibfnamefont
  {C.}~\bibnamefont {Gross}}, \ and\ \bibinfo {author} {\bibfnamefont
  {I.}~\bibnamefont {Bloch}},\ }\href {\doibase 10.1038/s41586-020-2463-x}
  {\bibfield  {journal} {\bibinfo  {journal} {Nature}\ }\textbf {\bibinfo
  {volume} {583}},\ \bibinfo {pages} {369} (\bibinfo {year}
  {2020})}\BibitemShut {NoStop}%
\bibitem [{\citenamefont {Browaeys}\ and\ \citenamefont
  {Lahaye}(2020)}]{Browaeys2020NPhys}%
  \BibitemOpen
  \bibfield  {author} {\bibinfo {author} {\bibfnamefont {A.}~\bibnamefont
  {Browaeys}}\ and\ \bibinfo {author} {\bibfnamefont {T.}~\bibnamefont
  {Lahaye}},\ }\href {\doibase 10.1038/s41567-019-0733-z} {\bibfield  {journal}
  {\bibinfo  {journal} {Nature Physics}\ }\textbf {\bibinfo {volume} {16}},\
  \bibinfo {pages} {132} (\bibinfo {year} {2020})}\BibitemShut {NoStop}%
\bibitem [{\citenamefont {Ebadi}\ \emph {et~al.}(2021)\citenamefont {Ebadi},
  \citenamefont {Wang}, \citenamefont {Levine}, \citenamefont {Keesling},
  \citenamefont {Semeghini}, \citenamefont {Omran}, \citenamefont {Bluvstein},
  \citenamefont {Samajdar}, \citenamefont {Pichler}, \citenamefont {Ho},
  \citenamefont {Choi}, \citenamefont {Sachdev}, \citenamefont {Greiner},
  \citenamefont {Vuleti{\'{c}}},\ and\ \citenamefont
  {Lukin}}]{Ebadi2021Nature}%
  \BibitemOpen
  \bibfield  {author} {\bibinfo {author} {\bibfnamefont {S.}~\bibnamefont
  {Ebadi}}, \bibinfo {author} {\bibfnamefont {T.~T.}\ \bibnamefont {Wang}},
  \bibinfo {author} {\bibfnamefont {H.}~\bibnamefont {Levine}}, \bibinfo
  {author} {\bibfnamefont {A.}~\bibnamefont {Keesling}}, \bibinfo {author}
  {\bibfnamefont {G.}~\bibnamefont {Semeghini}}, \bibinfo {author}
  {\bibfnamefont {A.}~\bibnamefont {Omran}}, \bibinfo {author} {\bibfnamefont
  {D.}~\bibnamefont {Bluvstein}}, \bibinfo {author} {\bibfnamefont
  {R.}~\bibnamefont {Samajdar}}, \bibinfo {author} {\bibfnamefont
  {H.}~\bibnamefont {Pichler}}, \bibinfo {author} {\bibfnamefont {W.~W.}\
  \bibnamefont {Ho}}, \bibinfo {author} {\bibfnamefont {S.}~\bibnamefont
  {Choi}}, \bibinfo {author} {\bibfnamefont {S.}~\bibnamefont {Sachdev}},
  \bibinfo {author} {\bibfnamefont {M.}~\bibnamefont {Greiner}}, \bibinfo
  {author} {\bibfnamefont {V.}~\bibnamefont {Vuleti{\'{c}}}}, \ and\ \bibinfo
  {author} {\bibfnamefont {M.~D.}\ \bibnamefont {Lukin}},\ }\href {\doibase
  10.1038/s41586-021-03582-4} {\bibfield  {journal} {\bibinfo  {journal}
  {Nature}\ }\textbf {\bibinfo {volume} {595}},\ \bibinfo {pages} {227}
  (\bibinfo {year} {2021})}\BibitemShut {NoStop}%
\bibitem [{\citenamefont {Scholl}\ \emph {et~al.}(2021)\citenamefont {Scholl},
  \citenamefont {Schuler}, \citenamefont {Williams}, \citenamefont
  {Eberharter}, \citenamefont {Barredo}, \citenamefont {Schymik}, \citenamefont
  {Lienhard}, \citenamefont {Henry}, \citenamefont {Lang}, \citenamefont
  {Lahaye}, \citenamefont {L{\"a}uchli},\ and\ \citenamefont
  {Browaeys}}]{Scholl2021Nature}%
  \BibitemOpen
  \bibfield  {author} {\bibinfo {author} {\bibfnamefont {P.}~\bibnamefont
  {Scholl}}, \bibinfo {author} {\bibfnamefont {M.}~\bibnamefont {Schuler}},
  \bibinfo {author} {\bibfnamefont {H.~J.}\ \bibnamefont {Williams}}, \bibinfo
  {author} {\bibfnamefont {A.~A.}\ \bibnamefont {Eberharter}}, \bibinfo
  {author} {\bibfnamefont {D.}~\bibnamefont {Barredo}}, \bibinfo {author}
  {\bibfnamefont {K.-N.}\ \bibnamefont {Schymik}}, \bibinfo {author}
  {\bibfnamefont {V.}~\bibnamefont {Lienhard}}, \bibinfo {author}
  {\bibfnamefont {L.-P.}\ \bibnamefont {Henry}}, \bibinfo {author}
  {\bibfnamefont {T.~C.}\ \bibnamefont {Lang}}, \bibinfo {author}
  {\bibfnamefont {T.}~\bibnamefont {Lahaye}}, \bibinfo {author} {\bibfnamefont
  {A.~M.}\ \bibnamefont {L{\"a}uchli}}, \ and\ \bibinfo {author} {\bibfnamefont
  {A.}~\bibnamefont {Browaeys}},\ }\href {\doibase 10.1038/s41586-021-03585-1}
  {\bibfield  {journal} {\bibinfo  {journal} {Nature}\ }\textbf {\bibinfo
  {volume} {595}},\ \bibinfo {pages} {233} (\bibinfo {year}
  {2021})}\BibitemShut {NoStop}%
\bibitem [{\citenamefont {Semeghini}\ \emph {et~al.}(2021)\citenamefont
  {Semeghini}, \citenamefont {Levine}, \citenamefont {Keesling}, \citenamefont
  {Ebadi}, \citenamefont {Wang}, \citenamefont {Bluvstein}, \citenamefont
  {Verresen}, \citenamefont {Pichler}, \citenamefont {Kalinowski},
  \citenamefont {Samajdar}, \citenamefont {Omran}, \citenamefont {Sachdev},
  \citenamefont {Vishwanath}, \citenamefont {Greiner}, \citenamefont
  {Vuleti\'{c}},\ and\ \citenamefont {Lukin}}]{Semeghini2021Science}%
  \BibitemOpen
  \bibfield  {author} {\bibinfo {author} {\bibfnamefont {G.}~\bibnamefont
  {Semeghini}}, \bibinfo {author} {\bibfnamefont {H.}~\bibnamefont {Levine}},
  \bibinfo {author} {\bibfnamefont {A.}~\bibnamefont {Keesling}}, \bibinfo
  {author} {\bibfnamefont {S.}~\bibnamefont {Ebadi}}, \bibinfo {author}
  {\bibfnamefont {T.~T.}\ \bibnamefont {Wang}}, \bibinfo {author}
  {\bibfnamefont {D.}~\bibnamefont {Bluvstein}}, \bibinfo {author}
  {\bibfnamefont {R.}~\bibnamefont {Verresen}}, \bibinfo {author}
  {\bibfnamefont {H.}~\bibnamefont {Pichler}}, \bibinfo {author} {\bibfnamefont
  {M.}~\bibnamefont {Kalinowski}}, \bibinfo {author} {\bibfnamefont
  {R.}~\bibnamefont {Samajdar}}, \bibinfo {author} {\bibfnamefont
  {A.}~\bibnamefont {Omran}}, \bibinfo {author} {\bibfnamefont
  {S.}~\bibnamefont {Sachdev}}, \bibinfo {author} {\bibfnamefont
  {A.}~\bibnamefont {Vishwanath}}, \bibinfo {author} {\bibfnamefont
  {M.}~\bibnamefont {Greiner}}, \bibinfo {author} {\bibfnamefont
  {V.}~\bibnamefont {Vuleti\'{c}}}, \ and\ \bibinfo {author} {\bibfnamefont
  {M.~D.}\ \bibnamefont {Lukin}},\ }\href {\doibase 10.1126/science.abi8794}
  {\bibfield  {journal} {\bibinfo  {journal} {Science}\ }\textbf {\bibinfo
  {volume} {374}},\ \bibinfo {pages} {1242} (\bibinfo {year}
  {2021})}\BibitemShut {NoStop}%
\bibitem [{\citenamefont {Satzinger}\ and\ \citenamefont {et.
  al.}(2021)}]{Satzinger2021Science}%
  \BibitemOpen
  \bibfield  {author} {\bibinfo {author} {\bibfnamefont {K.~J.}\ \bibnamefont
  {Satzinger}}\ and\ \bibinfo {author} {\bibnamefont {et. al.}},\ }\href
  {\doibase 10.1126/science.abi8378} {\bibfield  {journal} {\bibinfo  {journal}
  {Science}\ }\textbf {\bibinfo {volume} {374}},\ \bibinfo {pages} {1237}
  (\bibinfo {year} {2021})}\BibitemShut {NoStop}%
\bibitem [{\citenamefont {Saffman}\ \emph {et~al.}(2010)\citenamefont
  {Saffman}, \citenamefont {Walker},\ and\ \citenamefont
  {M\o{}lmer}}]{Saffman2010RMP}%
  \BibitemOpen
  \bibfield  {author} {\bibinfo {author} {\bibfnamefont {M.}~\bibnamefont
  {Saffman}}, \bibinfo {author} {\bibfnamefont {T.~G.}\ \bibnamefont {Walker}},
  \ and\ \bibinfo {author} {\bibfnamefont {K.}~\bibnamefont {M\o{}lmer}},\
  }\href {\doibase 10.1103/RevModPhys.82.2313} {\bibfield  {journal} {\bibinfo
  {journal} {Rev. Mod. Phys.}\ }\textbf {\bibinfo {volume} {82}},\ \bibinfo
  {pages} {2313} (\bibinfo {year} {2010})}\BibitemShut {NoStop}%
\bibitem [{\citenamefont {Lukin}\ \emph {et~al.}(2001)\citenamefont {Lukin},
  \citenamefont {Fleischhauer}, \citenamefont {Cote}, \citenamefont {Duan},
  \citenamefont {Jaksch}, \citenamefont {Cirac},\ and\ \citenamefont
  {Zoller}}]{Lukin2001PRL}%
  \BibitemOpen
  \bibfield  {author} {\bibinfo {author} {\bibfnamefont {M.~D.}\ \bibnamefont
  {Lukin}}, \bibinfo {author} {\bibfnamefont {M.}~\bibnamefont {Fleischhauer}},
  \bibinfo {author} {\bibfnamefont {R.}~\bibnamefont {Cote}}, \bibinfo {author}
  {\bibfnamefont {L.~M.}\ \bibnamefont {Duan}}, \bibinfo {author}
  {\bibfnamefont {D.}~\bibnamefont {Jaksch}}, \bibinfo {author} {\bibfnamefont
  {J.~I.}\ \bibnamefont {Cirac}}, \ and\ \bibinfo {author} {\bibfnamefont
  {P.}~\bibnamefont {Zoller}},\ }\href {\doibase 10.1103/PhysRevLett.87.037901}
  {\bibfield  {journal} {\bibinfo  {journal} {Phys. Rev. Lett.}\ }\textbf
  {\bibinfo {volume} {87}},\ \bibinfo {pages} {037901} (\bibinfo {year}
  {2001})}\BibitemShut {NoStop}%
\bibitem [{\citenamefont {Jaksch}\ \emph {et~al.}(2000)\citenamefont {Jaksch},
  \citenamefont {Cirac}, \citenamefont {Zoller}, \citenamefont {Rolston},
  \citenamefont {C\^ot\'e},\ and\ \citenamefont {Lukin}}]{Jaksch2000PRL}%
  \BibitemOpen
  \bibfield  {author} {\bibinfo {author} {\bibfnamefont {D.}~\bibnamefont
  {Jaksch}}, \bibinfo {author} {\bibfnamefont {J.~I.}\ \bibnamefont {Cirac}},
  \bibinfo {author} {\bibfnamefont {P.}~\bibnamefont {Zoller}}, \bibinfo
  {author} {\bibfnamefont {S.~L.}\ \bibnamefont {Rolston}}, \bibinfo {author}
  {\bibfnamefont {R.}~\bibnamefont {C\^ot\'e}}, \ and\ \bibinfo {author}
  {\bibfnamefont {M.~D.}\ \bibnamefont {Lukin}},\ }\href {\doibase
  10.1103/PhysRevLett.85.2208} {\bibfield  {journal} {\bibinfo  {journal}
  {Phys. Rev. Lett.}\ }\textbf {\bibinfo {volume} {85}},\ \bibinfo {pages}
  {2208} (\bibinfo {year} {2000})}\BibitemShut {NoStop}%
\bibitem [{\citenamefont {Wu}\ \emph {et~al.}(2021)\citenamefont {Wu},
  \citenamefont {Liang}, \citenamefont {Tian}, \citenamefont {Yang},
  \citenamefont {Chen}, \citenamefont {Liu}, \citenamefont {Tey},\ and\
  \citenamefont {You}}]{Wu2021CPB}%
  \BibitemOpen
  \bibfield  {author} {\bibinfo {author} {\bibfnamefont {X.}~\bibnamefont
  {Wu}}, \bibinfo {author} {\bibfnamefont {X.}~\bibnamefont {Liang}}, \bibinfo
  {author} {\bibfnamefont {Y.}~\bibnamefont {Tian}}, \bibinfo {author}
  {\bibfnamefont {F.}~\bibnamefont {Yang}}, \bibinfo {author} {\bibfnamefont
  {C.}~\bibnamefont {Chen}}, \bibinfo {author} {\bibfnamefont {Y.-C.}\
  \bibnamefont {Liu}}, \bibinfo {author} {\bibfnamefont {M.~K.}\ \bibnamefont
  {Tey}}, \ and\ \bibinfo {author} {\bibfnamefont {L.}~\bibnamefont {You}},\
  }\href {\doibase 10.1088/1674-1056/abd76f} {\bibfield  {journal} {\bibinfo
  {journal} {Chinese Physics B}\ }\textbf {\bibinfo {volume} {30}},\ \bibinfo
  {pages} {020305} (\bibinfo {year} {2021})}\BibitemShut {NoStop}%
\bibitem [{\citenamefont {Dudin}\ and\ \citenamefont
  {Kuzmich}(2012)}]{Dudin2012Science}%
  \BibitemOpen
  \bibfield  {author} {\bibinfo {author} {\bibfnamefont {Y.~O.}\ \bibnamefont
  {Dudin}}\ and\ \bibinfo {author} {\bibfnamefont {A.}~\bibnamefont
  {Kuzmich}},\ }\href {\doibase 10.1126/science.1217901} {\bibfield  {journal}
  {\bibinfo  {journal} {Science}\ }\textbf {\bibinfo {volume} {336}},\ \bibinfo
  {pages} {887} (\bibinfo {year} {2012})}\BibitemShut {NoStop}%
\bibitem [{\citenamefont {Paris-Mandoki}\ \emph
  {et~al.}(2017{\natexlab{b}})\citenamefont {Paris-Mandoki}, \citenamefont
  {Braun}, \citenamefont {Kumlin}, \citenamefont {Tresp}, \citenamefont
  {Mirgorodskiy}, \citenamefont {Christaller}, \citenamefont {B\"uchler},\ and\
  \citenamefont {Hofferberth}}]{Mandoki2017PRX}%
  \BibitemOpen
  \bibfield  {author} {\bibinfo {author} {\bibfnamefont {A.}~\bibnamefont
  {Paris-Mandoki}}, \bibinfo {author} {\bibfnamefont {C.}~\bibnamefont
  {Braun}}, \bibinfo {author} {\bibfnamefont {J.}~\bibnamefont {Kumlin}},
  \bibinfo {author} {\bibfnamefont {C.}~\bibnamefont {Tresp}}, \bibinfo
  {author} {\bibfnamefont {I.}~\bibnamefont {Mirgorodskiy}}, \bibinfo {author}
  {\bibfnamefont {F.}~\bibnamefont {Christaller}}, \bibinfo {author}
  {\bibfnamefont {H.~P.}\ \bibnamefont {B\"uchler}}, \ and\ \bibinfo {author}
  {\bibfnamefont {S.}~\bibnamefont {Hofferberth}},\ }\href {\doibase
  10.1103/PhysRevX.7.041010} {\bibfield  {journal} {\bibinfo  {journal} {Phys.
  Rev. X}\ }\textbf {\bibinfo {volume} {7}},\ \bibinfo {pages} {041010}
  (\bibinfo {year} {2017}{\natexlab{b}})}\BibitemShut {NoStop}%
\bibitem [{\citenamefont {Ripka}\ \emph {et~al.}(2018)\citenamefont {Ripka},
  \citenamefont {K{\"u}bler}, \citenamefont {L{\"o}w},\ and\ \citenamefont
  {Pfau}}]{Ripka2018Science}%
  \BibitemOpen
  \bibfield  {author} {\bibinfo {author} {\bibfnamefont {F.}~\bibnamefont
  {Ripka}}, \bibinfo {author} {\bibfnamefont {H.}~\bibnamefont {K{\"u}bler}},
  \bibinfo {author} {\bibfnamefont {R.}~\bibnamefont {L{\"o}w}}, \ and\
  \bibinfo {author} {\bibfnamefont {T.}~\bibnamefont {Pfau}},\ }\href {\doibase
  10.1126/science.aau1949} {\bibfield  {journal} {\bibinfo  {journal}
  {Science}\ }\textbf {\bibinfo {volume} {362}},\ \bibinfo {pages} {446}
  (\bibinfo {year} {2018})}\BibitemShut {NoStop}%
\bibitem [{\citenamefont {Li}\ \emph {et~al.}(2003)\citenamefont {Li},
  \citenamefont {Mourachko}, \citenamefont {Noel},\ and\ \citenamefont
  {Gallagher}}]{Li2003PRA}%
  \BibitemOpen
  \bibfield  {author} {\bibinfo {author} {\bibfnamefont {W.}~\bibnamefont
  {Li}}, \bibinfo {author} {\bibfnamefont {I.}~\bibnamefont {Mourachko}},
  \bibinfo {author} {\bibfnamefont {M.~W.}\ \bibnamefont {Noel}}, \ and\
  \bibinfo {author} {\bibfnamefont {T.~F.}\ \bibnamefont {Gallagher}},\ }\href
  {\doibase 10.1103/PhysRevA.67.052502} {\bibfield  {journal} {\bibinfo
  {journal} {Phys. Rev. A}\ }\textbf {\bibinfo {volume} {67}},\ \bibinfo
  {pages} {052502} (\bibinfo {year} {2003})}\BibitemShut {NoStop}%
\bibitem [{\citenamefont {Beterov}\ \emph {et~al.}(2009)\citenamefont
  {Beterov}, \citenamefont {Ryabtsev}, \citenamefont {Tretyakov},\ and\
  \citenamefont {Entin}}]{Beterov2009PRA}%
  \BibitemOpen
  \bibfield  {author} {\bibinfo {author} {\bibfnamefont {I.~I.}\ \bibnamefont
  {Beterov}}, \bibinfo {author} {\bibfnamefont {I.~I.}\ \bibnamefont
  {Ryabtsev}}, \bibinfo {author} {\bibfnamefont {D.~B.}\ \bibnamefont
  {Tretyakov}}, \ and\ \bibinfo {author} {\bibfnamefont {V.~M.}\ \bibnamefont
  {Entin}},\ }\href {\doibase 10.1103/PhysRevA.79.052504} {\bibfield  {journal}
  {\bibinfo  {journal} {Phys. Rev. A}\ }\textbf {\bibinfo {volume} {79}},\
  \bibinfo {pages} {052504} (\bibinfo {year} {2009})}\BibitemShut {NoStop}%
\bibitem [{\citenamefont {Roy}\ \emph {et~al.}(2017)\citenamefont {Roy},
  \citenamefont {Wilson},\ and\ \citenamefont {Firstenberg}}]{Roy2017RMP}%
  \BibitemOpen
  \bibfield  {author} {\bibinfo {author} {\bibfnamefont {D.}~\bibnamefont
  {Roy}}, \bibinfo {author} {\bibfnamefont {C.~M.}\ \bibnamefont {Wilson}}, \
  and\ \bibinfo {author} {\bibfnamefont {O.}~\bibnamefont {Firstenberg}},\
  }\href {\doibase 10.1103/RevModPhys.89.021001} {\bibfield  {journal}
  {\bibinfo  {journal} {Rev. Mod. Phys.}\ }\textbf {\bibinfo {volume} {89}},\
  \bibinfo {pages} {021001} (\bibinfo {year} {2017})}\BibitemShut {NoStop}%
\bibitem [{\citenamefont {Thompson}\ \emph {et~al.}(2013)\citenamefont
  {Thompson}, \citenamefont {Tiecke}, \citenamefont {de~Leon}, \citenamefont
  {Feist}, \citenamefont {Akimov}, \citenamefont {Gullans}, \citenamefont
  {Zibrov}, \citenamefont {Vuleti{\'{c}}},\ and\ \citenamefont
  {Lukin}}]{Thompson2013Science}%
  \BibitemOpen
  \bibfield  {author} {\bibinfo {author} {\bibfnamefont {J.~D.}\ \bibnamefont
  {Thompson}}, \bibinfo {author} {\bibfnamefont {T.~G.}\ \bibnamefont
  {Tiecke}}, \bibinfo {author} {\bibfnamefont {N.~P.}\ \bibnamefont {de~Leon}},
  \bibinfo {author} {\bibfnamefont {J.}~\bibnamefont {Feist}}, \bibinfo
  {author} {\bibfnamefont {A.~V.}\ \bibnamefont {Akimov}}, \bibinfo {author}
  {\bibfnamefont {M.}~\bibnamefont {Gullans}}, \bibinfo {author} {\bibfnamefont
  {A.~S.}\ \bibnamefont {Zibrov}}, \bibinfo {author} {\bibfnamefont
  {V.}~\bibnamefont {Vuleti{\'{c}}}}, \ and\ \bibinfo {author} {\bibfnamefont
  {M.~D.}\ \bibnamefont {Lukin}},\ }\href {\doibase 10.1126/science.1237125}
  {\bibfield  {journal} {\bibinfo  {journal} {Science}\ }\textbf {\bibinfo
  {volume} {340}},\ \bibinfo {pages} {1202} (\bibinfo {year}
  {2013})}\BibitemShut {NoStop}%
\bibitem [{\citenamefont {Arcari}\ \emph {et~al.}(2014)\citenamefont {Arcari},
  \citenamefont {S\"ollner}, \citenamefont {Javadi}, \citenamefont
  {Lindskov~Hansen}, \citenamefont {Mahmoodian}, \citenamefont {Liu},
  \citenamefont {Thyrrestrup}, \citenamefont {Lee}, \citenamefont {Song},
  \citenamefont {Stobbe},\ and\ \citenamefont {Lodahl}}]{Arcari2014PRL}%
  \BibitemOpen
  \bibfield  {author} {\bibinfo {author} {\bibfnamefont {M.}~\bibnamefont
  {Arcari}}, \bibinfo {author} {\bibfnamefont {I.}~\bibnamefont {S\"ollner}},
  \bibinfo {author} {\bibfnamefont {A.}~\bibnamefont {Javadi}}, \bibinfo
  {author} {\bibfnamefont {S.}~\bibnamefont {Lindskov~Hansen}}, \bibinfo
  {author} {\bibfnamefont {S.}~\bibnamefont {Mahmoodian}}, \bibinfo {author}
  {\bibfnamefont {J.}~\bibnamefont {Liu}}, \bibinfo {author} {\bibfnamefont
  {H.}~\bibnamefont {Thyrrestrup}}, \bibinfo {author} {\bibfnamefont {E.~H.}\
  \bibnamefont {Lee}}, \bibinfo {author} {\bibfnamefont {J.~D.}\ \bibnamefont
  {Song}}, \bibinfo {author} {\bibfnamefont {S.}~\bibnamefont {Stobbe}}, \ and\
  \bibinfo {author} {\bibfnamefont {P.}~\bibnamefont {Lodahl}},\ }\href
  {\doibase 10.1103/PhysRevLett.113.093603} {\bibfield  {journal} {\bibinfo
  {journal} {Phys. Rev. Lett.}\ }\textbf {\bibinfo {volume} {113}},\ \bibinfo
  {pages} {093603} (\bibinfo {year} {2014})}\BibitemShut {NoStop}%
\bibitem [{\citenamefont {Tiranov}\ \emph {et~al.}(2023)\citenamefont
  {Tiranov}, \citenamefont {Angelopoulou}, \citenamefont {van Diepen},
  \citenamefont {Schrinski}, \citenamefont {Sandberg}, \citenamefont {Wang},
  \citenamefont {Midolo}, \citenamefont {Scholz}, \citenamefont {Wieck},
  \citenamefont {Ludwig}, \citenamefont {S{\o}rensen},\ and\ \citenamefont
  {Lodahl}}]{Tiranov2023Science}%
  \BibitemOpen
  \bibfield  {author} {\bibinfo {author} {\bibfnamefont {A.}~\bibnamefont
  {Tiranov}}, \bibinfo {author} {\bibfnamefont {V.}~\bibnamefont
  {Angelopoulou}}, \bibinfo {author} {\bibfnamefont {C.~J.}\ \bibnamefont {van
  Diepen}}, \bibinfo {author} {\bibfnamefont {B.}~\bibnamefont {Schrinski}},
  \bibinfo {author} {\bibfnamefont {O.~A.~D.}\ \bibnamefont {Sandberg}},
  \bibinfo {author} {\bibfnamefont {Y.}~\bibnamefont {Wang}}, \bibinfo {author}
  {\bibfnamefont {L.}~\bibnamefont {Midolo}}, \bibinfo {author} {\bibfnamefont
  {S.}~\bibnamefont {Scholz}}, \bibinfo {author} {\bibfnamefont {A.~D.}\
  \bibnamefont {Wieck}}, \bibinfo {author} {\bibfnamefont {A.}~\bibnamefont
  {Ludwig}}, \bibinfo {author} {\bibfnamefont {A.~S.}\ \bibnamefont
  {S{\o}rensen}}, \ and\ \bibinfo {author} {\bibfnamefont {P.}~\bibnamefont
  {Lodahl}},\ }\href {\doibase 10.1126/science.ade9324} {\bibfield  {journal}
  {\bibinfo  {journal} {Science}\ }\textbf {\bibinfo {volume} {379}},\ \bibinfo
  {pages} {389} (\bibinfo {year} {2023})}\BibitemShut {NoStop}%
\bibitem [{\citenamefont {Kannan}\ \emph {et~al.}(2020)\citenamefont {Kannan},
  \citenamefont {Ruckriegel}, \citenamefont {Campbell}, \citenamefont
  {Frisk~Kockum}, \citenamefont {Braum{\"u}ller}, \citenamefont {Kim},
  \citenamefont {Kjaergaard}, \citenamefont {Krantz}, \citenamefont {Melville},
  \citenamefont {Niedzielski}, \citenamefont {Veps{\"a}l{\"a}inen},
  \citenamefont {Winik}, \citenamefont {Yoder}, \citenamefont {Nori},
  \citenamefont {Orlando}, \citenamefont {Gustavsson},\ and\ \citenamefont
  {Oliver}}]{Kannan2020Nature}%
  \BibitemOpen
  \bibfield  {author} {\bibinfo {author} {\bibfnamefont {B.}~\bibnamefont
  {Kannan}}, \bibinfo {author} {\bibfnamefont {M.~J.}\ \bibnamefont
  {Ruckriegel}}, \bibinfo {author} {\bibfnamefont {D.~L.}\ \bibnamefont
  {Campbell}}, \bibinfo {author} {\bibfnamefont {A.}~\bibnamefont
  {Frisk~Kockum}}, \bibinfo {author} {\bibfnamefont {J.}~\bibnamefont
  {Braum{\"u}ller}}, \bibinfo {author} {\bibfnamefont {D.~K.}\ \bibnamefont
  {Kim}}, \bibinfo {author} {\bibfnamefont {M.}~\bibnamefont {Kjaergaard}},
  \bibinfo {author} {\bibfnamefont {P.}~\bibnamefont {Krantz}}, \bibinfo
  {author} {\bibfnamefont {A.}~\bibnamefont {Melville}}, \bibinfo {author}
  {\bibfnamefont {B.~M.}\ \bibnamefont {Niedzielski}}, \bibinfo {author}
  {\bibfnamefont {A.}~\bibnamefont {Veps{\"a}l{\"a}inen}}, \bibinfo {author}
  {\bibfnamefont {R.}~\bibnamefont {Winik}}, \bibinfo {author} {\bibfnamefont
  {J.~L.}\ \bibnamefont {Yoder}}, \bibinfo {author} {\bibfnamefont
  {F.}~\bibnamefont {Nori}}, \bibinfo {author} {\bibfnamefont {T.~P.}\
  \bibnamefont {Orlando}}, \bibinfo {author} {\bibfnamefont {S.}~\bibnamefont
  {Gustavsson}}, \ and\ \bibinfo {author} {\bibfnamefont {W.~D.}\ \bibnamefont
  {Oliver}},\ }\href {\doibase 10.1038/s41586-020-2529-9} {\bibfield  {journal}
  {\bibinfo  {journal} {Nature}\ }\textbf {\bibinfo {volume} {583}},\ \bibinfo
  {pages} {775} (\bibinfo {year} {2020})}\BibitemShut {NoStop}%
\bibitem [{\citenamefont {Solomons}\ \emph {et~al.}(2024)\citenamefont
  {Solomons}, \citenamefont {Ben-Maimon},\ and\ \citenamefont
  {Shahmoon}}]{Solomons24}%
  \BibitemOpen
  \bibfield  {author} {\bibinfo {author} {\bibfnamefont {Y.}~\bibnamefont
  {Solomons}}, \bibinfo {author} {\bibfnamefont {R.}~\bibnamefont
  {Ben-Maimon}}, \ and\ \bibinfo {author} {\bibfnamefont {E.}~\bibnamefont
  {Shahmoon}},\ }\href {\doibase 10.1103/PRXQuantum.5.020329} {\bibfield
  {journal} {\bibinfo  {journal} {PRX Quantum}\ }\textbf {\bibinfo {volume}
  {5}},\ \bibinfo {pages} {020329} (\bibinfo {year} {2024})}\BibitemShut
  {NoStop}%
\bibitem [{\citenamefont {M{\o}lmer}\ \emph {et~al.}(1993)\citenamefont
  {M{\o}lmer}, \citenamefont {Castin},\ and\ \citenamefont
  {Dalibard}}]{Molmer1993JOSAB}%
  \BibitemOpen
  \bibfield  {author} {\bibinfo {author} {\bibfnamefont {K.}~\bibnamefont
  {M{\o}lmer}}, \bibinfo {author} {\bibfnamefont {Y.}~\bibnamefont {Castin}}, \
  and\ \bibinfo {author} {\bibfnamefont {J.}~\bibnamefont {Dalibard}},\ }\href
  {\doibase 10.1364/JOSAB.10.000524} {\bibfield  {journal} {\bibinfo  {journal}
  {J. Opt. Soc. Am. B}\ }\textbf {\bibinfo {volume} {10}},\ \bibinfo {pages}
  {524} (\bibinfo {year} {1993})}\BibitemShut {NoStop}%
\bibitem [{\citenamefont {Shen}\ and\ \citenamefont
  {Fan}(2007{\natexlab{a}})}]{Shen2007PRA}%
  \BibitemOpen
  \bibfield  {author} {\bibinfo {author} {\bibfnamefont {J.-T.}\ \bibnamefont
  {Shen}}\ and\ \bibinfo {author} {\bibfnamefont {S.}~\bibnamefont {Fan}},\
  }\href {\doibase 10.1103/PhysRevA.76.062709} {\bibfield  {journal} {\bibinfo
  {journal} {Phys. Rev. A}\ }\textbf {\bibinfo {volume} {76}},\ \bibinfo
  {pages} {062709} (\bibinfo {year} {2007}{\natexlab{a}})}\BibitemShut
  {NoStop}%
\bibitem [{\citenamefont {Shen}\ and\ \citenamefont
  {Fan}(2007{\natexlab{b}})}]{Shen2007PRL}%
  \BibitemOpen
  \bibfield  {author} {\bibinfo {author} {\bibfnamefont {J.-T.}\ \bibnamefont
  {Shen}}\ and\ \bibinfo {author} {\bibfnamefont {S.}~\bibnamefont {Fan}},\
  }\href {\doibase 10.1103/PhysRevLett.98.153003} {\bibfield  {journal}
  {\bibinfo  {journal} {Phys. Rev. Lett.}\ }\textbf {\bibinfo {volume} {98}},\
  \bibinfo {pages} {153003} (\bibinfo {year} {2007}{\natexlab{b}})}\BibitemShut
  {NoStop}%
\bibitem [{\citenamefont {Yang}\ \emph {et~al.}(2022)\citenamefont {Yang},
  \citenamefont {Lund}, \citenamefont {Pohl}, \citenamefont {Lodahl},\ and\
  \citenamefont {M{\o}lmer}}]{yang2022determin}%
  \BibitemOpen
  \bibfield  {author} {\bibinfo {author} {\bibfnamefont {F.}~\bibnamefont
  {Yang}}, \bibinfo {author} {\bibfnamefont {M.~M.}\ \bibnamefont {Lund}},
  \bibinfo {author} {\bibfnamefont {T.}~\bibnamefont {Pohl}}, \bibinfo {author}
  {\bibfnamefont {P.}~\bibnamefont {Lodahl}}, \ and\ \bibinfo {author}
  {\bibfnamefont {K.}~\bibnamefont {M{\o}lmer}},\ }\href {\doibase
  10.1103/PhysRevLett.128.213603} {\bibfield  {journal} {\bibinfo  {journal}
  {Phys. Rev. Lett.}\ }\textbf {\bibinfo {volume} {128}},\ \bibinfo {pages}
  {213603} (\bibinfo {year} {2022})}\BibitemShut {NoStop}%
\bibitem [{\citenamefont {Witthaut}\ \emph {et~al.}(2012)\citenamefont
  {Witthaut}, \citenamefont {Lukin},\ and\ \citenamefont
  {S{\o}rensen}}]{witthaut2012photon}%
  \BibitemOpen
  \bibfield  {author} {\bibinfo {author} {\bibfnamefont {D.}~\bibnamefont
  {Witthaut}}, \bibinfo {author} {\bibfnamefont {M.~D.}\ \bibnamefont {Lukin}},
  \ and\ \bibinfo {author} {\bibfnamefont {A.~S.}\ \bibnamefont
  {S{\o}rensen}},\ }\href {\doibase 10.1209/0295-5075/97/50007} {\bibfield
  {journal} {\bibinfo  {journal} {Europhys. Lett.}\ }\textbf {\bibinfo {volume}
  {97}},\ \bibinfo {pages} {50007} (\bibinfo {year} {2012})}\BibitemShut
  {NoStop}%
\bibitem [{\citenamefont {Ralph}\ \emph {et~al.}(2015)\citenamefont {Ralph},
  \citenamefont {S{\"o}llner}, \citenamefont {Mahmoodian}, \citenamefont
  {White},\ and\ \citenamefont {Lodahl}}]{ralph2015photon}%
  \BibitemOpen
  \bibfield  {author} {\bibinfo {author} {\bibfnamefont {T.}~\bibnamefont
  {Ralph}}, \bibinfo {author} {\bibfnamefont {I.}~\bibnamefont {S{\"o}llner}},
  \bibinfo {author} {\bibfnamefont {S.}~\bibnamefont {Mahmoodian}}, \bibinfo
  {author} {\bibfnamefont {A.}~\bibnamefont {White}}, \ and\ \bibinfo {author}
  {\bibfnamefont {P.}~\bibnamefont {Lodahl}},\ }\href {\doibase
  10.1103/PhysRevLett.114.173603} {\bibfield  {journal} {\bibinfo  {journal}
  {Phys. Rev. Lett.}\ }\textbf {\bibinfo {volume} {114}},\ \bibinfo {pages}
  {173603} (\bibinfo {year} {2015})}\BibitemShut {NoStop}%
\bibitem [{\citenamefont {Knill}\ \emph {et~al.}(2001)\citenamefont {Knill},
  \citenamefont {Raymond},\ and\ \citenamefont {Gerald}}]{Knill2001Nature}%
  \BibitemOpen
  \bibfield  {author} {\bibinfo {author} {\bibfnamefont {E.}~\bibnamefont
  {Knill}}, \bibinfo {author} {\bibfnamefont {L.}~\bibnamefont {Raymond}}, \
  and\ \bibinfo {author} {\bibfnamefont {J.~M.}\ \bibnamefont {Gerald}},\
  }\href {\doibase 10.1038/35051009} {\bibfield  {journal} {\bibinfo  {journal}
  {Nature}\ }\textbf {\bibinfo {volume} {409}},\ \bibinfo {pages} {46}
  (\bibinfo {year} {2001})}\BibitemShut {NoStop}%
\bibitem [{\citenamefont {Eckstein}\ \emph {et~al.}(2011)\citenamefont
  {Eckstein}, \citenamefont {Brecht},\ and\ \citenamefont
  {Silberhorn}}]{eckstein2011quantum}%
  \BibitemOpen
  \bibfield  {author} {\bibinfo {author} {\bibfnamefont {A.}~\bibnamefont
  {Eckstein}}, \bibinfo {author} {\bibfnamefont {B.}~\bibnamefont {Brecht}}, \
  and\ \bibinfo {author} {\bibfnamefont {C.}~\bibnamefont {Silberhorn}},\
  }\href {\doibase 10.1364/OE.19.013770} {\bibfield  {journal} {\bibinfo
  {journal} {Opt. Express}\ }\textbf {\bibinfo {volume} {19}},\ \bibinfo
  {pages} {13770} (\bibinfo {year} {2011})}\BibitemShut {NoStop}%
\bibitem [{\citenamefont {Ansari}\ \emph {et~al.}(2018)\citenamefont {Ansari},
  \citenamefont {Donohue}, \citenamefont {Allgaier}, \citenamefont {Sansoni},
  \citenamefont {Brecht}, \citenamefont {Roslund}, \citenamefont {Treps},
  \citenamefont {Harder},\ and\ \citenamefont
  {Silberhorn}}]{ansari2018tomography}%
  \BibitemOpen
  \bibfield  {author} {\bibinfo {author} {\bibfnamefont {V.}~\bibnamefont
  {Ansari}}, \bibinfo {author} {\bibfnamefont {J.~M.}\ \bibnamefont {Donohue}},
  \bibinfo {author} {\bibfnamefont {M.}~\bibnamefont {Allgaier}}, \bibinfo
  {author} {\bibfnamefont {L.}~\bibnamefont {Sansoni}}, \bibinfo {author}
  {\bibfnamefont {B.}~\bibnamefont {Brecht}}, \bibinfo {author} {\bibfnamefont
  {J.}~\bibnamefont {Roslund}}, \bibinfo {author} {\bibfnamefont
  {N.}~\bibnamefont {Treps}}, \bibinfo {author} {\bibfnamefont
  {G.}~\bibnamefont {Harder}}, \ and\ \bibinfo {author} {\bibfnamefont
  {C.}~\bibnamefont {Silberhorn}},\ }\href {\doibase
  10.1103/PhysRevLett.120.213601} {\bibfield  {journal} {\bibinfo  {journal}
  {Phys. Rev. Lett.}\ }\textbf {\bibinfo {volume} {120}},\ \bibinfo {pages}
  {213601} (\bibinfo {year} {2018})}\BibitemShut {NoStop}%
\bibitem [{\citenamefont {Chumak}\ \emph {et~al.}(2010)\citenamefont {Chumak},
  \citenamefont {Tiberkevich}, \citenamefont {Karenowska}, \citenamefont
  {Serga}, \citenamefont {Gregg}, \citenamefont {Slavin},\ and\ \citenamefont
  {Hillebrands}}]{chumak2010all}%
  \BibitemOpen
  \bibfield  {author} {\bibinfo {author} {\bibfnamefont {A.~V.}\ \bibnamefont
  {Chumak}}, \bibinfo {author} {\bibfnamefont {V.~S.}\ \bibnamefont
  {Tiberkevich}}, \bibinfo {author} {\bibfnamefont {A.~D.}\ \bibnamefont
  {Karenowska}}, \bibinfo {author} {\bibfnamefont {A.~A.}\ \bibnamefont
  {Serga}}, \bibinfo {author} {\bibfnamefont {J.~F.}\ \bibnamefont {Gregg}},
  \bibinfo {author} {\bibfnamefont {A.~N.}\ \bibnamefont {Slavin}}, \ and\
  \bibinfo {author} {\bibfnamefont {B.}~\bibnamefont {Hillebrands}},\ }\href
  {\doibase 10.1038/ncomms1142} {\bibfield  {journal} {\bibinfo  {journal}
  {Nat. Commun.}\ }\textbf {\bibinfo {volume} {1}},\ \bibinfo {pages} {141}
  (\bibinfo {year} {2010})}\BibitemShut {NoStop}%
\bibitem [{\citenamefont {Sivan}\ and\ \citenamefont
  {Pendry}(2011)}]{sivan2011time}%
  \BibitemOpen
  \bibfield  {author} {\bibinfo {author} {\bibfnamefont {Y.}~\bibnamefont
  {Sivan}}\ and\ \bibinfo {author} {\bibfnamefont {J.~B.}\ \bibnamefont
  {Pendry}},\ }\href {\doibase 10.1103/PhysRevLett.106.193902} {\bibfield
  {journal} {\bibinfo  {journal} {Phys. Rev. Lett.}\ }\textbf {\bibinfo
  {volume} {106}},\ \bibinfo {pages} {193902} (\bibinfo {year}
  {2011})}\BibitemShut {NoStop}%
\bibitem [{\citenamefont {Minkov}\ and\ \citenamefont
  {Fan}(2018)}]{minkov2018localization}%
  \BibitemOpen
  \bibfield  {author} {\bibinfo {author} {\bibfnamefont {M.}~\bibnamefont
  {Minkov}}\ and\ \bibinfo {author} {\bibfnamefont {S.}~\bibnamefont {Fan}},\
  }\href {\doibase 10.1103/PhysRevB.97.060301} {\bibfield  {journal} {\bibinfo
  {journal} {Phys. Rev. B}\ }\textbf {\bibinfo {volume} {97}},\ \bibinfo
  {pages} {060301} (\bibinfo {year} {2018})}\BibitemShut {NoStop}%
\bibitem [{\citenamefont {Gil}\ \emph {et~al.}(2014)\citenamefont {Gil},
  \citenamefont {Mukherjee}, \citenamefont {Bridge}, \citenamefont {Jones},\
  and\ \citenamefont {Pohl}}]{gil14}%
  \BibitemOpen
  \bibfield  {author} {\bibinfo {author} {\bibfnamefont {L.~I.~R.}\
  \bibnamefont {Gil}}, \bibinfo {author} {\bibfnamefont {R.}~\bibnamefont
  {Mukherjee}}, \bibinfo {author} {\bibfnamefont {E.~M.}\ \bibnamefont
  {Bridge}}, \bibinfo {author} {\bibfnamefont {M.~P.~A.}\ \bibnamefont
  {Jones}}, \ and\ \bibinfo {author} {\bibfnamefont {T.}~\bibnamefont {Pohl}},\
  }\href {\doibase 10.1103/PhysRevLett.112.103601} {\bibfield  {journal}
  {\bibinfo  {journal} {Phys. Rev. Lett.}\ }\textbf {\bibinfo {volume} {112}},\
  \bibinfo {pages} {103601} (\bibinfo {year} {2014})}\BibitemShut {NoStop}%
\bibitem [{\citenamefont {Zeiher}\ \emph {et~al.}(2016)\citenamefont {Zeiher},
  \citenamefont {van Bijnen}, \citenamefont {Schau{\ss}}, \citenamefont {Hild},
  \citenamefont {Choi}, \citenamefont {Pohl}, \citenamefont {Bloch},\ and\
  \citenamefont {Gross}}]{zeiher16}%
  \BibitemOpen
  \bibfield  {author} {\bibinfo {author} {\bibfnamefont {J.}~\bibnamefont
  {Zeiher}}, \bibinfo {author} {\bibfnamefont {R.}~\bibnamefont {van Bijnen}},
  \bibinfo {author} {\bibfnamefont {P.}~\bibnamefont {Schau{\ss}}}, \bibinfo
  {author} {\bibfnamefont {S.}~\bibnamefont {Hild}}, \bibinfo {author}
  {\bibfnamefont {J.-y.}\ \bibnamefont {Choi}}, \bibinfo {author}
  {\bibfnamefont {T.}~\bibnamefont {Pohl}}, \bibinfo {author} {\bibfnamefont
  {I.}~\bibnamefont {Bloch}}, \ and\ \bibinfo {author} {\bibfnamefont
  {C.}~\bibnamefont {Gross}},\ }\href {\doibase 10.1038/nphys3835} {\bibfield
  {journal} {\bibinfo  {journal} {Nature Physics}\ }\textbf {\bibinfo {volume}
  {12}},\ \bibinfo {pages} {1095} (\bibinfo {year} {2016})}\BibitemShut
  {NoStop}%
\bibitem [{\citenamefont {Walther}\ \emph {et~al.}(2022)\citenamefont
  {Walther}, \citenamefont {Zhang}, \citenamefont {Yelin},\ and\ \citenamefont
  {Pohl}}]{walther22}%
  \BibitemOpen
  \bibfield  {author} {\bibinfo {author} {\bibfnamefont {V.}~\bibnamefont
  {Walther}}, \bibinfo {author} {\bibfnamefont {L.}~\bibnamefont {Zhang}},
  \bibinfo {author} {\bibfnamefont {S.~F.}\ \bibnamefont {Yelin}}, \ and\
  \bibinfo {author} {\bibfnamefont {T.}~\bibnamefont {Pohl}},\ }\href {\doibase
  10.1103/PhysRevB.105.075307} {\bibfield  {journal} {\bibinfo  {journal}
  {Phys. Rev. B}\ }\textbf {\bibinfo {volume} {105}},\ \bibinfo {pages}
  {075307} (\bibinfo {year} {2022})}\BibitemShut {NoStop}%
\bibitem [{\citenamefont {Solomons}\ and\ \citenamefont
  {Shahmoon}(2023)}]{Solomons23}%
  \BibitemOpen
  \bibfield  {author} {\bibinfo {author} {\bibfnamefont {Y.}~\bibnamefont
  {Solomons}}\ and\ \bibinfo {author} {\bibfnamefont {E.}~\bibnamefont
  {Shahmoon}},\ }\href {\doibase 10.1103/PhysRevA.107.033709} {\bibfield
  {journal} {\bibinfo  {journal} {Phys. Rev. A}\ }\textbf {\bibinfo {volume}
  {107}},\ \bibinfo {pages} {033709} (\bibinfo {year} {2023})}\BibitemShut
  {NoStop}%
\bibitem [{\citenamefont {Schrinski}\ \emph {et~al.}(2022)\citenamefont
  {Schrinski}, \citenamefont {Lamaison},\ and\ \citenamefont
  {S\o{}rensen}}]{PhysRevLett.129.130502}%
  \BibitemOpen
  \bibfield  {author} {\bibinfo {author} {\bibfnamefont {B.}~\bibnamefont
  {Schrinski}}, \bibinfo {author} {\bibfnamefont {M.}~\bibnamefont {Lamaison}},
  \ and\ \bibinfo {author} {\bibfnamefont {A.~S.}\ \bibnamefont
  {S\o{}rensen}},\ }\href {\doibase 10.1103/PhysRevLett.129.130502} {\bibfield
  {journal} {\bibinfo  {journal} {Phys. Rev. Lett.}\ }\textbf {\bibinfo
  {volume} {129}},\ \bibinfo {pages} {130502} (\bibinfo {year}
  {2022})}\BibitemShut {NoStop}%
\bibitem [{\citenamefont {Graham}\ \emph {et~al.}(2022)\citenamefont {Graham},
  \citenamefont {Song}, \citenamefont {Scott}, \citenamefont {Poole},
  \citenamefont {Phuttitarn}, \citenamefont {Jooya}, \citenamefont {Eichler},
  \citenamefont {Jiang}, \citenamefont {Marra}, \citenamefont {Grinkemeyer},
  \citenamefont {Kwon}, \citenamefont {Ebert}, \citenamefont {Cherek},
  \citenamefont {Lichtman}, \citenamefont {Gillette}, \citenamefont {Gilbert},
  \citenamefont {Bowman}, \citenamefont {Ballance}, \citenamefont {Campbell},
  \citenamefont {Dahl}, \citenamefont {Crawford}, \citenamefont {Blunt},
  \citenamefont {Rogers}, \citenamefont {Noel},\ and\ \citenamefont
  {Saffman}}]{Graham2022Nature}%
  \BibitemOpen
  \bibfield  {author} {\bibinfo {author} {\bibfnamefont {T.~M.}\ \bibnamefont
  {Graham}}, \bibinfo {author} {\bibfnamefont {Y.}~\bibnamefont {Song}},
  \bibinfo {author} {\bibfnamefont {J.}~\bibnamefont {Scott}}, \bibinfo
  {author} {\bibfnamefont {C.}~\bibnamefont {Poole}}, \bibinfo {author}
  {\bibfnamefont {L.}~\bibnamefont {Phuttitarn}}, \bibinfo {author}
  {\bibfnamefont {K.}~\bibnamefont {Jooya}}, \bibinfo {author} {\bibfnamefont
  {P.}~\bibnamefont {Eichler}}, \bibinfo {author} {\bibfnamefont
  {X.}~\bibnamefont {Jiang}}, \bibinfo {author} {\bibfnamefont
  {A.}~\bibnamefont {Marra}}, \bibinfo {author} {\bibfnamefont
  {B.}~\bibnamefont {Grinkemeyer}}, \bibinfo {author} {\bibfnamefont
  {M.}~\bibnamefont {Kwon}}, \bibinfo {author} {\bibfnamefont {M.}~\bibnamefont
  {Ebert}}, \bibinfo {author} {\bibfnamefont {J.}~\bibnamefont {Cherek}},
  \bibinfo {author} {\bibfnamefont {M.~T.}\ \bibnamefont {Lichtman}}, \bibinfo
  {author} {\bibfnamefont {M.}~\bibnamefont {Gillette}}, \bibinfo {author}
  {\bibfnamefont {J.}~\bibnamefont {Gilbert}}, \bibinfo {author} {\bibfnamefont
  {D.}~\bibnamefont {Bowman}}, \bibinfo {author} {\bibfnamefont
  {T.}~\bibnamefont {Ballance}}, \bibinfo {author} {\bibfnamefont
  {C.}~\bibnamefont {Campbell}}, \bibinfo {author} {\bibfnamefont {E.~D.}\
  \bibnamefont {Dahl}}, \bibinfo {author} {\bibfnamefont {O.}~\bibnamefont
  {Crawford}}, \bibinfo {author} {\bibfnamefont {N.~S.}\ \bibnamefont {Blunt}},
  \bibinfo {author} {\bibfnamefont {B.}~\bibnamefont {Rogers}}, \bibinfo
  {author} {\bibfnamefont {T.}~\bibnamefont {Noel}}, \ and\ \bibinfo {author}
  {\bibfnamefont {M.}~\bibnamefont {Saffman}},\ }\href {\doibase
  10.1038/s41586-022-04603-6} {\bibfield  {journal} {\bibinfo  {journal}
  {Nature}\ }\textbf {\bibinfo {volume} {604}},\ \bibinfo {pages} {457}
  (\bibinfo {year} {2022})}\BibitemShut {NoStop}%
\bibitem [{\citenamefont {Scholl}\ \emph {et~al.}(2023)\citenamefont {Scholl},
  \citenamefont {Shaw}, \citenamefont {Tsai}, \citenamefont {Finkelstein},
  \citenamefont {Choi},\ and\ \citenamefont {Endres}}]{Scholl2023Nature}%
  \BibitemOpen
  \bibfield  {author} {\bibinfo {author} {\bibfnamefont {P.}~\bibnamefont
  {Scholl}}, \bibinfo {author} {\bibfnamefont {A.~L.}\ \bibnamefont {Shaw}},
  \bibinfo {author} {\bibfnamefont {R.~B.-S.}\ \bibnamefont {Tsai}}, \bibinfo
  {author} {\bibfnamefont {R.}~\bibnamefont {Finkelstein}}, \bibinfo {author}
  {\bibfnamefont {J.}~\bibnamefont {Choi}}, \ and\ \bibinfo {author}
  {\bibfnamefont {M.}~\bibnamefont {Endres}},\ }\href {\doibase
  10.1038/s41586-023-06516-4} {\bibfield  {journal} {\bibinfo  {journal}
  {Nature}\ }\textbf {\bibinfo {volume} {622}},\ \bibinfo {pages} {273}
  (\bibinfo {year} {2023})}\BibitemShut {NoStop}%
\bibitem [{\citenamefont {Singh}\ \emph {et~al.}(2023)\citenamefont {Singh},
  \citenamefont {Bradley}, \citenamefont {Anand}, \citenamefont {Ramesh},
  \citenamefont {White},\ and\ \citenamefont {Bernien}}]{Singh2023Science}%
  \BibitemOpen
  \bibfield  {author} {\bibinfo {author} {\bibfnamefont {K.}~\bibnamefont
  {Singh}}, \bibinfo {author} {\bibfnamefont {C.~E.}\ \bibnamefont {Bradley}},
  \bibinfo {author} {\bibfnamefont {S.}~\bibnamefont {Anand}}, \bibinfo
  {author} {\bibfnamefont {V.}~\bibnamefont {Ramesh}}, \bibinfo {author}
  {\bibfnamefont {R.}~\bibnamefont {White}}, \ and\ \bibinfo {author}
  {\bibfnamefont {H.}~\bibnamefont {Bernien}},\ }\href {\doibase
  10.1126/science.ade5337} {\bibfield  {journal} {\bibinfo  {journal}
  {Science}\ }\textbf {\bibinfo {volume} {380}},\ \bibinfo {pages} {1265}
  (\bibinfo {year} {2023})}\BibitemShut {NoStop}%
\bibitem [{\citenamefont {Evered}\ \emph {et~al.}(2023)\citenamefont {Evered},
  \citenamefont {Bluvstein}, \citenamefont {Kalinowski}, \citenamefont {Ebadi},
  \citenamefont {Manovitz}, \citenamefont {Zhou}, \citenamefont {Li},
  \citenamefont {Geim}, \citenamefont {Wang}, \citenamefont {Maskara},
  \citenamefont {Levine}, \citenamefont {Semeghini}, \citenamefont {Greiner},
  \citenamefont {Vuleti{\'{c}}},\ and\ \citenamefont
  {Lukin}}]{Evered2023Nature}%
  \BibitemOpen
  \bibfield  {author} {\bibinfo {author} {\bibfnamefont {S.~J.}\ \bibnamefont
  {Evered}}, \bibinfo {author} {\bibfnamefont {D.}~\bibnamefont {Bluvstein}},
  \bibinfo {author} {\bibfnamefont {M.}~\bibnamefont {Kalinowski}}, \bibinfo
  {author} {\bibfnamefont {S.}~\bibnamefont {Ebadi}}, \bibinfo {author}
  {\bibfnamefont {T.}~\bibnamefont {Manovitz}}, \bibinfo {author}
  {\bibfnamefont {H.}~\bibnamefont {Zhou}}, \bibinfo {author} {\bibfnamefont
  {S.~H.}\ \bibnamefont {Li}}, \bibinfo {author} {\bibfnamefont {A.~A.}\
  \bibnamefont {Geim}}, \bibinfo {author} {\bibfnamefont {T.~T.}\ \bibnamefont
  {Wang}}, \bibinfo {author} {\bibfnamefont {N.}~\bibnamefont {Maskara}},
  \bibinfo {author} {\bibfnamefont {H.}~\bibnamefont {Levine}}, \bibinfo
  {author} {\bibfnamefont {G.}~\bibnamefont {Semeghini}}, \bibinfo {author}
  {\bibfnamefont {M.}~\bibnamefont {Greiner}}, \bibinfo {author} {\bibfnamefont
  {V.}~\bibnamefont {Vuleti{\'{c}}}}, \ and\ \bibinfo {author} {\bibfnamefont
  {M.~D.}\ \bibnamefont {Lukin}},\ }\href {\doibase 10.1038/s41586-023-06481-y}
  {\bibfield  {journal} {\bibinfo  {journal} {Nature}\ }\textbf {\bibinfo
  {volume} {622}},\ \bibinfo {pages} {268} (\bibinfo {year}
  {2023})}\BibitemShut {NoStop}%
\bibitem [{\citenamefont {Bluvstein}\ \emph {et~al.}(2024)\citenamefont
  {Bluvstein}, \citenamefont {Evered}, \citenamefont {Geim}, \citenamefont
  {Li}, \citenamefont {Zhou}, \citenamefont {Manovitz}, \citenamefont {Ebadi},
  \citenamefont {Cain}, \citenamefont {Kalinowski}, \citenamefont {Hangleiter},
  \citenamefont {Bonilla~Ataides}, \citenamefont {Maskara}, \citenamefont
  {Cong}, \citenamefont {Gao}, \citenamefont {Sales~Rodriguez}, \citenamefont
  {Karolyshyn}, \citenamefont {Semeghini}, \citenamefont {Gullans},
  \citenamefont {Greiner}, \citenamefont {Vuleti{\'{c}}},\ and\ \citenamefont
  {Lukin}}]{Bluvstein2024Nature}%
  \BibitemOpen
  \bibfield  {author} {\bibinfo {author} {\bibfnamefont {D.}~\bibnamefont
  {Bluvstein}}, \bibinfo {author} {\bibfnamefont {S.~J.}\ \bibnamefont
  {Evered}}, \bibinfo {author} {\bibfnamefont {A.~A.}\ \bibnamefont {Geim}},
  \bibinfo {author} {\bibfnamefont {S.~H.}\ \bibnamefont {Li}}, \bibinfo
  {author} {\bibfnamefont {H.}~\bibnamefont {Zhou}}, \bibinfo {author}
  {\bibfnamefont {T.}~\bibnamefont {Manovitz}}, \bibinfo {author}
  {\bibfnamefont {S.}~\bibnamefont {Ebadi}}, \bibinfo {author} {\bibfnamefont
  {M.}~\bibnamefont {Cain}}, \bibinfo {author} {\bibfnamefont {M.}~\bibnamefont
  {Kalinowski}}, \bibinfo {author} {\bibfnamefont {D.}~\bibnamefont
  {Hangleiter}}, \bibinfo {author} {\bibfnamefont {J.~P.}\ \bibnamefont
  {Bonilla~Ataides}}, \bibinfo {author} {\bibfnamefont {N.}~\bibnamefont
  {Maskara}}, \bibinfo {author} {\bibfnamefont {I.}~\bibnamefont {Cong}},
  \bibinfo {author} {\bibfnamefont {X.}~\bibnamefont {Gao}}, \bibinfo {author}
  {\bibfnamefont {P.}~\bibnamefont {Sales~Rodriguez}}, \bibinfo {author}
  {\bibfnamefont {T.}~\bibnamefont {Karolyshyn}}, \bibinfo {author}
  {\bibfnamefont {G.}~\bibnamefont {Semeghini}}, \bibinfo {author}
  {\bibfnamefont {M.~J.}\ \bibnamefont {Gullans}}, \bibinfo {author}
  {\bibfnamefont {M.}~\bibnamefont {Greiner}}, \bibinfo {author} {\bibfnamefont
  {V.}~\bibnamefont {Vuleti{\'{c}}}}, \ and\ \bibinfo {author} {\bibfnamefont
  {M.~D.}\ \bibnamefont {Lukin}},\ }\href {\doibase 10.1038/s41586-023-06927-3}
  {\bibfield  {journal} {\bibinfo  {journal} {Nature}\ }\textbf {\bibinfo
  {volume} {626}},\ \bibinfo {pages} {58} (\bibinfo {year} {2024})}\BibitemShut
  {NoStop}%
\bibitem [{\citenamefont {Shah}\ \emph {et~al.}(2024)\citenamefont {Shah},
  \citenamefont {Patti}, \citenamefont {Rubies-Bigorda},\ and\ \citenamefont
  {Yelin}}]{Shah2024PRA}%
  \BibitemOpen
  \bibfield  {author} {\bibinfo {author} {\bibfnamefont {F.}~\bibnamefont
  {Shah}}, \bibinfo {author} {\bibfnamefont {T.~L.}\ \bibnamefont {Patti}},
  \bibinfo {author} {\bibfnamefont {O.}~\bibnamefont {Rubies-Bigorda}}, \ and\
  \bibinfo {author} {\bibfnamefont {S.~F.}\ \bibnamefont {Yelin}},\ }\href
  {\doibase 10.1103/PhysRevA.109.012613} {\bibfield  {journal} {\bibinfo
  {journal} {Phys. Rev. A}\ }\textbf {\bibinfo {volume} {109}},\ \bibinfo
  {pages} {012613} (\bibinfo {year} {2024})}\BibitemShut {NoStop}%
\bibitem [{\citenamefont {Nishad}\ \emph {et~al.}(2023)\citenamefont {Nishad},
  \citenamefont {Keselman}, \citenamefont {Lahaye}, \citenamefont {Browaeys},\
  and\ \citenamefont {Tsesses}}]{Nishad2023PRA}%
  \BibitemOpen
  \bibfield  {author} {\bibinfo {author} {\bibfnamefont {N.}~\bibnamefont
  {Nishad}}, \bibinfo {author} {\bibfnamefont {A.}~\bibnamefont {Keselman}},
  \bibinfo {author} {\bibfnamefont {T.}~\bibnamefont {Lahaye}}, \bibinfo
  {author} {\bibfnamefont {A.}~\bibnamefont {Browaeys}}, \ and\ \bibinfo
  {author} {\bibfnamefont {S.}~\bibnamefont {Tsesses}},\ }\href {\doibase
  10.1103/PhysRevA.108.053318} {\bibfield  {journal} {\bibinfo  {journal}
  {Phys. Rev. A}\ }\textbf {\bibinfo {volume} {108}},\ \bibinfo {pages}
  {053318} (\bibinfo {year} {2023})}\BibitemShut {NoStop}%
\bibitem [{\citenamefont {Chen}\ \emph {et~al.}(2023)\citenamefont {Chen},
  \citenamefont {Bornet}, \citenamefont {Bintz}, \citenamefont {Emperauger},
  \citenamefont {Leclerc}, \citenamefont {Liu}, \citenamefont {Scholl},
  \citenamefont {Barredo}, \citenamefont {Hauschild}, \citenamefont
  {Chatterjee}, \citenamefont {Schuler}, \citenamefont {L{\"a}uchli},
  \citenamefont {Zaletel}, \citenamefont {Lahaye}, \citenamefont {Yao},\ and\
  \citenamefont {Browaeys}}]{Chen2023Nature}%
  \BibitemOpen
  \bibfield  {author} {\bibinfo {author} {\bibfnamefont {C.}~\bibnamefont
  {Chen}}, \bibinfo {author} {\bibfnamefont {G.}~\bibnamefont {Bornet}},
  \bibinfo {author} {\bibfnamefont {M.}~\bibnamefont {Bintz}}, \bibinfo
  {author} {\bibfnamefont {G.}~\bibnamefont {Emperauger}}, \bibinfo {author}
  {\bibfnamefont {L.}~\bibnamefont {Leclerc}}, \bibinfo {author} {\bibfnamefont
  {V.~S.}\ \bibnamefont {Liu}}, \bibinfo {author} {\bibfnamefont
  {P.}~\bibnamefont {Scholl}}, \bibinfo {author} {\bibfnamefont
  {D.}~\bibnamefont {Barredo}}, \bibinfo {author} {\bibfnamefont
  {J.}~\bibnamefont {Hauschild}}, \bibinfo {author} {\bibfnamefont
  {S.}~\bibnamefont {Chatterjee}}, \bibinfo {author} {\bibfnamefont
  {M.}~\bibnamefont {Schuler}}, \bibinfo {author} {\bibfnamefont {A.~M.}\
  \bibnamefont {L{\"a}uchli}}, \bibinfo {author} {\bibfnamefont {M.~P.}\
  \bibnamefont {Zaletel}}, \bibinfo {author} {\bibfnamefont {T.}~\bibnamefont
  {Lahaye}}, \bibinfo {author} {\bibfnamefont {N.~Y.}\ \bibnamefont {Yao}}, \
  and\ \bibinfo {author} {\bibfnamefont {A.}~\bibnamefont {Browaeys}},\ }\href
  {\doibase 10.1038/s41586-023-05859-2} {\bibfield  {journal} {\bibinfo
  {journal} {Nature}\ }\textbf {\bibinfo {volume} {616}},\ \bibinfo {pages}
  {691} (\bibinfo {year} {2023})}\BibitemShut {NoStop}%
\bibitem [{\citenamefont {Shaw}\ \emph {et~al.}(2024)\citenamefont {Shaw},
  \citenamefont {Chen}, \citenamefont {Choi}, \citenamefont {Mark},
  \citenamefont {Scholl}, \citenamefont {Finkelstein}, \citenamefont {Elben},
  \citenamefont {Choi},\ and\ \citenamefont {Endres}}]{Shaw2024Nature}%
  \BibitemOpen
  \bibfield  {author} {\bibinfo {author} {\bibfnamefont {A.~L.}\ \bibnamefont
  {Shaw}}, \bibinfo {author} {\bibfnamefont {Z.}~\bibnamefont {Chen}}, \bibinfo
  {author} {\bibfnamefont {J.}~\bibnamefont {Choi}}, \bibinfo {author}
  {\bibfnamefont {D.~K.}\ \bibnamefont {Mark}}, \bibinfo {author}
  {\bibfnamefont {P.}~\bibnamefont {Scholl}}, \bibinfo {author} {\bibfnamefont
  {R.}~\bibnamefont {Finkelstein}}, \bibinfo {author} {\bibfnamefont
  {A.}~\bibnamefont {Elben}}, \bibinfo {author} {\bibfnamefont
  {S.}~\bibnamefont {Choi}}, \ and\ \bibinfo {author} {\bibfnamefont
  {M.}~\bibnamefont {Endres}},\ }\href {\doibase 10.1038/s41586-024-07173-x}
  {\bibfield  {journal} {\bibinfo  {journal} {Nature}\ }\textbf {\bibinfo
  {volume} {628}},\ \bibinfo {pages} {71} (\bibinfo {year} {2024})}\BibitemShut
  {NoStop}%
\bibitem [{\citenamefont {Eckner}\ \emph {et~al.}(2023)\citenamefont {Eckner},
  \citenamefont {Darkwah~Oppong}, \citenamefont {Cao}, \citenamefont {Young},
  \citenamefont {Milner}, \citenamefont {Robinson}, \citenamefont {Ye},\ and\
  \citenamefont {Kaufman}}]{Eckner2023Nature}%
  \BibitemOpen
  \bibfield  {author} {\bibinfo {author} {\bibfnamefont {W.~J.}\ \bibnamefont
  {Eckner}}, \bibinfo {author} {\bibfnamefont {N.}~\bibnamefont
  {Darkwah~Oppong}}, \bibinfo {author} {\bibfnamefont {A.}~\bibnamefont {Cao}},
  \bibinfo {author} {\bibfnamefont {A.~W.}\ \bibnamefont {Young}}, \bibinfo
  {author} {\bibfnamefont {W.~R.}\ \bibnamefont {Milner}}, \bibinfo {author}
  {\bibfnamefont {J.~M.}\ \bibnamefont {Robinson}}, \bibinfo {author}
  {\bibfnamefont {J.}~\bibnamefont {Ye}}, \ and\ \bibinfo {author}
  {\bibfnamefont {A.~M.}\ \bibnamefont {Kaufman}},\ }\href {\doibase
  10.1038/s41586-023-06360-6} {\bibfield  {journal} {\bibinfo  {journal}
  {Nature}\ }\textbf {\bibinfo {volume} {621}},\ \bibinfo {pages} {734}
  (\bibinfo {year} {2023})}\BibitemShut {NoStop}%
\bibitem [{\citenamefont {Bornet}\ \emph {et~al.}(2023)\citenamefont {Bornet},
  \citenamefont {Emperauger}, \citenamefont {Chen}, \citenamefont {Ye},
  \citenamefont {Block}, \citenamefont {Bintz}, \citenamefont {Boyd},
  \citenamefont {Barredo}, \citenamefont {Comparin}, \citenamefont {Mezzacapo},
  \citenamefont {Roscilde}, \citenamefont {Lahaye}, \citenamefont {Yao},\ and\
  \citenamefont {Browaeys}}]{Bornet2023Nature}%
  \BibitemOpen
  \bibfield  {author} {\bibinfo {author} {\bibfnamefont {G.}~\bibnamefont
  {Bornet}}, \bibinfo {author} {\bibfnamefont {G.}~\bibnamefont {Emperauger}},
  \bibinfo {author} {\bibfnamefont {C.}~\bibnamefont {Chen}}, \bibinfo {author}
  {\bibfnamefont {B.}~\bibnamefont {Ye}}, \bibinfo {author} {\bibfnamefont
  {M.}~\bibnamefont {Block}}, \bibinfo {author} {\bibfnamefont
  {M.}~\bibnamefont {Bintz}}, \bibinfo {author} {\bibfnamefont {J.~A.}\
  \bibnamefont {Boyd}}, \bibinfo {author} {\bibfnamefont {D.}~\bibnamefont
  {Barredo}}, \bibinfo {author} {\bibfnamefont {T.}~\bibnamefont {Comparin}},
  \bibinfo {author} {\bibfnamefont {F.}~\bibnamefont {Mezzacapo}}, \bibinfo
  {author} {\bibfnamefont {T.}~\bibnamefont {Roscilde}}, \bibinfo {author}
  {\bibfnamefont {T.}~\bibnamefont {Lahaye}}, \bibinfo {author} {\bibfnamefont
  {N.~Y.}\ \bibnamefont {Yao}}, \ and\ \bibinfo {author} {\bibfnamefont
  {A.}~\bibnamefont {Browaeys}},\ }\href {\doibase 10.1038/s41586-023-06414-9}
  {\bibfield  {journal} {\bibinfo  {journal} {Nature}\ }\textbf {\bibinfo
  {volume} {621}},\ \bibinfo {pages} {728} (\bibinfo {year}
  {2023})}\BibitemShut {NoStop}%
\bibitem [{\citenamefont {Ba\ss{}ler}\ \emph {et~al.}(2024)\citenamefont
  {Ba\ss{}ler}, \citenamefont {Aiello}, \citenamefont {Schmidt}, \citenamefont
  {Genes},\ and\ \citenamefont {Reitz}}]{Bassler2024PRL}%
  \BibitemOpen
  \bibfield  {author} {\bibinfo {author} {\bibfnamefont {N.~S.}\ \bibnamefont
  {Ba\ss{}ler}}, \bibinfo {author} {\bibfnamefont {A.}~\bibnamefont {Aiello}},
  \bibinfo {author} {\bibfnamefont {K.~P.}\ \bibnamefont {Schmidt}}, \bibinfo
  {author} {\bibfnamefont {C.}~\bibnamefont {Genes}}, \ and\ \bibinfo {author}
  {\bibfnamefont {M.}~\bibnamefont {Reitz}},\ }\href {\doibase
  10.1103/PhysRevLett.132.043602} {\bibfield  {journal} {\bibinfo  {journal}
  {Phys. Rev. Lett.}\ }\textbf {\bibinfo {volume} {132}},\ \bibinfo {pages}
  {043602} (\bibinfo {year} {2024})}\BibitemShut {NoStop}%
\bibitem [{\citenamefont {Sch\"affner}\ \emph {et~al.}(2024)\citenamefont
  {Sch\"affner}, \citenamefont {Schreiber}, \citenamefont {Lenz}, \citenamefont
  {Schlosser},\ and\ \citenamefont {Birkl}}]{Schaffner2024PRXQuantum}%
  \BibitemOpen
  \bibfield  {author} {\bibinfo {author} {\bibfnamefont {D.}~\bibnamefont
  {Sch\"affner}}, \bibinfo {author} {\bibfnamefont {T.}~\bibnamefont
  {Schreiber}}, \bibinfo {author} {\bibfnamefont {F.}~\bibnamefont {Lenz}},
  \bibinfo {author} {\bibfnamefont {M.}~\bibnamefont {Schlosser}}, \ and\
  \bibinfo {author} {\bibfnamefont {G.}~\bibnamefont {Birkl}},\ }\href
  {\doibase 10.1103/PRXQuantum.5.010311} {\bibfield  {journal} {\bibinfo
  {journal} {PRX Quantum}\ }\textbf {\bibinfo {volume} {5}},\ \bibinfo {pages}
  {010311} (\bibinfo {year} {2024})}\BibitemShut {NoStop}%
\bibitem [{\citenamefont {Kleinbeck}\ \emph {et~al.}(2023)\citenamefont
  {Kleinbeck}, \citenamefont {Busche}, \citenamefont {Stiesdal}, \citenamefont
  {Hofferberth}, \citenamefont {M\o{}lmer},\ and\ \citenamefont
  {B\"uchler}}]{PhysRevA.107.013717}%
  \BibitemOpen
  \bibfield  {author} {\bibinfo {author} {\bibfnamefont {K.}~\bibnamefont
  {Kleinbeck}}, \bibinfo {author} {\bibfnamefont {H.}~\bibnamefont {Busche}},
  \bibinfo {author} {\bibfnamefont {N.}~\bibnamefont {Stiesdal}}, \bibinfo
  {author} {\bibfnamefont {S.}~\bibnamefont {Hofferberth}}, \bibinfo {author}
  {\bibfnamefont {K.}~\bibnamefont {M\o{}lmer}}, \ and\ \bibinfo {author}
  {\bibfnamefont {H.~P.}\ \bibnamefont {B\"uchler}},\ }\href {\doibase
  10.1103/PhysRevA.107.013717} {\bibfield  {journal} {\bibinfo  {journal}
  {Phys. Rev. A}\ }\textbf {\bibinfo {volume} {107}},\ \bibinfo {pages}
  {013717} (\bibinfo {year} {2023})}\BibitemShut {NoStop}%
\bibitem [{\citenamefont {Lund}\ \emph {et~al.}(2024)\citenamefont {Lund},
  \citenamefont {Yang}, \citenamefont {Christiansen}, \citenamefont
  {Kornovan},\ and\ \citenamefont {M{\o}lmer}}]{lund2024subtraction}%
  \BibitemOpen
  \bibfield  {author} {\bibinfo {author} {\bibfnamefont {M.~M.}\ \bibnamefont
  {Lund}}, \bibinfo {author} {\bibfnamefont {F.}~\bibnamefont {Yang}}, \bibinfo
  {author} {\bibfnamefont {V.~R.}\ \bibnamefont {Christiansen}}, \bibinfo
  {author} {\bibfnamefont {D.}~\bibnamefont {Kornovan}}, \ and\ \bibinfo
  {author} {\bibfnamefont {K.}~\bibnamefont {M{\o}lmer}},\ }\href
  {https://arxiv.org/abs/2404.12328} {\bibfield  {journal} {\bibinfo  {journal}
  {arXiv:2404.12328}\ } (\bibinfo {year} {2024})}\BibitemShut {NoStop}%
\bibitem [{\citenamefont {Mahmoodian}\ \emph {et~al.}(2020)\citenamefont
  {Mahmoodian}, \citenamefont {Calaj\'o}, \citenamefont {Chang}, \citenamefont
  {Hammerer},\ and\ \citenamefont {S\o{}rensen}}]{Mohmoodian2020PRX}%
  \BibitemOpen
  \bibfield  {author} {\bibinfo {author} {\bibfnamefont {S.}~\bibnamefont
  {Mahmoodian}}, \bibinfo {author} {\bibfnamefont {G.}~\bibnamefont
  {Calaj\'o}}, \bibinfo {author} {\bibfnamefont {D.~E.}\ \bibnamefont {Chang}},
  \bibinfo {author} {\bibfnamefont {K.}~\bibnamefont {Hammerer}}, \ and\
  \bibinfo {author} {\bibfnamefont {A.~S.}\ \bibnamefont {S\o{}rensen}},\
  }\href {\doibase 10.1103/PhysRevX.10.031011} {\bibfield  {journal} {\bibinfo
  {journal} {Phys. Rev. X}\ }\textbf {\bibinfo {volume} {10}},\ \bibinfo
  {pages} {031011} (\bibinfo {year} {2020})}\BibitemShut {NoStop}%
\bibitem [{\citenamefont {Iversen}\ and\ \citenamefont
  {Pohl}(2021)}]{Iversen2021PRL}%
  \BibitemOpen
  \bibfield  {author} {\bibinfo {author} {\bibfnamefont {O.~A.}\ \bibnamefont
  {Iversen}}\ and\ \bibinfo {author} {\bibfnamefont {T.}~\bibnamefont {Pohl}},\
  }\href {\doibase 10.1103/PhysRevLett.126.083605} {\bibfield  {journal}
  {\bibinfo  {journal} {Phys. Rev. Lett.}\ }\textbf {\bibinfo {volume} {126}},\
  \bibinfo {pages} {083605} (\bibinfo {year} {2021})}\BibitemShut {NoStop}%
\bibitem [{\citenamefont {Iversen}\ and\ \citenamefont
  {Pohl}(2022)}]{Iversen2022PRR}%
  \BibitemOpen
  \bibfield  {author} {\bibinfo {author} {\bibfnamefont {O.~A.}\ \bibnamefont
  {Iversen}}\ and\ \bibinfo {author} {\bibfnamefont {T.}~\bibnamefont {Pohl}},\
  }\href {\doibase 10.1103/PhysRevResearch.4.023002} {\bibfield  {journal}
  {\bibinfo  {journal} {Phys. Rev. Res.}\ }\textbf {\bibinfo {volume} {4}},\
  \bibinfo {pages} {023002} (\bibinfo {year} {2022})}\BibitemShut {NoStop}%
\bibitem [{\citenamefont {Liedl}\ \emph {et~al.}(2024)\citenamefont {Liedl},
  \citenamefont {Tebbenjohanns}, \citenamefont {Bach}, \citenamefont {Pucher},
  \citenamefont {Rauschenbeutel},\ and\ \citenamefont
  {Schneeweiss}}]{PhysRevX.14.011020}%
  \BibitemOpen
  \bibfield  {author} {\bibinfo {author} {\bibfnamefont {C.}~\bibnamefont
  {Liedl}}, \bibinfo {author} {\bibfnamefont {F.}~\bibnamefont
  {Tebbenjohanns}}, \bibinfo {author} {\bibfnamefont {C.}~\bibnamefont {Bach}},
  \bibinfo {author} {\bibfnamefont {S.}~\bibnamefont {Pucher}}, \bibinfo
  {author} {\bibfnamefont {A.}~\bibnamefont {Rauschenbeutel}}, \ and\ \bibinfo
  {author} {\bibfnamefont {P.}~\bibnamefont {Schneeweiss}},\ }\href {\doibase
  10.1103/PhysRevX.14.011020} {\bibfield  {journal} {\bibinfo  {journal} {Phys.
  Rev. X}\ }\textbf {\bibinfo {volume} {14}},\ \bibinfo {pages} {011020}
  (\bibinfo {year} {2024})}\BibitemShut {NoStop}%
\bibitem [{\citenamefont {Cardenas-Lopez}\ \emph {et~al.}(2023)\citenamefont
  {Cardenas-Lopez}, \citenamefont {Masson}, \citenamefont {Zager},\ and\
  \citenamefont {Asenjo-Garcia}}]{PhysRevLett.131.033605}%
  \BibitemOpen
  \bibfield  {author} {\bibinfo {author} {\bibfnamefont {S.}~\bibnamefont
  {Cardenas-Lopez}}, \bibinfo {author} {\bibfnamefont {S.~J.}\ \bibnamefont
  {Masson}}, \bibinfo {author} {\bibfnamefont {Z.}~\bibnamefont {Zager}}, \
  and\ \bibinfo {author} {\bibfnamefont {A.}~\bibnamefont {Asenjo-Garcia}},\
  }\href {\doibase 10.1103/PhysRevLett.131.033605} {\bibfield  {journal}
  {\bibinfo  {journal} {Phys. Rev. Lett.}\ }\textbf {\bibinfo {volume} {131}},\
  \bibinfo {pages} {033605} (\bibinfo {year} {2023})}\BibitemShut {NoStop}%
\end{thebibliography}%
\end{document}